\documentclass{nature3}
\usepackage{graphicx}
\usepackage[version=4]{mhchem} 
\usepackage{float}

\usepackage{subfigure}

\usepackage{titlesec}
\titleformat{\section}{\normalfont\large\bfseries}{\thesection.}{3pt}{\space}[]
\titlespacing*{\section}{0em}{1ex}{1em}[0em]

\titleformat{\subsection}{\bfseries}{\thesubsection.}{3pt}{\space}[]
\titlespacing*{\subsection}{0em}{1ex}{1em}[0em]

\usepackage[symbol]{footmisc}

\usepackage[colorlinks,breaklinks=true,plainpages=false,citecolor=blue,linkcolor=blue,urlcolor=blue,bookmarksopen=true,bookmarksnumbered=false,bookmarksdepth=5]{hyperref}

\usepackage{amsmath}
\usepackage{amssymb}
\usepackage{amsbsy}

\newcommand{\project}[1]{\textsl{#1}}                               
\newcommand{\JWST}{\project{JWST}}                               
\newcommand{\HST}{\project{HST}} 
\newcommand{\Spitzer}{\project{Spitzer}}

\title{Nightside clouds and disequilibrium chemistry on the hot Jupiter WASP-43b}

\author{Taylor J. Bell$^{1,2}$,
Nicolas Crouzet$^{3}$,
Patricio E. Cubillos$^{4,5}$,
Laura Kreidberg$^{6}$,
Anjali A. A. Piette$^{7}$,
Michael T. Roman$^{8,9}$,
Joanna K. Barstow$^{10}$,
Jasmina Blecic$^{11,12}$,
Ludmila Carone$^{5}$,
Louis-Philippe Coulombe$^{13}$,
Elsa Ducrot$^{14}$,
Mark Hammond$^{15}$,
Jo\~ao M. Mendon\c ca$^{16}$,
Julianne I. Moses$^{17}$,
Vivien Parmentier$^{18}$,
Kevin B. Stevenson$^{19}$,
Lucas Teinturier$^{20,21}$,
Michael Zhang$^{22}$,
Natalie M. Batalha$^{23}$,
Jacob L. Bean$^{22}$,
Bj\"{o}rn Benneke$^{13}$,
Benjamin Charnay$^{20}$,
Katy L. Chubb$^{24}$,
Brice-Olivier Demory$^{25,26}$,
Peter Gao$^{7}$,
Elspeth K. H. Lee$^{25}$,
Mercedes López-Morales$^{27}$,
Giuseppe Morello$^{28,29,30}$,
Emily Rauscher$^{31}$,
David K. Sing$^{32,33}$,
Xianyu Tan$^{34,35,15}$,
Olivia Venot$^{36}$,
Hannah R. Wakeford$^{37}$,
Keshav Aggarwal$^{38}$,
Eva-Maria Ahrer$^{39,40}$,
Munazza K. Alam$^{7}$,
Robin Baeyens$^{41}$,
David Barrado$^{42}$,
Claudio Caceres$^{43,44,45}$,
Aarynn L. Carter$^{23}$,
Sarah L. Casewell$^{8}$,
Ryan C. Challener$^{31}$,
Ian J. M. Crossfield$^{46}$,
Leen Decin$^{47}$,
Jean-Michel D\'esert$^{41}$,
Ian Dobbs-Dixon$^{11}$,
Achr\`ene Dyrek$^{14}$,
N\'estor Espinoza$^{48,33}$,
Adina D. Feinstein$^{22,49}$,
Neale P. Gibson$^{50}$,
Joseph Harrington$^{51}$,
Christiane Helling$^{5}$,
Renyu Hu$^{52,53}$,
Nicolas Iro$^{54}$,
Eliza M.-R. Kempton$^{55}$,
Sarah Kendrew$^{56}$,
Thaddeus D. Komacek$^{55}$,
Jessica Krick$^{57}$,
Pierre-Olivier Lagage$^{14}$,
J\'er\'emy Leconte$^{58}$,
Monika Lendl$^{59}$,
Neil T. Lewis$^{60}$,
Joshua D. Lothringer$^{61}$,
Isaac Malsky$^{31}$,
Luigi Mancini$^{62,63,6}$,
Megan Mansfield$^{64}$,
Nathan J. Mayne$^{65}$,
Thomas Mikal-Evans$^{6}$,
Karan Molaverdikhani$^{66,67}$,
Nikolay K. Nikolov$^{48}$,
Matthew C. Nixon$^{55}$,
Enric Palle$^{28}$,
Dominique J. M. Petit dit de la Roche$^{59}$,
Caroline Piaulet$^{13}$,
Diana Powell$^{27}$,
Benjamin V.\ Rackham$^{68,69}$,
Aaron D. Schneider$^{47,70}$,
Maria E. Steinrueck$^{6}$,
Jake Taylor$^{15,13}$,
Luis Welbanks$^{71}$,
Sergei N. Yurchenko$^{72}$,
Xi Zhang$^{73}$,
Sebastian Zieba$^{6,3}$}

\begin{document}
\renewcommand{\figurename}{\hspace{-4pt}}
\renewcommand{\thefigure}{Fig.~\arabic{figure}}
\renewcommand{\theHfigure}{Fig.~\arabic{figure}}
\renewcommand{\tablename}{\hspace{-4pt}}
\renewcommand{\thetable}{Table \arabic{table}}
\renewcommand{\theHtable}{Table \arabic{table}}
\setcounter{figure}{0}
\setcounter{table}{0}

\maketitle

\begin{affiliations}
\item
BAER Institute, NASA Ames Research Center, Moffet Field, CA, USA
\item
Space Science and Astrobiology Division, NASA Ames Research Center, Moffett Field, CA, USA
\item
Leiden Observatory, University of Leiden, Leiden, The Netherlands
\item
INAF – Osservatorio Astrofisico di Torino, Pino Torinese, Italy
\item
Space Research Institute, Austrian Academy of Sciences, Graz, Austria
\item
Max Planck Institute for Astronomy, Heidelberg, Germany
\item
Earth and Planets Laboratory, Carnegie Institution for Science, Washington, DC, USA
\item
School of Physics and Astronomy, University of Leicester, Leicester, UK
\item
Universidad Adolfo Ibáñez: Peñalolén, Santiago, Chile
\item
School of Physical Sciences, The Open University, Milton Keynes, UK
\item
Department of Physics, New York University Abu Dhabi, Abu Dhabi, UAE
\item
Center for Astro, Particle and Planetary Physics (CAP3), New York University Abu Dhabi, Abu Dhabi, UAE
\item
Department of Physics and Trottier Institute for Research on Exoplanets, Université de Montréal, Montreal, QC, Canada
\item
Université Paris-Saclay, Université Paris Cité, CEA, CNRS, AIM, Gif-sur-Yvette, France
\item
Atmospheric, Oceanic and Planetary Physics, Department of Physics, University of Oxford, Oxford, UK
\item
DTU Space, Technical University of Denmark, Kgs. Lyngby, Denmark
\item
Space Science Institute, Boulder, CO, USA
\item
Université Côte d'Azur, Observatoire de la Côte d'Azur, CNRS, Laboratoire Lagrange, France
\item
Johns Hopkins APL, Laurel, MD, USA
\item
LESIA, Observatoire de Paris, Universit\'{e} PSL, Sorbonne Université, Universit\'{e} Paris Cit\'{e}, CNRS, Meudon, France
\item
Laboratoire de Météorologie Dynamique, IPSL, CNRS, Sorbonne Université, Ecole Normale Supérieure, Université PSL, Ecole Polytechnique, Institut Polytechnique de Paris, Paris, France
\item
Department of Astronomy \& Astrophysics, University of Chicago, Chicago, IL, USA
\item
Department of Astronomy \& Astrophysics, University of California, Santa Cruz, Santa Cruz, CA, USA
\item
Centre for Exoplanet Science, University of St Andrews, St Andrews, UK
\item
Center for Space and Habitability, University of Bern, Bern, Switzerland
\item
Space and Planetary Sciences, Institute of Physics, University of Bern, Bern, Switzerland
\item
Center for Astrophysics | Harvard \& Smithsonian, Cambridge, MA, USA
\item
Instituto de Astrofísica de Canarias (IAC), Tenerife, Spain
\item
INAF- Palermo Astronomical Observatory, Piazza del Parlamento, Palermo, Italy
\item
Department of Space, Earth and Environment, Chalmers University of Technology, Gothenburg, Sweden
\item
Department of Astronomy, University of Michigan, Ann Arbor, MI, USA
\item
Department of Earth and Planetary Sciences, Johns Hopkins University, Baltimore, MD, USA
\item
Department of Physics \& Astronomy, Johns Hopkins University, Baltimore, MD, USA
\item
Tsung-Dao Lee Institute, Shanghai Jiao Tong University, Shanghai, People’s Republic of China
\item
School of Physics and Astronomy, Shanghai Jiao Tong University, Shanghai, People’s Republic of China
\item
Université Paris Cité and Univ Paris Est Creteil, CNRS, LISA, Paris, France
\item
School of Physics, University of Bristol, Bristol, UK
\item
Indian Institute of Technology, Indore, India
\item
Centre for Exoplanets and Habitability, University of Warwick, Coventry, UK
\item
Department of Physics, University of Warwick, Coventry, UK
\item
Anton Pannekoek Institute for Astronomy, University of Amsterdam, Amsterdam, The Netherlands
\item
Centro de Astrobiología (CAB, CSIC-INTA), Departamento de Astrofísica, ESAC campus, Villanueva de la Cañada (Madrid), Spain
\item
Instituto de Astrofisica, Facultad Ciencias Exactas, Universidad Andres Bello, Santiago, Chile
\item
Centro de Astrofisica y Tecnologias Afines (CATA), Casilla 36-D, Santiago, Chile
\item
Nucleo Milenio de Formacion Planetaria (NPF), Chile
\item
Department of Physics \& Astronomy, University of Kansas, Lawrence, KS, USA
\item
Institute of Astronomy, Department of Physics and Astronomy, KU Leuven, Leuven, Belgium
\item
Space Telescope Science Institute, Baltimore, MD, USA
\item
Department of Astrophysical and Planetary Sciences, University of Colorado, Boulder, CO, USA
\item
School of Physics, Trinity College Dublin, Dublin, Ireland
\item
Planetary Sciences Group, Department of Physics and Florida Space Institute, University of Central Florida, Orlando, Florida, USA
\item
Astrophysics Section, Jet Propulsion Laboratory, California Institute of Technology, Pasadena, CA, USA
\item
Division of Geological and Planetary Sciences, California Institute of Technology, Pasadena, CA, USA
\item
Institute of Planetary Research - Extrasolar planets and atmospheres, German Aerospace Center (DLR), Berlin, Germany
\item
Department of Astronomy, University of Maryland, College Park, MD, USA
\item
European Space Agency, Space Telescope Science Institute, Baltimore, MD, USA
\item
California Institute of Technology, IPAC, Pasadena, CA, USA
\item
Laboratoire d'Astrophysique de Bordeaux, Université de Bordeaux, Pessac, France
\item
Département d'Astronomie, Université de Genève, Sauverny, Switzerland
\item
Department of Mathematics and Statistics, University of Exeter, Exeter, Devon, UK
\item
Department of Physics, Utah Valley University, Orem, UT, USA
\item
Department of Physics, University of Rome ``Tor Vergata", Rome, Italy
\item
INAF - Turin Astrophysical Observatory, Pino Torinese, Italy
\item
Steward Observatory, University of Arizona, Tucson, AZ, USA
\item
Department of Physics and Astronomy, Faculty of Environment, Science and Economy, University of Exeter, Exeter, UK
\item
Universitäts-Sternwarte, Ludwig-Maximilians-Universität München, München, Germany
\item
Exzellenzcluster Origins, Garching, Germany
\item
Department of Earth, Atmospheric and Planetary Sciences, Massachusetts Institute of Technology, Cambridge, MA, USA
\item
Kavli Institute for Astrophysics and Space Research, Massachusetts Institute of Technology, Cambridge, MA, USA
\item
Centre for ExoLife Sciences,  Niels Bohr Institute, Copenhagen, Denmark
\item
School of Earth and Space Exploration, Arizona State University, Tempe, AZ, USA
\item
Department of Physics and Astronomy, University College London, UK
\item
Department of Earth and Planetary Sciences, University of California Santa Cruz, Santa Cruz, California, USA
\end{affiliations}

\begin{abstract}
Hot Jupiters are among the best-studied exoplanets, but it is still poorly understood how their chemical composition and cloud properties vary with longitude. Theoretical models predict that clouds may condense on the nightside and that molecular abundances can be driven out of equilibrium by zonal winds. Here we report a phase-resolved emission spectrum of the hot Jupiter \mbox{WASP-43b} measured from $\pmb{5\textbf{--}12\,\mu}$m with JWST's Mid-Infrared Instrument (MIRI). The spectra reveal a large day--night temperature contrast (with average brightness temperatures of $\pmb{1524\pm35}$ and $\pmb{863\pm23}$\,Kelvin, respectively) and evidence for water absorption at all orbital phases. Comparisons with three-dimensional atmospheric models show that both the phase curve shape and emission spectra strongly suggest the presence of nightside clouds which become optically thick to thermal emission at pressures greater than $\pmb{\sim}$100\,mbar. The dayside is consistent with a cloudless atmosphere above the mid-infrared photosphere. Contrary to expectations from equilibrium chemistry but consistent with disequilibrium kinetics models, methane is not detected on the nightside (2$\pmb{\sigma}$ upper limit of 1--6 parts per million, depending on model assumptions).
\end{abstract}


\section*{Main Text}
\subsection*{Introduction}

Hot Jupiters are tidally synchronized to their host stars, with vast differences in irradiation between the dayside and nightside. Previous observations with the Hubble (HST) and Spitzer Space Telescopes show that these planets have cooler nightsides and weaker hotspot offsets than expected from cloud-free three-dimensional models\cite{Keating2019,Beatty2019,Kataria2015, mendonca2018a,Parmentier2018a}.
The main mechanism believed to be responsible for this behavior is the presence of nightside clouds, which would hide the thermal flux of the planet and lead to a sharp longitudinal gradient in brightness temperature\cite{Kataria2015,hu15, shporer15,mendonca2018a,Roman2021,tan2021a,Parmentier2021}. Other mechanisms have been proposed, such as the presence of atmospheric drag due to hydrodynamic instabilities or magnetic coupling\cite{perna10, Komacek2017,Rogers2017}, super-stellar atmospheric metallicity\cite{kataria14, zhang17}, or interaction between the deep winds and the photosphere\cite{Carone2020}, but these mechanisms are less universal than the cloud hypothesis\cite{Showman2020,Helling2021}.

\mbox{WASP-43b}, a hot Jupiter with an orbital period of just 19.5 hours\cite{Hellier2011}, is an ideal target for thermal phase curve observations. Its host star is a K7 main sequence star 87 pc away with metallicity close to solar and weak variability\cite{Scandariato2022}. Previous measurements of the planet's orbital phase curve in the near-infrared revealed a large temperature contrast between the day- and nightside hemispheres, broadly consistent with the presence of nightside clouds\cite{Stevenson2014, Kataria2015, Stevenson2017}, which could be composed of magnesium silicates (Mg$_2$SiO$_4$/MgSiO$_2$) and other minerals (e.g., MnS, Na$_2$S, metal oxides)\cite{VenotEtal2020-JWST-WASP-43b,helling2020}. Due to the low nightside flux, the exact temperature and cloud properties were challenging to determine from previous observations\cite{mendonca2018a, Morello2019, murphy2023}.  With the mid-infrared capabilities of JWST, we have the opportunity to measure the phase-resolved thermal spectrum with unprecedented sensitivity, particularly on the cold nightside. We observed a full orbit of \mbox{WASP-43b} in the 5--12~$\mu$m range with JWST MIRI\cite{Wright_2023} in Low Resolution Spectroscopy (LRS\cite{Kendrew2015}) slitless mode on December 1 and 2, 2022, as part of the Transiting Exoplanet Community Early Release Science Program (JWST-ERS-1366). This continuous observation lasted 26.5 hours at a cadence of 10.34 s (9216 integrations) and included a full phase curve with one transit and two eclipses. 

\subsection*{Results}

\begin{figure*}[!htbp]
    \centering
    \includegraphics[width=\textwidth]{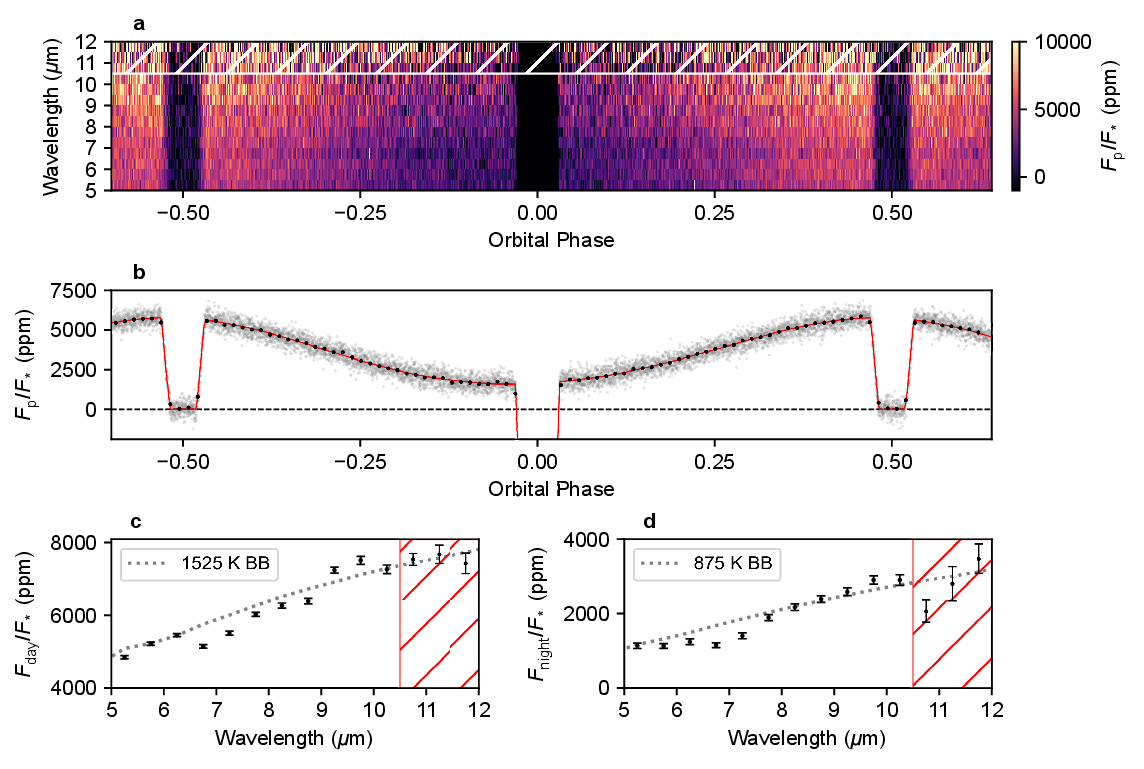}
    \caption{\textbf{A visualization of the observed light curves and the resulting emission spectra.} \textbf{a,} The observed spectroscopic light curves binned to a 0.5\,$\mu$m wavelength resolution and after systematic noise removal, following the \texttt{Eureka!}\ v1 methods. The first 779 integrations have been removed from this figure and our fits as they were impacted by strongly decreasing flux. Wavelengths longer than 10.5\,$\mu$m marked with a hatched region were affected by the ``shadowed region effect" (see Methods) and could not be reliably reduced. \textbf{b,} The observed band-integrated light curve after systematic noise removal (grey points) and binned data with a cadence of 15 minutes (black points, with error bars smaller than the point sizes), compared to the best-fitting astrophysical model (red line). The measured dayside (\textbf{c}) and nightside (\textbf{d}) emission spectra are shown with black points and 1$\sigma$ error bars, and blackbody curves (dotted line denoted as ``BB'', assuming a PHOENIX\cite{allard1995, hauschildt1999, husser2013} model for the star) are shown to emphasize planetary spectral features with blackbody temperatures estimated by eye to match the continuum flux levels. Wavelengths longer than 10.5\,$\mu$m were affected by the shadowed region effect and are unreliable.}
    \label{fig:data}
\end{figure*}

We performed multiple independent reductions and fits to these observations (see the Data Reduction Pipelines section and the Light Curve Fitting section in the Methods) to ensure robust conclusions. Our analyses all identified a strong systematic noise feature from 10.6--11.8\,$\mu$m, the source of which is still unclear, and we were unable to adequately detrend these 10.6--11.8\,$\mu$m data (see the ``Shadowed Region Effect'' section in the Methods). As shown in \ref{fig:binningUncertainties}, we also found that larger wavelength bins were required to accurately estimate our final spectral uncertainties (see the ``Spectral Binning'' section in the Methods). As a result, our final analyses only consider the 5--10.5\,$\mu$m data which we split into 11 channels with a constant 0.5~$\mu$m wavelength spacing.  Similar to MIRI commissioning observations, the data show a strong downward exponential ramp in the first ${\sim}$60 minutes and a weaker ramp throughout the observation\cite{Bouwman2023MiriTsoCommissioning} (see \ref{fig:rampAmplitudes}). To minimize correlations with the phase variations, we removed the initial strong ramp by excluding the first 779 integrations (134.2 minutes) and then fitted a single exponential ramp model to the remaining data. A single ramp effectively removed the systematic noise, with the broadband light curve showing scatter ${\sim}1.25\times$ the expected photon noise, while the spectroscopic light curves reach as low as ${\sim}1.1\times$ the photon limit, likely due to improved decorrelation of wavelength-dependent systematics. \ref{fig:data} shows the spectral light curves, broadband light curve, dayside spectra, and nightside spectra from our fiducial reduction and fit.

From our \texttt{Eureka!}\ v1 analysis (see Methods), we measure a broadband (5--10.5\,$\mu$m) peak-to-trough phase variation of 4180$\pm$33 parts per million (ppm) with an eclipse depth of 5752$\pm$19\,ppm and a nightside flux of 1636$\pm$37\,ppm. Assuming a PHOENIX stellar model and marginalizing over the published stellar and system parameters\cite{Esposito2017GapsWASP43b}, the broadband dayside brightness temperature is 1524$\pm$35\,K while the nightside is 863$\pm$23\,K. This corresponds to a day--night brightness temperature contrast of 659$\pm$19\,K, in agreement with the large contrasts previously observed\cite{Stevenson2014, Stevenson2017, mendonca2018a, Morello2019}. The phase variations are well-fitted by a sum of two sinusoids (the first and second harmonics), with two sinusoids preferred over a single sinusoid at 16$\sigma$ (see the Determining the Number of Sinusoid Harmonics section in the Methods) for the broadband light curve. The peak brightness of the broadband phase curve occurs at 7.34$\pm$0.38 degrees east from the substellar point (although individual reductions find offsets ranging from 7.34 to 9.60 degrees), while previous works have found offsets of 12.3$\pm$1.0\,$^{\circ}$E for \HST/WFC3's 1.1--1.7\,$\mu$m bandpass\cite{Stevenson2014}, offsets ranging from 4.4 to 12.2\,$^{\circ}$E for \Spitzer/IRAC's 3.6\,$\mu$m filter\cite{Stevenson2017,Morello2019}, and offsets ranging from 10.4 to 21.1\,$^{\circ}$E for \Spitzer/IRAC's 4.5\,$\mu$m filter\cite{Stevenson2017,mendonca2018a,Morello2019,May2020_wasp43b,Bell2021,murphy2023}.
Overall, these broadband data represent roughly an order of magnitude in improved precision on the eclipse depth (6$\times$), phase curve amplitude (6$\times$), and phase curve offset (10$\times$) over individual \Spitzer/IRAC 4.5\,$\mu$m observations of the system\cite{Stevenson2017,Bell2021,murphy2023}; this improvement is largely driven by \JWST's larger mirror (45${\times}$), about $12{\times}$ less pointing jitter (per-axis), about $4{\times}$ improved PSF-width stability (per-axis), and MIRI's much broader bandpass.

\subsubsection*{Model Interpretation}

To interpret the measurements, we compared the observations to synthetic phase curves and emission spectra derived from general circulation models (GCMs). Simulations were gathered from five different modelling groups, amounting to thirty-one separate GCM realisations exploring a range of approaches and assumptions. Notably, in addition to cloud-free simulations, a majority of the GCMs modelled clouds with spatial distributions that were either fully predicted\cite{showman2009,Parmentier2018a,murphy2023} or simply limited to the planet's nightside\cite{mendonca2018a}.  For the predictive cloud models, simulations favored warmer, clearer daysides with cooler, cloudier nightsides, but the precise distributions varied with assumptions regarding cloud physics and compositions. In general, models with smaller cloud particles or extended vertical distributions tended to produce thicker clouds at the pressures sensed by the observations. Details of the different models are provided in the Methods section. 

\begin{figure}[!b]
    \centering
    \includegraphics[width=0.80\textwidth]{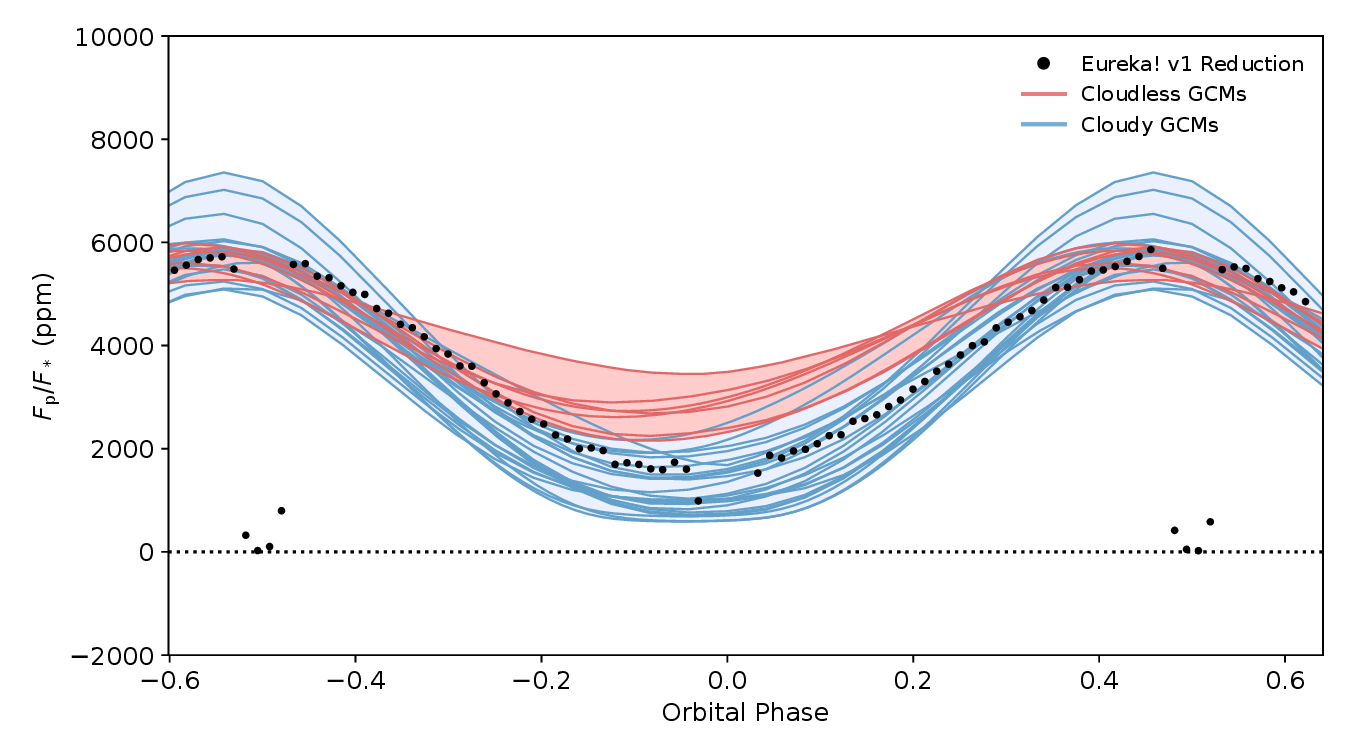}
    \caption{\textbf{A comparison of the observed 5--10.5~$\pmb{\mu}$m light curve to GCM simulations.} The black points show the temporally binned broadband light curve. The solid lines represent modelled phase curves derived from the 31 GCM simulations, integrated over the same wavelength range as the data, and separated into two groups based on the inclusion of clouds. The cloudless GCMs (red lines) simulated completely cloud-free skies, whereas the cloudy GCMs (blue lines) included at least some clouds on the nightside of the planet. The red and blue shaded areas span the range of all the cloudless and cloudy simulations, respectively, with the spread of values owing to differences in the various model assumptions and parameterisations. On average, the cloudless GCM phase curves have a maximum planet-to-star flux ratio of 5703 ppm and a minimum of 2681 ppm. This matches the observed maximum of the phase curve well but does not match its observed minimum at 1636$\pm$37 ppm. On average, the cloudy GCM phase curves have a maximum of 5866 ppm and a minimum of 1201 ppm, in better agreement with the observed night-side emission, but their spread of maximum values is much larger than the cloudless simulations. The cloudy models are able to suppress the nightside emission and better match the data; however, not all cloud models fit equally well and those with the optically thickest nightside clouds suppress too much emission. The models do not include the eclipse signals (phases -0.5 and 0.5) or transit signal (phase 0.0).}\label{fig:gcmPCs}
\end{figure}

Despite fundamental differences in the models and the parameterisations they employ, simulated phase curves derived from models that include cloud opacity on the planet's nightside provide a better match to the observed nightside flux compared to the clear simulations (see \ref{fig:gcmPCs}). In contrast, the observed dayside fluxes (180$^{\circ}$ orbital phase) were matched similarly well by models with and without clouds.  This implies the presence of widespread clouds preferentially on the planet's nightside with cloud optical thicknesses sufficient to suppress thermal emission and cool the thermal photosphere. Specifically, models with integrated mid-infrared cloud opacities of roughly 2--4 above the 300-mbar level (i.e., blocking ${\sim}$87--98\% of the underlying emission), best match the observed nightside flux.

Including nightside clouds also improved the agreement with the measured hot spot offset (7.34$\,\pm\,$0.38 degrees east). While cloudless models all produced eastward offsets greater than 16.6$^{\circ}$ (25.5$^{\circ}$ on average), simulations with clouds had offsets as low as 7.6$^{\circ}$ (with a mean of 16.4$^{\circ}$). These reduced offsets were associated with decreases in the eastward jet speeds of up to several km s$^{-1}$, with maximum winds of roughly 2.0-2.5 km s$^{-1}$ providing the best match (See \ref{tab:gcm_list_full} for further details). This modelled jet-speed reduction is likely due to a disruption in the equator-ward momentum transport\cite{Showman2011} brought about by nightside clouds\cite{mendonca2018a,Roman&Rauscher2017,Roman&Rauscher2019}. However, the resulting range of offsets seen in the suite of models suggests that this mechanism is quite sensitive to the details of cloud models, and other modelling factors (e.g. atmospheric drag\cite{perna10,Komacek2017,Carone2020}, radiative time scales\cite{cowan2012thermal,kataria14, zhang17}) likely still play an important role.

\begin{figure*}
    \centering
    \includegraphics[width=0.795\textwidth, trim=.0in 0.5in 0in 0in]{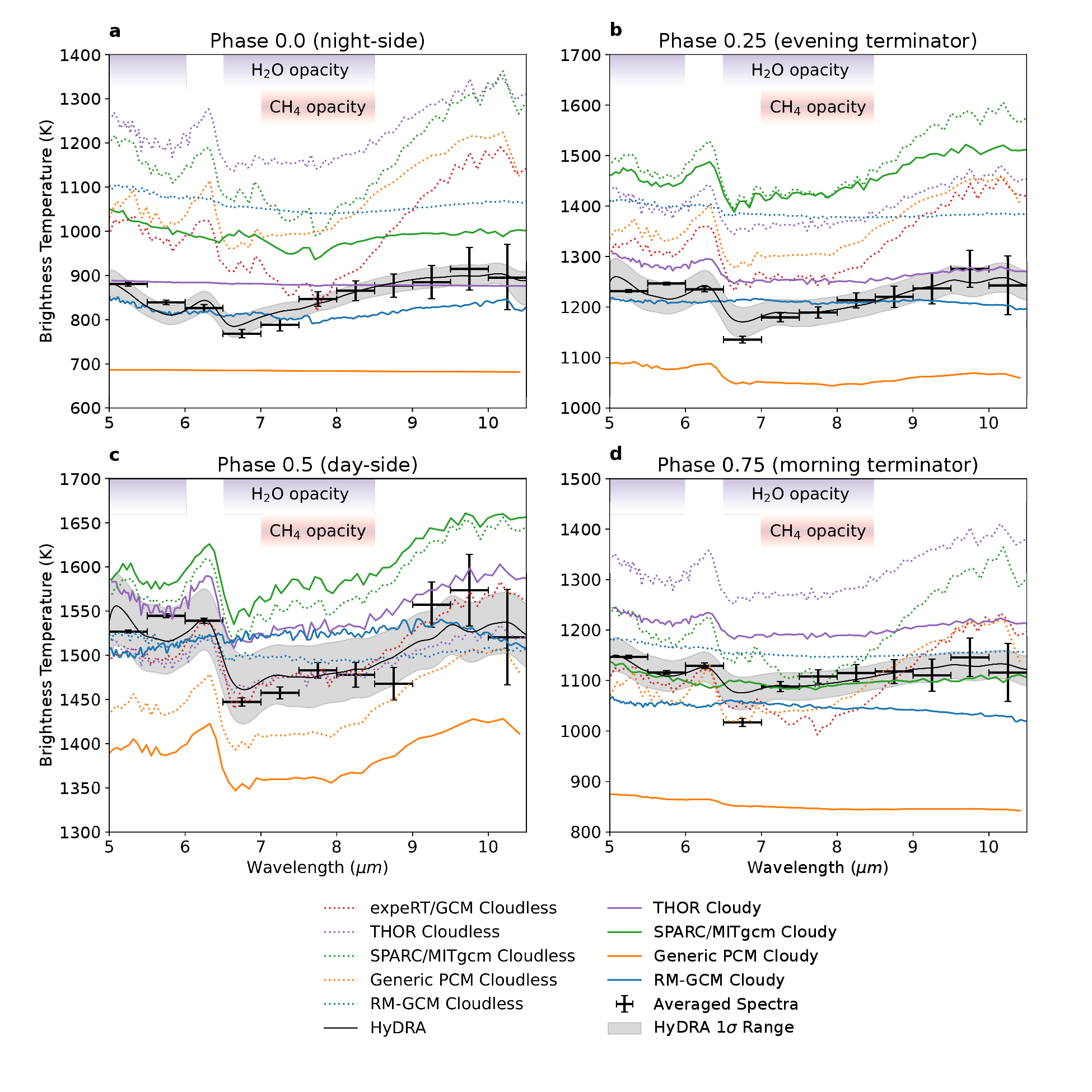}
    \caption{\textbf{A comparison of the observed and modelled spectra at different phases.} The panels show the observed emission spectrum with 1$\sigma$ error bars at phases 0.0 (\textbf{a}), 0.25 (\textbf{b}), 0.5 (\textbf{c}) and 0.75 (\textbf{d}), along with select modelled spectra derived from different cloudy and cloudless GCMs (described in the Methods section, and listed in \ref{tab:gcm_list_full}). Although absolute brightness temperatures differ significantly between models owing to various GCM assumptions, differences in the relative shape of the spectra are strongly dependent on the cloud and temperature structure found in the GCMs (see \ref{fig:compare_gcm_retrieval}). Models with more isothermal profiles (i.e., RM-GCM) or thick clouds at pressures of $\lesssim$10--100 mbar (i.e., THOR-cloudy, Generic PCM with 0.1-$\mu$m cloud particles) produce flatter spectra, while clearer skies yield stronger absorption features.  The observed spectra from the nightside and terminators appear muted in comparison to the clear-model spectra, suggesting the presence of at least some clouds or weak vertical temperature gradients at pressures of $\lesssim$10--100 mbar. In contrast, the spectral structure produced by water vapour opacity (indicated by the purple shading) appears more consistent with models lacking clouds at these low pressures on the dayside. Under equilibrium chemistry, methane also would exhibit an absorption feature at ${\sim}$7.5--8.5~$\mu$m (shaded pink) for the colder models at phases 0.0 and 0.75. Finally, the median retrieved spectrum and 1-sigma contours from the HyDRA retrieval are shown in grey.}\label{fig:gcmSpectra}
\end{figure*}

A comparison of the observed and modelled emission spectra further suggests that a majority of the cloud thermal opacity must be confined to pressures greater than ${\sim}$10--100 mbar, because the presence of substantial cloud opacity at lower pressures dampens the modeled spectral signature amplitude below what is observed (see \ref{fig:gcmSpectra}). No distinct spectral signatures indicative of the cloud composition were evident in the observations.
While no single GCM can match the emission spectra at all phases, spectra corresponding to nightside, morning, and evening terminators appear qualitatively similar to GCM results that are intermediate between clear and cloudy simulations. In contrast, the absorption features indicative of water vapour (between ${\sim}$5 and 8.5\,$\mu$m) seen in dayside emission spectra are more consistent with an absence of cloud opacity at these mid-infrared wavelengths. Altogether, these findings represent new constraints on the spatial distribution and opacity of \mbox{WASP-43b}'s clouds.

\begin{figure}     
    \centering
    \includegraphics[width=1.0\textwidth]{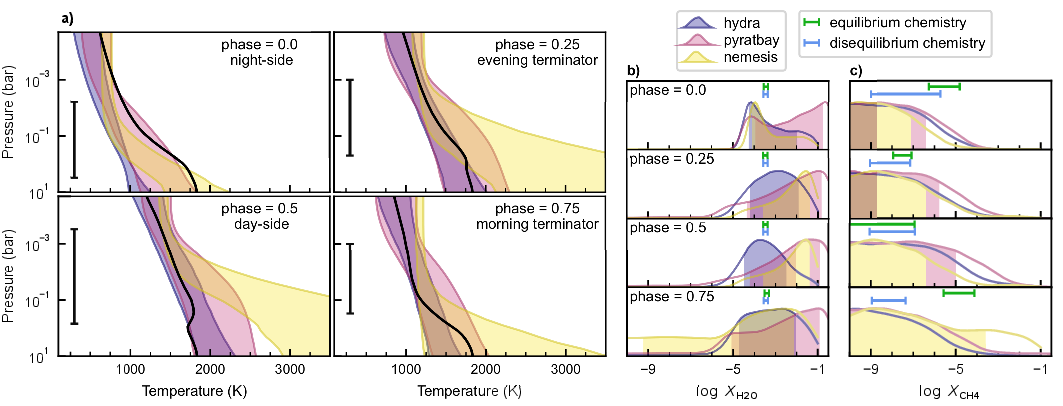}
    \caption{{\bf Atmospheric spectral retrievals for frameworks with free chemistry.}
      The subpanels of {\bf a} show the temperature profile contours (68\% confidence) constrained by the retrievals at each orbital phase (see legends).  All frameworks produced consistent non-inverted thermal profiles that are consistent with 2D radiative--convective equilibrium and photochemical models along the equator\cite{venot20} (black curves) over the range of pressures probed by the observations (black bars).
      The subpanels of {\bf b} show the H$_{2}$O abundance posterior distributions (volume mixing ratios). The
      shaded areas denote the span of the 68\% confidence intervals.  The green and blue bars on each panel denote the abundances predicted by equilibrium and
      disequilibrium-chemistry solar-abundance models\cite{venot20},
      respectively, at the pressures probed by the observations
      ($1-10^{-3}$ bar, approximately). The subpanels of {\bf c} are the same as the subpanels {\bf b} but for CH$_{4}$. The retrieved water abundances are consistent with either equilibrium or disequilibrium-chemistry estimations for solar composition (500 ppm), whereas the retrieved upper limits to the CH$_{4}$ abundance are
      more consistent with disequilibrium-chemistry predictions.}\label{fig:retrievals}
\end{figure}

We further characterized the chemical composition of \mbox{WASP-43b}'s atmosphere by applying a suite of atmospheric retrieval frameworks to the phase-resolved emission spectra. The retrievals spanned a broad range of model assumptions, including free chemical abundances versus equilibrium chemistry, different temperature profile parameterizations and different cloud models (see the Atmospheric Retrieval Models section in the Methods). Despite these differences, the retrievals yielded consistent results for both the chemical and thermal constraints. We detected water vapour across the dayside, nightside, morning and evening hemispheres, with detection significances of up to ${\sim}$3–4$\sigma$ (see \ref{fig:retrieved_spectra}, \ref{tab:ret_abundances}, \ref{tab:ret_chi_sqr}). The retrieved abundances of H$_2$O largely lie in the 10--10$^5$~ppm range for all four phases and for all the retrieval frameworks (\ref{fig:retrievals} and \ref{fig:equilibrium_retrievals}),  broadly consistent with the value expected for a solar composition (500~ppm) as well as previous observations\cite{Stevenson2017}.

We also searched for signatures of disequilibrium chemistry in the atmosphere of \mbox{WASP-43b}. While CH$_4$ is expected to be present on the nightside under thermochemical equilibrium conditions, we did not detect CH$_4$ at any phase (\ref{fig:retrievals}). In the pressure range probed by the nightside spectrum (1--10$^{-3}$ bar, \ref{fig:contribution_functions}), the equilibrium abundance of CH$_4$ is expected to vary between ${\sim}$1--100 ppm for a solar C/O ratio\cite{VenotEtal2020-JWST-WASP-43b}, compared to our 95\% upper limits of 1--6~ppm (\ref{tab:ret_abundances}).
The upper limits we place on the nightside CH$_4$ abundance are more consistent with disequilibrium models that account for vertical and horizontal transport\cite{VenotEtal2020-JWST-WASP-43b,helling2020,VenotEtal2020-JWST-WASP-43b,mendonca2018b}. In particular,  2D photochemical models and GCMs predict the strongest depletion of CH$_4$ on the nightside due to strong zonal winds ($>$ 1 km s$^{-1}$) transporting gas-phase constituents around the planet faster than the chemical reactions can maintain thermochemical equilibrium, thus ``quenching'' and homogenizing the global composition at values more representative of dayside conditions (see also refs.\cite{Agundez2014,tsai21k218b,moses22,baeyens22}). We note, however, that a low atmospheric C/O ratio and/or clouds at photospheric pressures could also lead to a non-detection of CH$_4$. We also searched for signatures of NH$_3$, which is predicted to have a volume mixing ratio less than 0.1--1 ppm in both equilibrium and disequilibrium chemistry models, and find that the results are inconclusive and model-dependent with the current retrieval frameworks.

Given the strong evidence for clouds from comparison with GCMs, we also searched for signatures of clouds in the atmospheric retrieval. Formally, the retrievals do not detect clouds with statistical significance, indicating that strong spectral features uniquely attributable to condensates are not visible in the data (see the Atmospheric Retrieval Models section in Methods and \ref{fig:nightside_clouds}). However, the retrievals may mimic the effects of cloud opacity with a more isothermal temperature profile, as both tend to decrease the amplitude of spectral features, but the cloud-free, more isothermal temperature profile requires fewer free parameters and is therefore statistically favored. Indeed, while the retrieved temperature profiles on the dayside and evening hemispheres agree well with the hemispherically averaged temperature profiles from the GCMs, they are more isothermal than the GCM predictions for the nightside and morning hemispheres (see \ref{fig:compare_gcm_retrieval}). This discrepancy may hint at the presence of clouds on the nightside and morning hemispheres, consistent with the locations of clouds found in the GCMs.

\subsection*{Discussion}

Taken together, our results highlight the unique capabilities of \textit{JWST}/MIRI for exoplanet atmosphere characterization. 
Combined with a range of atmospheric models, the observed phase curve and emission spectroscopy provide strong evidence that the atmospheric chemistry of \mbox{WASP-43b} is shaped by complex disequilibrium processes and new constraints on the optical thickness and pressure of nightside clouds. However, while cloudy GCM predictions match the data better than cloud-free models, none of the simulations simultaneously reproduced the observed phase curve and spectra within measured uncertainties. These remaining discrepancies underscore the importance of further exploring the effects of clouds and disequilibrium chemistry in numerical models, as JWST continues to place unprecedented observational constraints on smaller and cooler planets.


\clearpage

\clearpage
\begin{methods}
\renewcommand{\figurename}{\hspace{-4pt}}
\renewcommand{\thefigure}{Extended Data Fig.~\arabic{figure}}
\renewcommand{\theHfigure}{Extended Data Fig.~\arabic{figure}}
\renewcommand{\tablename}{\hspace{-4pt}}
\renewcommand{\thetable}{Extended Data Table \arabic{table}}
\renewcommand{\theHtable}{Extended Data Table \arabic{table}}
\setcounter{figure}{0}
\setcounter{table}{0}

\section*{Observations and Quality of the Data}

We observed a full orbit of \mbox{WASP-43b} with the JWST MIRI LRS slitless (SL) mode as a part of JWST-ERS-1366. We performed Target Acquisition with the F1500W filter and used the SLITLESSPRISM subarray for the science observation. The science observation was taken between 2022-12-01 at 00:54:30 UT and 2022-12-02 at 03:23:36 UT, for a total of 26.5 hours. We acquired 9216 integrations that were split into 3 exposures and 10 segments per exposure. Each integration lasts 10.34~s and is composed of 64 groups, with one frame per group. The LRS slitless mode reads an array of $416{\times}72$ pixels on the detector (the SLITLESSPRISM subarray) and uses the FASTR1 readout mode, which introduces an additional reset between integrations.

Due to the long duration of the observation, two high-gain antenna moves occurred 8.828 h and 17.661 h after the start of the science observation. They affect only a couple of integrations that we removed from the light curves. A cross-shaped artifact is present on the 2D images at the short wavelength end due to light scattered by detector pixels\cite{Andras2021}. It is stable over the duration of the observation but it contaminates the background and the spectral trace up to ${\sim}6\,\mu$m. This ``cruciform'' artifact is observed in all MIRI LRS observations; a dedicated analysis is underway to estimate and mitigate its impact.

In the broadband light curve, the flux decays by ${\sim}0.1\,\%$ during the first 60 minutes and continues to decay throughout the observation. This ramp is well modeled with one or two exponential functions after trimming the initial ${\sim}780$ integrations. Without trimming any data, at least two ramps are needed. In addition, a downward linear trend in flux is observed over the whole observation with a slope of -39 ppm per hour.
These two types of drifts also appear in the spectroscopic light curves. The exponential ramp amplitude in the first 60 minutes changes with wavelength from -0.67\% in the 5--5.5 micron bin (downward ramp) to +0.26\% in the 10--10.5 micron bin (upward ramp). The ramp becomes upward at wavelengths longer than 7.5 $\mu$m and its timescale increases to more than one hour at wavelengths longer than 10.5 $\mu$m. The slopes as a function of wavelength vary from -16 to -52 ppm, all downward.
Such drifts (initial ramp and linear or polynomial trend) are also observed in other MIRI LRS time-series observations\cite{Bouwman2023MiriTsoCommissioning} but the strength of the trends differ for each observation. In these \mbox{WASP-43b} observations, we note that their characteristic parameters vary smoothly with wavelength which may help identify their cause and build correction functions. 

Over the course of the observation, the position of the spectral trace on the detector varies by 0.0036 pixels RMS (0.027 pixels peak-to-peak) in the spatial direction, and the Gaussian standard deviation of the spatial PSF varies by 0.00069 pixels RMS (0.0084 pixels peak-to-peak) following a sharp increase by 0.022 pixels during the first 600 integrations. Depending on the wavelength bin, that spatial drift causes noise at the level of 7--156 ppm, while variations in the PSF width cause noise at the level of 4--54 ppm (these numbers are obtained from a linear decorrelation). Overall, the MIRI instrument used in LRS slitless mode remains remarkably stable over this 26.5\,hr-long continuous observation and the data are of exquisite quality.

The noise in the light curve increases sharply at wavelengths beyond 10.5 $\mu$m and the transit depths obtained at these long wavelengths by different reduction pipelines are discrepant. These wavelengths were not used in the retrieval analyses and the final broadband light curve. The cause is unknown but it might be related to the fact that this region of the detector receives different illumination before the observation\cite{bell2023arxiv} (see paragraph ``Shadowed Region Effect" below for more details).

\section*{Data Reduction Pipelines}

\subsection*{\texttt{Eureka!}\ v1 Reduction}
The \texttt{Eureka!}\ v1 reduction made use of version 0.9 of the \texttt{Eureka!}\ pipeline\cite{Bell2022Eureka}, CRDS version 11.16.16 and context 1018, and \texttt{jwst} package version 1.8.3\cite{stsci2022jwst}. The gain value of 5.5 obtained from these CRDS reference files is known to be incorrect, and the actual gain is estimated to be ${\sim}$3.1 electrons/DN although the gain may be wavelength dependent (priv.\ comm., Sarah Kendrew). A new reference file reflecting the updated gain is under development at STScI which will improve the accuracy of photon-noise calculations. For the rest of this analysis, we will assume a constant gain of 3.1 electrons/DN. The \texttt{Eureka!}\ Control Files and \texttt{Eureka!}\ Parameter Files files used in these analyses are available for download (\url{https://zenodo.org/doi/10.5281/zenodo.10525170}) and are summarized below.

\texttt{Eureka!}\ makes use of the \texttt{jwst} pipeline for Stages 1 and 2, and both stages were run with their default settings, with the exception of increasing the Stage 1 jump step's rejection threshold to 8.0 and skipping the photom step in Stage 2 because it is not necessary and can introduce additional noise for relative time series observations. In \texttt{Eureka!}'s Stage 3, we then rotated the MIRI/LRS slitless spectra 90 degrees counter-clockwise so that wavelength increases from left to right like the other JWST instruments to allow for easier reuse of \texttt{Eureka!}\ functions. We then extracted pixels 11--61 in the new y-direction (the spatial direction) and 140--393 in the new x-direction (spectral direction); pixels outside of these ranges primarily contain noise not useful for our reduction. Pixels marked as ``DO\textunderscore NOT\textunderscore USE'' in the DQ array were then masked as were any other unflagged NaN or inf pixels. A centroid was then fit to each integration by summing along the spectral direction and fitting the resulting 1D profile with a Gaussian function; the centroid from the first integration was used for determining aperture locations, while the centroids and PSF widths from all integrations were saved to be used as covariates when fitting the observations.

Our background subtraction method is tailored to mitigate several systematic effects unique to the MIRI instrument. First, MIRI/LRS observations exhibit a ``cruciform artifact''\cite{gaspar2021} at short wavelengths caused by scattered lights within the optics; this causes bright rays of scattered light which must be sigma-clipped in order to avoid over-subtracting the background. Additionally, MIRI/LRS observations exhibit periodic noise in the background flux which drifts with time\cite{Bouwman2023MiriTsoCommissioning} as well as $1/f$~noise\cite{Schlawin2020OneOverF} which leads to correlated noise in the cross-dispersion direction; as a result, background subtraction must be performed independently for each integration and column (row in MIRI's rotated reference frame). Further, in both these observations and the dedicated background calibration observations from JWST-COM/MIRI-1053, we found that there was a linear trend in the background flux, with the background flux increasing with increasing row index (column index in MIRI's rotated reference frame). To robustly remove this feature, we found that it was important to either (1) use the mean from an equal number of pixels on either side of the spectral trace for each column and integration, or (2) use a linear background model for each column and integration; we adopted the former as it resulted in less noisy light curves. To summarize, for each column in each integration we subtracted the mean of the pixels separated by $\ge$11 pixels from the center of the spectral trace after first masking 5$\sigma$ outliers in that column.

To compute the spatial profile for the optimal extraction of the source flux we calculated a median frame, sigma-clipping 5$\sigma$ outliers along the time axis and smoothing along the spectral direction using a 7-pixel-wide boxcar filter. During optimal extraction, we only used the pixels within 5 pixels of the fitted centroid and masked pixels that were 10$\sigma$ discrepant with the spatial profile. Background exclusion regions ranging from 9--13 pixels and source aperture regions ranging from 4--6 pixels were considered, but our values of 11 and 5 pixels were selected as they produced the lowest median absolute deviation light curves before fitting.

\subsection*{\texttt{Eureka!}\ v2 Reduction}
The \texttt{Eureka!}\ v2 reduction followed the same procedure as the \texttt{Eureka!}\ v1 reduction except for the following differences. First, this reduction made use of version 1.8.1 of the \texttt{jwst} pipeline. For Stage 1, we instead used a cosmic ray detection threshold of 5 and used a uniform ramp fitting weighting. For Stage 2, we performed background subtraction using columns away from the trace on the left and on the right and subtracted the background for each integration\cite{Bouwman2023MiriTsoCommissioning}. Stage 3 was then identical to \texttt{Eureka!}\ v1 reduction.

\subsection*{TEATRO Reduction}
We processed the data using the Transiting Exoplanet Atmosphere Tool for Reduction of Observations (TEATRO) that runs the \texttt{jwst} package, extracts and cleans the stellar spectra and light curves, and runs light curve fits.
We used the \texttt{jwst} package version 1.8.4, CRDS version 11.16.14 and context 1019. We started from the `uncal' files and ran Stages 1 and 2 of the pipeline. For Stage 1, we set a jump rejection threshold of 6, turned off the `jump.flag\_4\_neighbors' parameter, and used the default values for all other parameters. For Stage 2, we ran only the `AssignWcsStep', `FlatFieldStep', and `SourceTypeStep'; no photometric calibration was applied. The next steps were made using our own routines. We computed the background using two rectangular regions, one on each side of the spectral trace, between pixels 13--27 and 53--72 in the spatial direction respectively. We computed the background value for each row (rows are along the spatial direction) in each region using a biweight location, averaged the two values, and subtracted it from the full row. This background subtraction was done for each integration. Then, we extracted the stellar flux using aperture photometry by averaging pixels between 33--42 in each row to obtain the stellar spectrum at each integration. We also averaged pixels between 33--42 in the spatial direction and between 5--10.5~$\mu$m in the spectral direction to obtain the broadband flux. We averaged the spectra in eleven 0.5\,$\mu$m-wide wavelength channels. For each channel and for the broadband light curve, we normalised the light curve using the second eclipse as a reference flux, computed a running median filter using a 100-point window size, and rejected points that were more than 3$\sigma$ away from that median using a five iteration sigma-clipping. To limit the impact of the initial ramp on the fitting, we trim the first 779 integrations from the broadband light curve and a similar number of integrations for each channel (the exact number depends on the  channel). Finally, we subtracted 1 from the normalized light curves to have the secondary eclipse flux centered on zero. These cleaned light curves were used for phase curve, eclipse, and transit fits.  

\subsection*{SPARTA Reduction}
We reduce the data with the open source Simple Planetary Atmosphere Reduction Tool for Anyone (SPARTA), first introduced in ref.\cite{kempton2023} to analyze the MIRI LRS phase curve of GJ 1214b. We start from the uncalibrated data and proceed all the way to the final results without using any code from the \texttt{jwst} or \texttt{Eureka!}\ pipelines. In Stage 1, we start by discarding the first 5 groups as well as the last group, because these groups exhibit anomalies due to the reset switch charge decay and the last-frame effect.  We fit a slope to the up-the-ramp reads in every pixel of every integration in every exposure.  We calculate the residuals of these linear fits, and for every pixel, we compute a median residual for every group across all integrations. This ``median residual'' array has dimensions $N_{\rm grp} \times N_{\rm rows} \times N_{\rm cols}$. This array is subtracted from the original uncalibrated data and the up-the-ramp fit is redone, this time without discarding any groups except those that are more than 5$\sigma$ away from the best-fit line. Such outliers, which may be due to cosmic rays, are discarded and the fit recomputed until convergence. This procedure straightens out any nonlinearity in the up-the-ramp reads that is consistent across integrations, such as the reset switch charge decay, the last-frame effect, or the significant inaccuracies in the nonlinearity coefficients. After up-the-ramp fitting, we remove the background by removing the mean of columns 10--24 and 47--61 (inclusive, zero-indexed) for every row of every integration. Because these two regions are of equal size and equally distant from the trace, any linear spatial trend in the background is naturally removed.

In the next step, we compute a pixel-wise median image over all integrations. This median image is used as a template to determine the position of the trace in each integration, by shifting and scaling the template until it matches the integration (and minimizes the $\chi^2$). It is also used as the point spread profile for optimal extraction, after shifting in the spatial direction by the amount calculated in the previous step. Outliers more than 5$\sigma$ discrepant from the model image (which may be cosmic rays) are masked, and the optimal extraction is repeated until convergence. The z-scores image (image minus model image all divided by expected error, including photon noise and read noise) have a typical standard deviation of 0.88, compared to a theoretical minimum value of 1, indicating that the errors are being overestimated.

After optimal extraction, we gather together all the spectra and positions into one file. To reject outliers, we create a broadband light curve, detrend it by subtracting a median filter with a width 100 times less than the total data length, and reject integrations greater than 4$\sigma$ away from 0 (which may be cosmic rays). Sometimes only certain wavelengths of an integration are bad, not the entire integration. We repair these by detrending the light curve at each wavelength, identifying 4$\sigma$ outliers, and replacing them with the average of their two immediate temporal neighbors.

\section*{Spectral Binning}
To investigate the effects of spectral binning, we utilized the MIRI time-series commissioning observations of the transit of L168-9b (JWST-COM/MIRI-1033; ref.\cite{Bouwman2023MiriTsoCommissioning}). L168-9b was chosen to have a clear transit signal while also having no detectable atmospheric signatures in its transmission spectrum; as a result, the observed scatter in the transmission spectrum can be used as an independent measurement of the uncertainties in the transit depths. The same procedure can't be done on our \mbox{WASP-43b} science observations as there may be detectable atmospheric signatures.

Following the \texttt{Eureka!}\ reduction methods described by ref.\cite{Bouwman2023MiriTsoCommissioning}, we tried binning the L168-9b spectroscopic light curves at different resolutions and compared the observed standard deviation of the transmission spectrum to the median of the transit depth uncertainties estimated from fitting the spectral light curves. As shown in \ref{fig:binningUncertainties}, the uncertainties in the pixel-level light curves underestimate the scatter in the transmission spectrum by a factor of ${\sim}$2. Because pairs of rows (in MIRI's rotated reference frame) are reset together, it is reasonable to assume that there could be odd-even effects that would average out if combining pairs of rows; indeed, there do appear to be differences in the amplitude of the initial exponential ramp feature between odd and even rows. However, combining pairs of rows still leads to a significant underestimation of the scatter in the transmission spectrum. Interestingly, the underestimation of the uncertainties appears to decrease with decreasing wavelength resolution. This is likely explained by wavelength-correlated noise which gets averaged out with coarse binning. A likely culprit for this wavelength-correlated noise may be the 390\,Hz periodic noise observed in several MIRI subarrays which causes clearly structured noise with a period of ${\sim}$9 rows (ref.\cite{Bouwman2023MiriTsoCommissioning}; priv.\ comm., Michael Ressler); this noise source is believed to be caused by MIRI's electronics and possible mitigation strategies are still under investigation. Until the source of the noise is definitively determined and a noise mitigation method is developed, we recommend that MIRI/LRS observations should be binned to a fairly coarse spectral resolution as this gives better estimates of the uncertainties and also gives spectra that are closer to the photon-limited noise regime. However, we caution against quantitative extrapolations of the uncertainty underestimation to other datasets; because we don't know the source of the excess noise, we don't know how it might change with different parameters like groups per integration or stellar magnitude.

Ultimately, for each reduction method we binned the spectra down to a constant 0.50~$\mu$m wavelength grid spanning 5--12~$\mu$m, giving a total of 14 spectral channels. However, as is described below, we only end up using the 11 spectral channels spanning 5--10.5~$\mu$m for science. This 0.5\,$\mu$m binning scheme combines 7 wavelengths for the shortest bin and 25 wavelengths for the longest bin which has the added benefit of binning down the noise at longer wavelengths where there are fewer photons. However, even for this coarse of a binning scheme, we do expect there to be some additional noise beyond our estimated uncertainties on the spectrum of \mbox{WASP-43b} (\ref{fig:binningUncertainties}). Since the structure of this noise source is not well understood nor is the extent to which our error bars are underestimated, our best coarse of action was to consider error inflation when performing spectroscopic inferences (described in more detail below).

\section*{Light Curve Fitting}

\subsection*{Detrending the Initial Exponential Ramp}
As with other MIRI/LRS observations\cite{Bouwman2023MiriTsoCommissioning}, our spectroscopic light curves showed a strong exponential ramp at the start of the observations. As shown in \ref{fig:rampAmplitudes}, the strength and sign of the ramp varies with wavelength, changing from a strong downward ramp at 5\,$\mu$m to a nearly flat trend around 8\,$\mu$m, and then becoming an upward ramp toward longer wavelengths. From 10.6--11.8\,$\mu$m, the ramp timescale became significantly longer and the amplitude of the ramp became much stronger; this region of the detector was previously discussed\cite{bell2023arxiv} and is discussed in more detail below. In general, most of the ramp's strength had decayed within ${\sim}$30--60 minutes, but at the precision of our data the residual ramp signature still had an important impact on our nightside flux measurements due to the similarity of the ramp timescale with the orbital period. Unlike in the case of the MIRI/LRS commissioning observations of L168-9b\cite{Bouwman2023MiriTsoCommissioning}, we were not able to safely fit the entire dataset with a small number of exponential ramps. When fitting the entire dataset, we found that non-trivial choices about the priors for the ramp amplitudes and timescales resulted in significantly different spectra at phases 0.75 (morning hemisphere) and especially 0.0 (nightside); because the dayside spectrum is measured again near the end of the observations, it was less affected by this systematic noise.

Ultimately, we decided to conservatively discard the first 779 integrations (134.2 minutes), leaving only one transit-duration of baseline before the first eclipse ingress began. After removing the initial 779 integrations, we found that a single exponential ramp model with broad priors that varied freely with wavelength was adequate to remove the signature. In particular, after removing the first 779 integrations we found that our dayside and nightside emission spectra were not significantly affected by (1) fitting two exponential ramps instead of one, (2) adjusting our priors on the ramp timescale to exclude rapidly decaying ramps with timescales $>$15\,days$^{-1}$ instead of $>$100\,days$^{-1}$, (3) putting a uniform prior on the inverse timescale instead of the timescale, or (4) altering the functional form of the ramp by fitting for an exponential to which the time was raised. After removing the first ${\sim}$2 hours, we also found that the ramp amplitude and timescale did not vary strongly with wavelength (excluding the ``shadowed region'' described below), although fixing these parameters to those fitted to the broadband light curve affected several points in the nightside spectrum by more than 1$\sigma$; we ultimately decided to leave the timescale and amplitude to vary freely with wavelength as there is no a priori reason to assume that they should be equal. With careful crafting of priors, it appeared possible to get results similar to our final spectra while removing only the first few integrations, but trimming more integrations and only using a single exponential ramp model required fewer carefully tuned prior assumptions for which we have little physical motivation.

\subsection*{``Shadowed Region Effect"}
As was described in ref.\cite{bell2023arxiv}, we also identified a strong discontinuity in the spectroscopic light curves spanning pixel rows 156--220 (10.6--11.8\,$\mu$m) in these observations. In this range, the temporal behavior of the detector abruptly changes to a large-amplitude, long-timescale, upward ramp that appears to slightly overshoot before decaying back down and approaching an equilibrium. These pixels coincide with a region of the detector between the Lyot coronagraph region and the four-quadrant phase mask region which is unilluminated except when the dispersive element is in the optical path; as a result, we have taken to calling this unusual behavior as the ``shadowed region effect''. Strangely, not all MIRI/LRS observations demonstrate this behavior, with the MIRI/LRS TSO commissioning observations\cite{Bouwman2023MiriTsoCommissioning} and the GJ 1214b phase curve observations\cite{kempton2023} showing no such effect. In fact, we only know of two other observations which exhibit a similar behavior: the observation of the transit of WASP-80b (JWST-GTO-1177; priv.\ comm., Taylor Bell) and the observation of the phase curve of GJ~367b (JWST-GO-2508; priv.\ comm., Michael Zhang). Informatively, the eclipse observation of WASP-80b taken 36 hours after the WASP-80b transit using the same observing procedure (JWST-GTO-1177; priv.\ comm., Taylor Bell) did not show the same ``shadowed region effect'', indicating that the effect is unlikely to be caused by stray light from nearby stars or any other factors which stayed the same between those two observations. Our best guess at this point is that the effect is related to the illumination history of the detector and the filter used by the previous MIRI observation (because the detector is illuminated at all times, even when it is not in use), but this is still under investigation and at present there is no way of predicting whether or not an observation will be impacted by the ``shadowed region effect''. It is important to note, however, that from our limited knowledge at present that the ``shadowed region effect'' appears to either be present or not, with observations either strongly affected or seemingly completely unaffected.

Using the general methods described in the \texttt{Eureka!}\ v1 fit, we attempted to model the ``shadowed region effect'' with a combination of different ramp models, but nothing we tried was able to cleanly separate the effect from the phase variations, and there was always some clear structure left behind in the residuals of the fit. Another diagnostic that our detrending attempts were unsuccessful was that the phase offset as a function of wavelength smoothly varied around ${\sim}$10 degrees \textit{East} in the unaffected region of the detector, but in the ``shadowed region'' the phase offset would abruptly change to ${\sim}$5 degrees \textit{West}; such a sharp change in a suspect region of the detector seems highly unlikely to be astrophysical in nature. As a result, we ultimately chose to exclude the three spectral bins spanning 10.5--12\,$\mu$m from our retrieval efforts.

\subsection*{Determining the Number of Sinusoid Harmonics}
In order to determine the complexity of the phase curve model required to fit the data, we used the \texttt{Eureka!}\ v1 reduction and most of the \texttt{Eureka!}\ v1 fitting methods described below, with the exception of using the \texttt{dynesty}\cite{dynesty_v1_2_3} nested sampling algorithm (which computes the Bayesian evidence, $\mathcal{Z}$) and a \texttt{batman} transit and eclipse model. Within \texttt{dynesty}, we used 256 live points, `multi' bounds, `rwalk' sampling, and ran until the estimated $d\ln(\mathcal{Z})$ reached 0.1. We then evaluated a 1st-, 2nd-, and 4th-order model for the broadband light curve, excluding all 3rd-order sinusoidal terms from the 4th-order model as these terms are not likely to be produced by the planet's thermal radiation\cite{cowan2008,cowan2017}. We then compared the Bayesian evidences of the different models following refs.\cite{Trotta2008,Welbanks2021} and found that the 2nd-order model was significantly preferred over the 1st-order model at 16$\sigma$ ($\Delta\ln(\mathcal{Z})=128$), while the 2nd-order model was insignificantly preferred over the 4th-order model at 2.2$\sigma$ ($\Delta\ln(\mathcal{Z})=1.3$). This is also confirmed by eye, where the 1st-order model leaves clear phase variation signatures in the residuals, while the residuals from the 2nd-order model leave no noticeable phase variations behind. Finally, we also compared the phase-resolved spectra obtained from different order phase curve models; we found that our spectra significantly changed going from a 1st- to 2nd-order model (altering one or more spectral points by ${>}1\sigma$), but the 4th-order model did not significantly change the resulting phase-resolved spectra compared to the 2nd-order. As a result, the final fits from all reductions used a 2nd-order model. The broadband light curves obtained from the four reductions and the associated phase curve models are shown in Supplementary Fig.~1.

\subsection*{\texttt{Eureka!}\ v1 Fitting Methods}\label{sec:Eureka_v1}
We first sigma-clipped any data points that were 4$\sigma$ discrepant from a smoothed version of the data (made using a box-car filter with a width of 20 integrations) to remove any obviously errant data points while preserving the astrophysical signals like the transit.

Our astrophysical model consisted of a \texttt{starry}\cite{starry_v1.2.0} transit and eclipse model, as well as a second-order sinusoidal phase variation model. The complete astrophysical model had the form
\begin{equation}
	A(t) = F_*(t) + F_{\rm day}E(t)\Psi(\phi), \label{eqn:fluxEureka}
\end{equation}
where $F_*$ is the received stellar flux (and includes the starry transit model), $F_{\rm day}$ is the planetary flux at mid-eclipse, $E(t)$ is the starry eclipse model (neglecting eclipse mapping signals for the purposes of this paper), and $\Psi(\phi)$ is the phase variation model of the form
\begin{align}
\Psi(\phi) = 1 +& \text{AmpCos1}*(\cos(\phi)-1) + \text{AmpSin1}*\sin(\phi)\nonumber\\
  +& \text{AmpCos2}*(\cos(2\phi)-1) + \text{AmpSin2}*\sin(2\phi), \label{eqn:phaseVarsEureka}
\end{align}
where $\phi$ is the orbital phase in radians \textit{with respect to eclipse} and AmpCos1, AmpSin1, AmpCos2, and AmpSin2 are all fitted coefficients. The second-order phase variation terms allow for thermal variations across the face of the planet that are more gradual or steep than a simple first-order sinusoid would allow. We numerically computed dayside, morning, nightside, and evening spectra using the above $\Psi(\phi)$ function at $\phi$= 0, $\pi/2$, $\pi$, and $3\pi/2$, respectively. In order to allow the \texttt{starry} eclipse function to account for light travel time, we used a stellar radius of 0.667 $R_{\rm \odot}$ (ref.\cite{Bonomo2017}) to convert the fitted $a/R_{\rm *}$ to physical units. For our transit model, we used a reparameterized version of the quadratic limb-darkening model\cite{Kipping2013LimbDarkening}.

Our systematics model consisted of a single exponential ramp in time to account for the idle-recovery drift documented for MIRI/LRS TSO observations\cite{Bouwman2023MiriTsoCommissioning}, a linear trend in time, and a linear trend with the spatial position and PSF width. The full systematics model can be written as
\begin{equation}
	S(t, y, s_y) = L(t_l)*R(t_r)*Y(y)*SY(s_y),
\end{equation}
The linear trend in time is modelled as
\begin{equation}
L(t_l) = c_0 + c_1t_l, \label{eqn:linearEureka}
\end{equation}
where $t_l$ is the time with respect to the mid-point of the observations and where $c_0$ and $c_1$ are coefficients. The exponential ramp is modelled as
\begin{equation}
R(t_r) = 1 + r_0e^{r_1t_r} \label{eqn:rampEureka}
\end{equation}
where $t_r$ is the time with respect to the first integration and where $r_0$ and $r_1$ are coefficients. The linear trends as a function of spatial position, $y$, are PSF width $s_y$ are modelled as
\begin{equation}
Y(y) = 1 + fy \label{eqn:yEureka}
\end{equation}
and
\begin{equation}
SY(s_y) = 1 + gs_y, \label{eqn:syEureka}
\end{equation}
where $f$ and $g$ are coefficients. The linear trends as a function of spatial position and PSF width are with respect to the mean-subtracted spatial position and PSF width. Finally, we also fitted a multiplier to the estimated Poisson noise level for each integration to allow us to account for any noise above the photon-limit as well as an incorrect value for the gain applied in Stage 3.

With an initial fit to the broadband light curve (5--10.5\,$\mu$m), we assumed a zero eccentricity and placed a Gaussian prior on the planet's orbital parameters ($P$, $t_0$, $i$, $a/R_{\rm *}$) based on previously published values for the planet\cite{Esposito2017GapsWASP43b}. For the fits to the spectroscopic phase curves, we then fixed these orbital parameters to the estimated best-fit from the broadband light fit to avoid variations in these wavelength-independent values causing spurious features in the final spectra. We fitted the observations using the No U-Turns Sampler (NUTS) from \texttt{pymc3}\cite{pymc3} with 3 chains, each taking 6,000 tuning steps and 6,000 production draws with a target acceptance rate of 0.95. We used the Gelman-Rubin statistic\cite{GelmanRubin1992} to ensure the chains had converged. We then used the 16th, 50th, and 84th percentiles from the \texttt{pymc3} samples to estimate the best-fit values and their uncertainties.

\subsection*{\texttt{Eureka!}\ v2 Fitting Methods}
For the second fit made with \texttt{Eureka!}, we proceeded very similarly to the \texttt{Eureka!}\ v1 fit. We clipped the light curves using a box-car filter of 20 integrations wide with a maximum of 20 iterations and a rejection threshold of 4$\sigma$ to reject these outliers. We also modelled the phase curve using a second-order sinusoidal function but we modelled the transit and eclipse using \texttt{batman}\cite{Kreidberg2015Batman} instead of \texttt{starry}. Like in the \texttt{Eureka!}\ v1 Fit, we modeled instrumental systematics with a linear polynomial model in time (Eqn. \ref{eqn:linearEureka}), an exponential ramp (Eqn. \ref{eqn:rampEureka}), a first-order polynomial in $y$ position (Eqn. \ref{eqn:yEureka}) and a first-order polynomial in PSF width in the $s_y$ direction (Eqn. \ref{eqn:syEureka}). 

We fitted the data using the \texttt{emcee} sampler\cite{foreman-mackey_emcee_2013} instead of NUTS, with 500 walkers and 1500 steps. The jump parameters that we used were: $R_{\rm p}/R_{\rm \star}$, $F_{\rm p}/F_{\rm \star}$, $u_1$, $u_2$, AmpCos1, AmpSin1, AmpCos2, AmpSin2, $c_0$, $c_1$, $r_0$, $r_1$, $y_{pos}$, $y_{width}$ and scatter$_{\rm multi}$ (multiplier to the estimated Poisson noise level for each integration like in the \texttt{Eureka!}\ v1 Fit). We used uniform priors on $u_1$ and $u_2$ from 0 to 1, uniform priors on AmpCos1, AmpSin1, AmpCos2, AmpSin2 from -1.5 to 1.5 and broad normal priors and all other jump parameters. Convergence, mean values and uncertainties were computed in the same way as for the \texttt{Eureka!}\ v1 Fit.

\subsection*{TEATRO Fitting Methods}

To measure the planet's emission as a function of longitude, we modeled the light curves with a phase variation model, an eclipse model, a transit model, and an instrument systematics model. The phase curve model $\Psi(t)$ consists of two sinusoids: one at the planet's orbital period $P$ and one at $P/2$ to account for second-order variations. The eclipse model $E(t)$ and transit model $T(t)$ are computed with the \texttt{exoplanet}\cite{exoplanet:joss} package that uses the \texttt{starry} package\cite{starry_v1.2.0}. We save the eclipse depth $\delta_e$ and normalize $E(t)$ to a value of 0 during the eclipse and 1 out of the eclipse, which we then call $E_N(t)$. We used published transit ephemerides\cite{Ivshina2022}, a null eccentricity, and published stellar parameters\cite{Gillon2012}. The planet-to-star radius ratio $R_p / R_s$, impact parameter $b$, and mid-transit time $t_0$ are obtained from a fit to the broadband light curve. The systematics model $S(t)$ is composed of a linear function to account for a downward trend and an exponential function to account for the initial ramp. The full model is expressed as:
\vspace{10pt}
\begin{align}
F(t) &= (\Psi(t) - \Psi(t_e) + \delta_e) \times E_N(t) + T(t) + S(t) \label{eqn:fluxCrouzet}\\[10pt]
\Psi(t) &= a_\Psi \, sin(2\pi\,t/P - b_\Psi) + c_\Psi \, cos(4\pi\,t/P - d_\Psi) \\[10pt]
S(t) &= a_S \, e^{-b_S\,t} + c_S \, t + d_S
\end{align}
where $\Psi(t_e)$ is the value of $\Psi$ at the mid-eclipse time $t_e$. 

We fit our model to the data using an MCMC procedure based on the \texttt{PyMC3} package\cite{pymc3} and gradient-based inference methods as implemented in the \texttt{exoplanet} package\cite{exoplanet:joss}. We set normal priors for $t_0$, $R_p / R_s$, the stellar density $\rho_s$, $a_\Psi$, $b_\Psi$, $c_S$, and $d_S$ with mean values obtained from an initial non-linear least-squares fit, a normal prior for $a_S$ with a zero mean, uniform priors for the surface brightness ratio between the planet's dayside and the star $s$, $b$, $c_\Psi$, and $d_\Psi$, uninformative priors for the quadratic limb darkening parameters\cite{Kipping2013LimbDarkening}, and allowed for wide search ranges. We ran two MCMC chains with 5,000 tuning steps and 100,000 posterior samples. Convergence was obtained for all parameters (except in one case where $a_S$ was negligible and $b_S$ was unconstrained). We merged the posterior distributions of both chains and used their median and standard deviation to infer final values and uncertainties for the parameters. We also verified that the values obtained from each chain were consistent.

For the spectroscopic light curve fits, we fixed all physical parameters to those obtained from the broadband light curve fit except the surface brightness ratio $s$ that sets the eclipse depth, we masked the transit part of the light curve, and used a similar procedure. After the fits, we calculated the eclipse depth $\delta_e$ as $s \times (R_p/R_s)^2$, where $R_p$ and $R_s$ are the planet and star radius respectively, and computed $\Psi(t)$ for the final parameters, $\Psi_f(t)$. The planetary flux is $\Psi_f(t) - \Psi_f(t_e) + \delta_e$.
We computed the uncertainty on the eclipse depth in two ways: from the standard deviation of the posterior distribution of $s \times (R_p/R_s)^2$ and from the standard deviation of the in-eclipse points divided by $\sqrt{N_e}$, where $N_e$ is the number of in-eclipse points, and took the maximum of the two.
To estimate the uncertainty on the planet's flux, we computed the $1\sigma$ interval of $\Psi(t)$ based on the posterior distributions of its parameters, computed the $1\sigma$ uncertainty of $d_S$, and added them in quadrature to the uncertainty on the eclipse depth in order to obtain more conservative uncertainties.

The spectra presented in this paper and used in the combined spectra are based on system parameters that were derived from a broadband light curve obtained in the 5--12\,$\mu$m range, a transit fit in which the stellar mass and radius were fixed, a simpler additive model in which the phase curve was not turned off during the eclipse, and a MCMC run that consisted in two chains of 10,000 tuning steps and 10,000 posterior draws.
Updated spectra based on system parameters derived from the broadband light curve obtained in the 5--10.5\,$\mu$m range, a transit fit that has the stellar density as a free parameter, the light curve model shown in Eq. \ref{eqn:fluxCrouzet}, and two MCMC chains of 5,000 tuning steps and 100,000 posterior draws are consistent within 1 sigma at every point with those shown here. As we average four reductions and inflate the uncertainties during the retrievals, the impact of these updates on our results are expected to be marginal.

\subsection*{SPARTA Fitting Methods}

We use \texttt{emcee}\cite{foreman-mackey_emcee_2013} to fit a broadband light curve with the transit time, eclipse time, eclipse depth, 4 phase curve parameters (2 for the first-order and 2 for the second-order sinusoids), transit depth, $a/R_*$, $b$, flux normalization, error inflation factor, instrumental ramp amplitude and 1/timescale, linear slope with time, and linear slope with trace $y$ position as free parameters. The best-fit transit and eclipse times, $a/Rs$, and $b$ are fixed for the spectroscopic light curves.

For the spectroscopic fits, we then use \texttt{emcee} to fit the free parameters: the eclipse depth, 4 phase curve parameters, error inflation factor, flux normalization, instrumental ramp amplitude and 1/timescale, linear slope with time, and linear slope with trace $y$ position. All parameters are given uniform priors. 1/timescale is given a prior of 5--100 $days^{-1}$, but the other priors are unconstraining. In summary, the instrumental model is:
\begin{equation}
    S = F_* (1 + A\exp{(-t/\tau)} + c_y y + m(t-\overline{t})),
\end{equation}
while the planetary flux model is:
\begin{equation}
    F_p = E + C_1(\cos(\omega t) - 1) + D_1 \sin(\omega t) + C_2(\cos(2\omega t) - 1) + D_2 \sin(2\omega t).
    \label{eq:fluxSPARTA}
\end{equation}
Note that the phase variations were set to be zero during eclipse.

\subsection*{Combining Independent Spectra}

Comparing the phase-resolved spectra from each reduction (Supplementary Fig.~2), we see that for wavelengths below 10.5\,$\mu$m the spectra are typically consistent, while larger differences arise in the $>$10.5\,$\mu$m region affected by the ``shadowed region effect''. For our final, fiducial spectrum, we decided to use the mean spectrum and inflated our uncertainties to account for disagreements between different reductions. The mean phase-resolved spectra were computed by taking the mean $F_{\rm p}/F_{\rm *}$ per wavelength. The uncertainties were computed by taking the mean uncertainty per-wavelength, and then adding in quadrature the RMS between the individual reductions and the mean spectrum; this minimally affects the uncertainties where there is minimal disagreement and appreciably increases the uncertainties where the larger disagreements arise.

Each reduction also computed a transmission spectrum (Supplementary Fig.~2) which appears quite flat (within uncertainties) and minimal differences between reductions. \mbox{WASP-43b} is not an excellent target for transmission spectroscopy, however, and these transmission spectra are not expected to be overly constraining.

\section*{Atmospheric Forward Models}\label{sec:GCMdetails}

General Circulation Models (GCMs) were used to simulate atmospheric conditions, from which synthetic phase curves and emission spectra were forward modelled and compared to the observations. The GCMs used in this study are listed in Supplementary Table 1, and details of each simulation are provided in \ref{tab:gcm_list_full} and the following sections.

\subsection*{Generic PCM}

The Generic Planetary Climate Model (\texttt{Generic PCM}) is a 3D Global Climate model designed for modelling the atmosphere of exoplanets and for paleoclimatic studies. The model has been used for the study of planetary atmospheres of the Solar System\cite{Turbet2021,Spiga2020,charnay14}, terrestrial exoplanets\cite{turbet_habitability_2016}, mini-Neptunes\cite{charnay_3d_2015} and Hot Jupiters\cite{Teinturier_2023_pcm}. The dynamical core solves the primitive equations of meteorology on a Arakawa C grid. The horizontal resolution is 64$\times$48 (i.e., 5.625$\times$3.75$^\circ$) with 40 vertical levels, equally spaced in logarithmic scale between 10 Pa and 800 bars. Along with the various parameterizations of physical processes described in refs.\cite{Turbet2021,Spiga2020,charnay14,charnay_3d_2015,turbet_habitability_2016}, the \texttt{Generic PCM} treats clouds as radiatively active tracers of fixed radii. 

The model is initialised using temperature profiles from the radiative-convective 1D model Exo-REM\cite{blain_2021}. The radiative data are computed offline using the out-of-equilibrium chemical profiles of the Exo-REM run. We use 27 frequency bins in the stellar channel (0.261 to 10.4 $\mu$m) and 26 in the planetary channel (0.625 to 324 $\mu$m), all bins including 16 k-coefficients. We start the model from a rest state (no winds), with a horizontally homogeneous temperature profile. Models are integrated for 2000 days, which is long enough to complete the spin-up phase of the simulation above the photosphere. We do not include Rayleigh drag in our models. The simulations are performed including clouds of \ce{Mg2SiO4}, with varying cloud radii ($0.1, 0.5, 1, 3, 5 \mathrm{\mu}$m). We also computed cloudless and \ce{Mg2SiO4} models with a 10x solar metallicity and the same radii for the cloud particles. Regardless of the composition and size of the clouds, our model clearly indicates that there is no cloud formation on the day-side. Asymmetric limbs are a natural result of our model with the eastern terminator being warmer while the western limb is cloudier and cooler. Spectral phase curves were produced using the Pytmosph3R code\cite{falco_toward_2022}.

\subsection*{SPARC/MITgcm with Radiative Transfer Post-Processing by gCMCRT} 
SPARC/MITgcm couples a state-of-the-art non-grey, radiative transfer code with the \texttt{MITgcm}\cite{showman2009}. The \texttt{MITgcm} solves the primitive equations of dynamical meteorology on a cube-sphere grid\cite{Adcroft2004}. It is coupled to the non-grey radiative transfer scheme based on the plane-parallel radiative transfer code of ref.\cite{Marley1999}. The stellar irradiation incident on \mbox{WASP-43b} is computed with a PHOENIX model\cite{allard1995, hauschildt1999, husser2013}. We use previously published opacities\cite{Freedman2008}, including more recent updates\cite{Freedman2014,Marley2021}, and the molecular abundances are calculated assuming local chemical equilibrium\cite{Visscher2010}. In the GCM simulations, the radiative transfer calculations are performed on 11 frequency bins ranging from 0.26 to 300~$\mu$m, with 8 k-coefficients per bin statistically representing the complex line-by-line opacities\cite{Kataria2015}. The strong visible absorbers TiO and VO are excluded in our k tables similar to our previous GCMs of \mbox{WASP-43b}\cite{Kataria2015,VenotEtal2020-JWST-WASP-43b} that best match the observed dayside emission spectrum and photometry. 

Clouds in the GCM are modeled as tracers that are advected by the flow\cite{parmentier2013} and can settle under gravity. Their formation and evaporation are subjected to chemical equilibrium predictions, i.e., the condensation curves of various minerals described in ref.\cite{Visscher2010}. The conversion between the condensable ``vapour'' and clouds is treated as a simple linear relaxation over a short relaxation timescale of 100 s. The scattering and absorption of the spatial- and time-dependent clouds are included in both the thermal and visible wavelengths of the radiative transfer. A similar dynamics-cloud-radiative coupling has been developed in our previous GCMs with simplified radiative transfer and has been used to study the atmospheric dynamics of brown dwarfs\cite{tan2021a,tan2021b} and ultra-hot Jupiters\cite{Komacek2022}. Clouds are assumed to follow a log-normal size distribution\cite{AckermanMarley2001apj} which is described by the reference radius $r_0$ and a nondimensional deviation $\sigma$: $n(r) = \frac{\mathcal{N}}{\sqrt{2\pi}\sigma r}\exp\left(-\frac{[\ln (r/r_0)]^2}{2\sigma^2}\right)$, where $n(r)$ is the number density per radius bin of $r$ and $\mathcal{N}$ is the total number density. $\sigma$ and $r_0$ are free parameters and the local $\mathcal{N}$ is obtained from the local mass mixing ratio of clouds. The size distribution is held fixed throughout the model and is the same for all types of clouds.

Our GCMs do not explicitly impose a uniform radiative heat flux at the bottom boundary but rather relax the temperature of the lowest model layer (i.e., the highest pressure layer) to a certain value over a short timescale of 100 s. This assumes that the deep GCM layer reaches the convective zone and the temperature there is set by the interior convection that ties to the interior structure of the planet. This lowest-layer temperature is in principle informed by internal structure models of \mbox{WASP-43b} which are run by MESA hot Jupiter evolution modules\cite{Komacek2017} to match the present radius of \mbox{WASP-43b}. In most models, this lowest-layer temperature is about 2509 K at about 510 bars.
The horizontal resolution of our GCMs is typically C48, equivalent to about $1.88^{\circ}$ per grid cell. The vertical domain is from $2\times 10^{-4}$ bar at the top to 700 bars at the bottom and is discretized to 53 vertical layers. We typically run the simulation for over 1200 days and average all physical quantities over the last 100 days of the simulations.  

All our GCMs assume  a solar composition. We performed a baseline cloudless model and  one case with only MnS and Na$_2$S clouds with $r_0=3$ microns, and then a few cases with MnS, Na$_2$S and MgSiO$_3$ clouds with $r_0=1$, 1.5, 2, and 3 microns.  The $\sigma$ is held fixed at 0.5 in all our cloudy GCMs.

We post-process our GCM simulations with the state-of-the-art gCMCRT code, which is a publicly available hybrid Monte Carlo radiative transfer (MCRT) and ray-tracing radiative transfer code. The model is described in detail in ref.\cite{Lee2022} and has been applied to a range of exoplanet atmospheres\cite{Lee2021,Komacek2022}. gCMCRT can natively compute albedo, transmission, and emission spectra at both low and high spectral resolution.  gCMCRT uses custom k tables, which take cross-section data from both HELIOS-K\cite{Grimm2021} and EXOplanet Pressure-broadened LINES\cite{Gharib2021}. Here, we apply gCMCRT to compute low-resolution emission spectra and phase curves at R $\approx$ 300 from our GCM simulations. We use the three-dimensional temperature and condensate cloud tracer mixing ratio from the time-averaged end-state of each case. We assume the same cloud particle size distribution as our GCMs. 

\subsection*{expeRT/MITgcm}
The GCM expeRT/MITgcm uses the same dynamical core as SPARC/MITgcm and solves the hydrostatic primitive equations on a C32 cubed-sphere grid\cite{Adcroft2004}. It resolves the atmosphere above 100~bar on 41 log spaced cells between $1 \times 10^{-5}$ bar and $100$~bar.Below 100~bar, six linearly spaced grid cells between $100$~bar and $700$~bar are added. The model expeRT/MITgm thus resolves deep dynamics in non-inflated hot Jupiters like \mbox{WASP-43b}\cite{Carone2020,Schneider2022}.  

The GCM is coupled to a non-grey radiative transfer scheme based on petitRADTRANS\cite{Molliere2019}. Fluxes are recalculated every 4th dynamical timestep. Stellar irradiation is described by the spectral fluxes  from the PHOENIX model atmosphere suite\cite{allard1995, hauschildt1999, husser2013}. The GCM operates on a pre-calculated grid of correlated k-binned opacities. Opacities from the ExoMol database\cite{Chubb2021} are precalculated offline on a grid of 1000 logarithmically spaced temperature points between 100 K and 4000 K for every vertical layer. We further include the same species as shown in ref.\cite{Schneider2022} except TiO and VO to avoid the formation of a temperature inversion in the upper atmosphere. These are: \ce{H2O}\cite{Polyansky2018}, \ce{CH4}\cite{yurchenko17}, \ce{CO2}\cite{20YuMeFr.co2}, \ce{NH3}\cite{Coles_2019_NH3}, \ce{CO}\cite{li2015}, \ce{H2S}\cite{azzam2016}, \ce{HCN}\cite{Barber2014}, \ce{PH3}\cite{Sousa_Silva_2014}, \ce{FeH}\cite{Wende2010}, Na\cite{Piskunov1995,allard1995}, and K\cite{Piskunov1995,allard1995}. For Rayleigh scattering, the opacities are \ce{H2}\cite{Dalgarno1962} and \ce{He}\cite{Chan1965}and we add the following CIA opacities: \ce{H2}-\ce{H2}\cite{Richard2012}, and \ce{H2}-{He}\cite{Richard2012}. We use for radiative transfer calculations in the GCM the same wavelength resolution as SPARC/MITgcm (S1), but incorporate 16 instead of 8 k-coefficients. Two cloud-free \mbox{WASP-43b} GCM simulations were performed, one with solar and one 10x solar element abundances. Each simulation ran for 1500~days to ensure that the deep wind jet has fully developed. The GCM results used in this paper were time averaged over the last 100~simulation days.

Spectra and phase curves are produced from our GCM results in post-processing with petitRADTRANS\cite{Molliere2019} and prt\_phasecurve\cite{Schneider2022} using a spectral resolution of $R=100$ for both the phase curve and the spectra.

\subsection*{RM-GCM}\label{sec:rmgcm}
Originally adapted from the GCM of ref.\cite{Hoskins1975} by refs.\cite{MenouRauscher2009,RauscherMenou2010,RauscherMenou2012}, the RM-GCM has been applied to the numerous investigations of hot Jupiters  and mini-Neptunes\cite{rauscher2014atmospheric,Roman&Rauscher2017,may2020super,beltz2022magnetic}. The GCM's dynamical core solves the primitive equations of meteorology using a spectral representation of the domain, and it is coupled to a two-stream, double-grey radiative transfer scheme based on ref.\cite{toon1989}. Recent updates have added aerosol scattering\cite{Roman&Rauscher2017} with radiative feedback\cite{Roman&Rauscher2019,Roman2021}. 

Following ref.\cite{Roman2021}, aerosols are representative of condensate clouds and are treated as purely temperature-dependent sources of opacity, with constant mixing ratios set by the assumed solar elemental abundances. The optical thicknesses of the clouds are determined by converting the relative molecular abundances (or partial pressures) of each species into particles with prescribed densities and radii\cite{Roman2021}. The model includes up to 13 different cloud species of various condensation temperatures, abundances, and scattering properties. Places where clouds overlap have mixed properties, weighted by the optical thickness of each species. 

Simulations from this GCM included a clear atmosphere and two sets of cloudy simulations. Following ref.\cite{Roman2021}, one set of cases included 13 different species: KCl, ZnS, Na$_2$S, MnS, Cr$_2$O$_3$, SiO$_2$, Mg$_2$SiO$_4$, VO, Ni, Fe, Ca$_2$SiO$_4$, CaTiO$_2$, and Al$_2$O$_3$; the other set omitted ZnS, Na$_2$S, MnS, Fe, and Ni, based on considerations of nucleation efficiency\cite{Gao2020}. For both cloud composition scenarios, the models explored the observational consequences of variations in the cloud deck's vertical thickness through a series of simulations with clouds tops truncated over a range of heights at 5-layer intervals (roughly a scale height), ranging from 5 to 45 layers of the 50-layer model. This effectively mimics a range of vertical mixing strengths. From the complete set published in ref.\cite{murphy2023}, we selected a subset, with clouds of maximum vertical extents between two and nine scale heights from each of the two cloud composition scenarios. 

Simulations were initialized with clear skies, no winds, and no horizontal temperature gradients. We ran the simulations for over 3500 planetary orbits, assuming tidal synchronization. Resulting temperature, wind, and cloud fields of the GCM were then post-processed\cite{Zhang2017, Malsky2021} to yield corresponding emission phase curves. 

\subsection*{THOR}
\texttt{THOR}\cite{mendonca2016,deitrick2020} is an open-source GCM developed to study the atmospheres and climates of exoplanets, free from Earth- or Solar System-centric tunings. The core that solves the fluid flow equations, the dynamical core, solves the non-hydrostatic compressible Euler equations on an icosahedral grid\cite{tomita2004, mendonca2016}. \texttt{THOR} has been validated and used to simulate the atmosphere of Earth\cite{mendonca2016,mendonca2022}, Solar System planets\cite{mendonca2020b,shao2022} and exoplanets\cite{mendonca2016,deitrick2020,deitrick2022}. 

For this work, \texttt{THOR} used the same configuration as with previously published simulations to study the atmospheric temperature structure, cloud cover and chemistry of \mbox{WASP-43b}\cite{mendonca2018a,mendonca2018b,tsai2018}. Two simulations were conducted, one with a clear atmosphere and another with a cloud structure on the nightside of the planet. To represent the radiative processes, \texttt{THOR} uses a simple two-band formulation calibrated to reproduce the results from more complex non-grey models on \mbox{WASP-43b}\cite{Kataria2015,malik2017}. A simple cloud distribution on the nightside of the planet and optical cloud properties are parameterized\cite{mendonca2018a} and adapted to reproduce previous HST\cite{Stevenson2014} and Spitzer\cite{Stevenson2017,mendonca2018a} observations. These simulations on \mbox{WASP-43b} with \texttt{THOR} have also been used to test the performance of future Ariel phase curves observations\cite{charnay2022}.

Both simulations, with clear and cloudy atmospheres, started with isothermal atmospheres (1440 K, equilibrium temperature) and integrated for roughly 9400 planetary orbits (assuming a tidally locked configuration) until a statistically steady state of the deep atmosphere thermal structure was reached. The long integration avoids biasing the results toward the set initial conditions\cite{mendonca2020b}. 

The multiwavelength spectra are obtained from post-processing the 3D simulations with a multiwavelength radiative transfer model\cite{mendonca2015}. The disc-averaged planet spectrum is calculated at each orbital phase by projecting the outgoing intensity for each geographical location of the observed hemisphere.  The spectra include cross-sections of the main absorbers in the infrared, drawn from the ExoMOL (\ce{H2O}\cite{Polyansky2018}, \ce{CH4}\cite{Yurchenko2014}, \ce{NH3}\cite{Yurchenko2011}, \ce{HCN}\cite{Harris2006} and \ce{H2S}\cite{azzam2016}), HITEMP\cite{Rothman2010} (\ce{CO2} and \ce{CO}), and HITRAN\cite{rothman2013} (\ce{C2H2}) databases. The Na and K resonance lines\cite{draine2011} are also added, and \ce{H2}-\ce{H2} \ce{H2}-He CIA\cite{Richard2012}. The atmospheric bulk composition was assumed to have solar abundance (consistent with HST-WFC3 spectrum observations), and each chemical species concentration was calculated with the FastChem model\cite{stock2018}. The PHOENIX models\cite{allard1995, hauschildt1999, husser2013} were used for the WASP-43 star spectrum.

\section*{Atmospheric Retrieval Models}

We perform atmospheric retrievals on the phase-resolved emission spectra using six different retrieval frameworks, each described in turn below. The chemical constraints from these analyses are summarized in \ref{tab:ret_abundances} and \ref{tab:ret_chi_sqr}, and the spectral fits obtained are shown in \ref{fig:retrieved_spectra}. Across the six retrieval analyses, we use an error inflation parameter to account for the effects of unknown data and/or model uncertainties. This free parameter is wavelength-independent and multiplies the 1$\sigma$ error bars in the calculation of the likelihood function in the Bayesian sampling algorithm.

\subsection*{HyDRA Retrieval Framework}

The \textsc{H}y\textsc{DRA} atmospheric retrieval framework\cite{Gandhi2018} consists of a parametric atmospheric forward model coupled to \textsc{PyMultiNest}\cite{Feroz2009,buchner14}, a Nested Sampling Bayesian parameter estimation algorithm\cite{Skilling2006}. \textsc{H}y\textsc{DRA} has been applied to hydrogen-rich atmospheres\cite{Gandhi2020, Piette2020b}, and further adapted for secondary atmospheres\cite{Piette2022} and high-resolution spectroscopy in both 1D and 2D\cite{Gandhi2019,Gandhi2022}. The input parameters for the atmospheric forward model include constant-with-depth abundances for each of the chemical species considered, six temperature profile parameters corresponding to the temperature profile model of ref.\cite{Madhusudhan2009}, and a constant-with-wavelength multiplicative error inflation parameter to account for model uncertainties. We additionally include a dilution parameter, $A_{\rm HS}$, for the dayside, morning and evening hemispheres, which multiplies the emission spectrum by a constant factor $<1$ and accounts for temperature inhomogeneities in each hemisphere\cite{Taylor_2020}.

We consider opacity contributions from the chemical species which are expected to be present in hot Jupiter atmospheres and which have opacity in the MIRI LRS wavelength range: H$_2$O\cite{Rothman2010}, CH$_4$\cite{Yurchenko2013,Yurchenko2014}, NH$_3$\cite{Yurchenko2011}, HCN\cite{Harris2006,Barber2014}, CO\cite{Rothman2010}, CO$_2$\cite{Rothman2010}, C$_2$H$_2$\cite{rothman2013,Gordon2017}, SO$_2$\cite{Underwood2016mnras}, H$_2$S\cite{azzam2016,Chubb2018} and collision-induced absorption (CIA) due to H$_2$-H$_2$ and H$_2$-He\cite{Richard2012}. The line-by-line absorption cross sections for these species are calculated following the methods described in ref.\cite{Gandhi2018}, using data from each of the references listed. We further explore retrievals with a simple silicate cloud model, which includes the modal particle size, cloud particle abundance, cloud base pressure and a pressure exponent for the drop-off of cloud particle number density with decreasing pressure. The opacity structure of the cloud is calculated using the absorption cross sections of ref.\cite{Pinhas2017}.

Given the input chemical abundances, temperature profile and cloud parameters, the forward model calculates line-by-line radiative transfer to produce the thermal emission spectrum at a resolution of $R{\sim}15000$. The spectrum is then convolved to a resolution of 100, binned to the same wavelength bins as the observations and compared to the observed spectrum to calculate the likelihood of the model instance. The Nested Sampling algorithm explores the parameter space using 2000 live points, and further calculates the Bayesian Evidence of the retrieval model, which can be used to compare different models\cite{Trotta2008}. In particular, we calculate the detection significance of a particular chemical species by comparing retrievals which include/exclude that species, fixing the value of the error inflation parameter to be the median retrieved value found with the full retrieval model.

Across the four phases, the only chemical species detected with statistical significance ($\gtrsim3\sigma$) is H$_2$O. The retrieved H$_2$O abundances are in the range ${\sim}$30--10$^4$~ppm (1-$\sigma$ uncertainties), with detection significances varying between ${\sim}$3-4$\sigma$ (see \ref{tab:ret_abundances}). We do not detect CH$_4$ at any phase, and place an upper limit of 16~ppm on the nightside CH$_4$ abundance, potentially indicating disequilibrium chemistry processes as described in the main text. We do not detect NH$_3$ at any phase either, consistent with the very low NH$_3$ abundances predicted by both chemical equilibrium and disequilibrium models\cite{VenotEtal2020-JWST-WASP-43b}. The retrievals do not favor cloudy models over clear models with statistical significance, with extremely weak preferences of $<1\sigma$ at all phases. Additionally, the posterior probability distributions for the cloud parameters are unconstrained. \ref{fig:contribution_functions} shows the pressure ranges of the atmospheric model probed by the observations.

\subsection*{\textsc{PyratBay} Retrieval Framework}

\textsc{PyratBay} is an open-source framework that enables atmospheric modeling, spectral synthesis, and atmospheric retrievals of exoplanet observations\cite{CubillosBlecic2021-PyratBay}. The atmospheric model consists of parametric temperature, composition, and altitude profiles as a function of pressure, for which emission and transmission spectra can be generated. The radiative transfer module considers opacity from alkali lines\cite{BurrowsEtal2000apjBDspectra}, Rayleigh scattering\cite{Kurucz1970saorsAtlas, LecavelierEtal2008aaRayleighHD189733b}, Exomol and HITEMP molecular line lists\cite{TennysonEtal2016jmsExomol, RothmanEtal2010jqsrtHITEMP} pre-processed with the \textsc{repack} package\cite{Cubillos2017apjRepack} to extract the dominant line transitions, collision-induced absorption\cite{BorysowEtal2001jqsrtH2H2highT}, and cloud opacities. The \textsc{PyratBay} retrieval framework has the ability to stagger model complexity and explore a hierarchy of different model assumptions. Temperature models range from an isothermal profile to physically motivated parameterized models\cite{ParmentierGuillot2014aapTmodel, MadhusudhanSeager2009apjRetrieval}. Composition profiles range from the simpler constant-with-altitude ``free abundance" to the more complex ``chemically consistent" retrievals, the latter accomplished via the TEA code\cite{BlecicEtal2016apsjTEA}; while cloud condensate prescriptions range from the classic ``power law+grey" to a ``single-particle-size" haze profile, a partial-coverage factor ``patchy clouds"\cite{LineParmentier2016-patchy}, and the complex parameterized Mie-scattering thermal stability cloud (\textsc{TSC}) model (Blecic et al., in prep). The \textsc{TSC} cloud prescription, initially inspired by refs.\cite{benneke_strict_2015, AckermanMarley2001apj}, has additional flexibility in the location of the cloud base and was further improved for this analysis (see below). The formulation utilizes a parameterized cloud shape, effective particle size, and gas number density below the cloud deck, while the atmospheric mixing and settling are wrapped up inside the cloud extent and the condensate mole fraction as free parameters. This cloud model was applied to \mbox{WASP-43b} {\em MIRI/JWST} phase curve simulations\cite{VenotEtal2020-JWST-WASP-43b}, generated during the {\em JWST} preparatory phase, in anticipation of the actual \mbox{WASP-43b} {\em MIRI/JWST} observations. We showed that the \textsc{TSC} model has the ability to distinguish between MgSiO$_{3}$ and MnS clouds on the nightside of the planet.

For this analysis, we conducted a detailed investigation using various model assumptions. We started by exploring simple temperature prescriptions and gradually moved toward more complex ones. Initially, we considered opacity contributions from all chemical species expected to be observed in the {\em MIRI} wavelength range (H$_2$O, CH$_4$, NH$_3$, HCN, CO, CO$_2$, C$_2$H$_2$, SO$_2$, H$_2$S), but eventually focused only on those that are fit by the data. We also implemented the dilution parameter\cite{Taylor_2020} and an error inflation factor, which account for some additional model and data uncertainties. The constraints on H$_2$O (together with the detection significance\cite{Benneke2013}) and the upper limit for CH$_4$ for all phases are given in \ref{tab:ret_abundances}. The abundances of these species across all phases were largely model independent. However, the tentative constraints on NH$_3$, which we saw in multiple phases, were strongly model dependent, and were completely erased with the inclusion of the dilution parameter and the error inflation, thus we do not report them here. WASP-43\,b emission spectra were computed at the resolution of $R{\sim}15000$ utilizing opacity sampling of high-resolution pre-computed cross sections ($R{\sim}10^6$) of considered species. 
Furthermore, we thoroughly examined the possibility of detecting clouds in each of the four-quadrant phases, with a special emphasis on the nightside of the planet. To do this, we employed the \textsc{TSC} model, as in our previous analysis\cite{VenotEtal2020-JWST-WASP-43b}, and explored a range of cloud species, MgSiO$_{3}$, MnS, ZnS, KCl, which would condense under the temperature regimes expected for \mbox{WASP-43b}\cite{MorleyEtal2012} (\ref{fig:nightside_clouds}, left panel). We also introduced the effective standard deviation of the log-normal distribution\cite{AckermanMarley2001apj} as a free parameter ($\sigma_{log}$), allowing for even more flexibility in our cloud model (\ref{fig:nightside_clouds}, right, last sub-panel). To thoroughly explore the parameter space, we used two Bayesian samplers, the differential-evolution Markov Chain Monte Carlo (MCMC) algorithm\cite{terBraak2008SnookerDEMC}, implemented following ref.\cite{CubillosEtal2017apjRednoise}, and the Nested-sampling algorithm, implemented through PyMultiNest\cite{feroz09, buchner14}, utilizing 15 million models and 2000 live points, respectively. Our investigation did not provide constraints on any of the cloud parameters for any of the explored cloud condensates at any of the planetary phases, indicating the absence of detectable spectral features from clouds in the observations (\ref{fig:nightside_clouds}, right panels).

\subsection*{\textsc{NEMESIS} Retrieval Framework}
\textsc{NEMESIS}\cite{Irwin2008,kt2018} is a free retrieval framework that uses a fast correlated-k\cite{lacisoinas} forward model, combined with either an optimal estimation or nested sampling retrieval algorithm. It has been used to perform retrievals on spectra of numerous planetary targets, both inside and outside the Solar System\cite{barstow20,Irwin2020}. In this work, we use the \textsc{PyMultiNest} sampler\cite{buchner14} with 500 live points. The retrieval model presented includes four spectrally active gases, H$_2$O\cite{Polyansky2018}, CO\cite{li2015}, CH$_4$\cite{yurchenko17} and NH$_3$\cite{Coles_2019_NH3} , with k-tables calculated as in ref.\cite{Chubb2021}; we did not include CO$_2$ or H$_2$S after initial tests indicated these were not required to fit the spectrum. All gases are assumed to be well-mixed in altitude. Collision-induced absorption from H$_2$ and He is taken from refs.\cite{BorysowEtal2001jqsrtH2H2highT,Borysow2002}. The spectrum is calculated at the resolution of the observation, using optimised channel integrated ktables generated from original ktables with a resolving power R=1000. The temperature profile is modelled as a 3-parameter Guillot profile, after ref.\cite{ParmentierGuillot2014aapTmodel}, with free parameters $\kappa$, $\gamma$ and $\beta$ ($\alpha$ is fixed to be zero). We include a well-mixed, spectrally grey cloud with a scalable total optical depth with a cloud top at 12.5 mbar. The other retrieved parameters are a hotspot dilution factor for phases 0.25, 0.5 and 0.75, following ref.\cite{Taylor_2020}, and an error inflation term.  

To calculate the detection significance for H$_2$O, we run the retrieval with and without H$_2$O, with all other aspects of the run identical. We then take the difference of the PyMultinest global log evidence values for the two scenarios, and convert from log(Bayesian Evidence) to sigma following\cite{Trotta2008}. The 99\% upper limit for CH$_4$ is calculated from the equally weighted posterior distribution.  We also attempt to retrieve CO and NH$_3$ abundances. CO is generally poorly constrained, and NH$_3$ is unconstrained for phases 0 and 0.75; for log(NH$_3$), we recover a 99\% upper limit of -2.2 at phase 0.25, and -3.9 at phase 0.5.  The cloud opacity is also generally unconstrained, with the total optical depth able to span several orders of magnitude. We stress that this model is very crude since it has only one variable cloud parameter, and further exploration of suitable cloud models for mid-infrared phase curves is warranted in future work. 

\subsection*{\textsc{SCARLET} Retrieval Framework}

We perform atmospheric retrievals on the four phase-resolved spectra using the SCARLET framework\cite{benneke_strict_2015,benneke_water_2019}. The planetary disk-integrated thermal emission $F_p$ is modeled for a given set of atomic/molecular abundances, temperature-pressure profile, and cloud properties. We compare our model spectra to the observations by normalizing the thermal emission $F_p$ using a PHOENIX\cite{allard1995, hauschildt1999, husser2013} stellar model spectrum with effective temperature $T_{\rm eff}$ = 4300~K and surface gravity $\log g$ = 4.50. The model spectra are computed at a resolving power of R = 15,625, convolved to the resolving power of MIRI/LRS and then binned to the 11 spectral bins ($<$10.5~$\mu$m) considered in the analysis, assuming the throughput to be uniform over a single bin. 

The atmospheric analysis is performed considering thermochemical equilibrium, where the metallicity [M/H] ($\mathcal{U}[-3,3]$) and carbon-to-oxygen ratio ($\mathcal{U}[0,3]$) are free parameters that dictate the overall atmospheric composition. We use a free parametrization of the temperature-pressure profile\cite{pelletier_where_2021} by fitting for $N$ = 4 temperature points ($\mathcal{U}[100,4400]~K$) with a constant spacing in log-pressure. The temperature-pressure profile is interpolated to the 50 layers (P = $10^2$ -- $10^{-6}$ bar) considered in the model using a spline function to produce a smooth profile. We use a grid of chemical equilibrium abundances produced with FastChem2\cite{fastchem2} to interpolate the abundance of species as a function of temperature and pressure for given values of [M/H] and C/O. The species considered in the equilibrium chemistry are H, H$^-$\cite{bell1987,john1988}, H$_2$, He, H$_2$O\cite{Polyansky2018}, OH\cite{Rothman2010}, CH$_4$\cite{Yurchenko2014}, C$_2$H$_2$\cite{Chubb_2020}, CO\cite{Rothman2010}, CO$_2$\cite{Rothman2010}, NH$_3$\cite{Coles_2019_NH3}, HCN\cite{Barber2014}, PH$_3$\cite{Sousa_Silva_2014}, TiO\cite{McKemmish2019}, and VO\cite{McKemmish2016}. All opacities for these species are considered when computing the thermal emission. We account for potential spatial atmospheric inhomogeneities in the planetary disk that are observed at a given phase by including an area fraction parameter $A_{HS}$ ($\mathcal{U}[0,1]$), which is meant to represent the possibility of a fraction of the disk contributing to most of the observed thermal emission\cite{Taylor_2020}. This parameter is considered for all phases with the exception of the nightside, which is expected to be relatively uniform. Finally, we fit for an error inflation parameter $k_\sigma$ ($\mathcal{U}[0.1,10]$) to account for potential model and data uncertainty, which results in a total of 8 (7 for the nightside) free parameters. We consider 8 walkers per free parameter for the retrievals which are run for 30,000 steps. The first 18,000 steps are discarded when producing the posterior distributions of the free parameters.

\subsection*{\textsc{PLATON} Retrieval Framework}
\textsc{PLATON}\cite{zhang_2020}, PLanetary Atmosphere Tool for Observer Noobs, is a Bayesian retrieval tool which assumes equilibrium chemistry. We adopt the T/P profile parameterization of ref.\cite{line_2013}, and use the \texttt{dynesty} nested sampler to retrieve the following free parameters: stellar radius, stellar temperature, log(Z), C/O, 5 T/P parameters (log($\kappa_{th}$), log($\gamma$), log($\gamma_2$), $\alpha$, $\beta$), and error multiple.  The stellar radius and temperature are given Gaussian priors with means and standard deviations set by the measurements in ref.\cite{Bonomo2017}: $4400 \pm 200$ K and $0.667 \pm 0.011 R_\odot$ respectively.  The combination of the two have a similar effect to the dilution parameter of other retrieval codes, which multiplies the emission spectrum by a constant. For phase 0, we obtain a significantly better fit when methane opacity is set to zero (thus removing all spectral features from methane). We therefore adopt this as the fiducial model, whereas for other phases, we do not zero out any opacities.

For all retrievals, we use nested sampling with 1000 live points.  The opacities (computed at R=10,000) and the line lists used to compute them are listed in ref.\cite{zhang_2020}.  We include all 31 species in retrieval, notably including H$_2$O, CO, CO$_2$, CH$_4$ (except on the nightside), H$_2$S, and NH$_3$.

\subsection*{\textsc{ARCiS} Retrieval Framework}

\textsc{ARCiS} (ARtful modelling Code for exoplanet Science) is an atmospheric modelling and Bayesian retrieval code\cite{18OrMi,20MiOrCh} which utilises the Multinest\cite{feroz09} Monte Carlo nested sampling algorithm.
The code was used in previous retrievals of the atmosphere of \mbox{WASP-43b} in transmission\cite{20ChMiKa}, using the observations of ref.\cite{Kreidberg2014}, and in phase-resolved emission\cite{22ChMi}, using the observations of refs.\cite{Stevenson2014,Stevenson2017,14BlHaMa,Morello2019}. Ref.\cite{20ChMiKa} found some evidence that AlO improves the fit of the transmission spectra of \mbox{WASP-43b} in the 1.1~-~1.6~$\mu$m region. We therefore include in our models for this work the following set of molecules in our free molecular retrievals:
H$_2$O\cite{Polyansky2018}, CO\cite{li2015}, CO$_2$\cite{20YuMeFr.co2}, NH$_3$\cite{Coles_2019_NH3}, CH$_4$\cite{yurchenko17}, and AlO\cite{ExoMol_AlO}. The molecular line lists are from the ExoMol\cite{TennysonEtal2016jmsExomol,20TeYuAl} or HITEMP\cite{Rothman2010} databases as specified, and k-tables from the ExoMolOP opacity database\cite{Chubb2021}. Collision-induced absorption for H$_2$ and He are taken from refs.\cite{BorysowEtal2001jqsrtH2H2highT,Borysow2002}. We explore the inclusion of a variety of additional molecules which have available line list data with spectral features in the region of our observations, including HCN\cite{Barber2014}, SiO\cite{ExoMol_SiO}, and N$_2$O\cite{Rothman2010}. We use the Bayes factor, which is the difference between the Nested Sampling Global Log-Evidence (log($E$)) between two models, in order to assess whether the inclusion of a particular parameter is statistically significant. For this, we run a retrieval with the base set of species only and another with the base set plus the molecule being assessed. The difference in log($E$) between the two models is converted to a significance in terms of $\sigma$ using the metric of ref.\cite{08Trotta}. We explore the inclusion of a simple grey, patchy cloud model, which parametrises cloud top pressure and degree of cloud coverage (from 0 for completely clear to 1 for completely covered). We use 1000 live points and a sampling efficiency of 0.3 in Multinest for all retrievals. 

We run retrievals both including and not including a retrieved error-inflation parameter. The error-inflation parameter is implemented as per ref.\cite{15LiTeBu} to account for underestimated uncertainties and/or unknown missing forward model parameters. All phases apart from 0 retrieved a parameter which increases the observational error bars by between 2~-~3~$\times$ their original values. The pressure-temperature profile 
parametrisation of ref.\cite{10Guillot} is used in all cases. We find evidence for the inclusion of H$_2$O for all four phases, although this evidence goes from strong to weak when error inflation is included for the morning phase (0.75). We find no strong evidence for CH$_4$ at any phase, with 95$\%$ confidence upper limits on the log of the volume mixing ratio (VMR) of -4.9, -2.9, -3.2, and -2.2
for phases 0, 0.25, 0.5, and 0.75, respectively. We find some model-dependent hints of moderate evidence (based on the metric of ref.\cite{08Trotta}) of 4.4$\sigma$ for NH$_3$ at phase 0.5 (constrained to log(VMR)~=~-4.5$^{+0.7}_{-0.5}$), 3.1$\sigma$ for CO at phase 0.5 (log(VMR)~=~-1.7$^{+0.5}_{-0.7}$), and 2.6$\sigma$ for CO at phase 0.25 (log(VMR)~=~-4.0$^{+0.3}_{-0.4}$). However, these disappear when the error-inflation parameter is introduced. We are not able to constrain any of the cloud parameters for any phase, and so do not find a statistical reason to include our simple cloud parametrisation in the models in order to better fit the observations.

\section*{Data Availability}
The data used in this paper are associated with JWST DD-ERS program 1366 (PIs Batalha, Bean, and Stevenson; observation 11) and are publicly available from the Mikulski Archive for Space Telescopes (\url{https://mast.stsci.edu}). Additional intermediate and final results from this work will be archived on Zenodo at \url{https://zenodo.org/doi/10.5281/zenodo.10525170} upon final publication.

\section*{Code Availability}
We used the following codes to process, extract, reduce and analyse the data: STScI's JWST Calibration pipeline\cite{stsci2022jwst}, Eureka!\cite{Bell2022Eureka}, TEATRO, SPARTA\cite{kempton2023}, Generic PCM\cite{Turbet2021,Spiga2020,charnay14,charnay_3d_2015,turbet_habitability_2016}, SPARC/MITgcm\cite{showman2009,parmentier2013,tan2021a,tan2021b}, expeRT/GCM\cite{Adcroft2004,Carone2020,Schneider2022}, RM-GCM\cite{MenouRauscher2009,RauscherMenou2010,RauscherMenou2012,Roman&Rauscher2017,Roman2021}, THOR\cite{mendonca2015,mendonca2016,mendonca2018a,malik2017,malik2019,deitrick2020}, HyDRA\cite{Gandhi2018}, PyratBay\cite{CubillosBlecic2021-PyratBay}, NEMESIS\cite{Irwin2008,kt2018}, SCARLET\cite{benneke_strict_2015,benneke_water_2019}, PLATON\cite{zhang_2020}, starry\cite{starry_v1.2.0}, exoplanet\cite{exoplanet:joss}, PyMC3\cite{pymc3}, emcee\cite{foreman-mackey_emcee_2013}, dynesty\cite{dynesty_v1_2_3}, numpy\cite{numpy}, astropy\cite{astropy2013, astropy2018}, and matplotlib\cite{matplotlib}.

\begin{addendum}

 \item[Acknowledgments]
 TJB acknowledges funding support from the NASA Next Generation Space Telescope Flight Investigations program (now JWST) via WBS 411672.07.05.05.03.02. JB acknowledges the support received in part from the NYUAD IT High Performance Computing resources, services, and staff expertise. ED acknowledges funding as a Paris Region Fellow through the Marie Sklodowska-Curie Action. MZ and BVR acknowledge funding from the 51 Pegasi b Fellowship. ADF acknowledges support from the NSF Graduate Research Fellowship Program. MM, DP, and LW acknowledge funding from the NHFP Sagan Fellowship Program. PEC is funded by the Austrian Science Fund (FWF) Erwin Schroedinger Fellowship program J4595-N. KLC acknowledges funding from STFC, under project number ST/V000861/1. LD acknowledges funding from the KU Leuven Interdisciplinary Grant (IDN/19/028), the European Union H2020-MSCA-ITN-2019 under Grant no.\ 860470 (CHAMELEON) and the FWO research grant G086217N. OV acknowledges funding from the ANR project `EXACT' (ANR-21-CE49-0008-01) and from the Centre National d'\'{E}tudes Spatiales (CNES). LT and BC acknowledge access to the HPC resources of MesoPSL financed by the Region Ile de France and the project Equip@Meso (reference ANR-10-EQPX-29-01) of the programme Investissements d'Avenir supervised by the Agence Nationale pour la Recherche.

 \item[Author Contributions Statement] All authors played a significant role in one or more of the following: development of
the original proposal, management of the project, definition of the target list and observation plan, analysis
of the data, theoretical modeling, and preparation of this manuscript. Some specific contributions are listed
as follows. NMB, JLB, and KBS provided overall program leadership and management. LK and NC coordinated the MIRI working group. LK, VP, KBS, DS, EK, OV, and PC made
significant contributions to the design of the program and the observing proposal. KBS generated the observing plan with input from
the team. AD, PL, RC, AC, GM, and MM
led or co-led working groups and/or contributed to significant strategic planning efforts like the design and
implementation of the pre-launch Data Challenges. PC, DS, RC, PL, and JB generated simulated data
for pre-launch testing of methods. LK, TJB, MR, NC, VP, AP, and JM contributed significantly to the writing
of this manuscript. TJB, NC, MZ, and ED contributed to the development of data analysis
pipelines and/or provided the data analysis products used in this analysis i.e., reduced the data, modeled the
light curves, and/or produced the planetary spectrum.  AP coordinated the atmospheric retrieval analysis with contributions from JB, LPC, MZ, KC, and JB. MR coordinated the GCM results and intepretation with contributions from XT, LT, LC, JM, IM, and MR. TJB, NC, PC, JB, LPC, and MH generated figures for this manuscript. MN, XZ, BR, JK, MLM, BC, SC, and RH provided significant feedback to the manuscript, and GM and KC coordinated comments from all authors.                                                 

\end{addendum}

\noindent \textbf{Competing Interests Statement} The authors declare no competing interests.\\

\noindent\textbf{Additional information}\newline
\textbf{Correspondence and requests for materials} should be addressed to \href{mailto:bell@baeri.org}{Taylor J. Bell}.\newline
\textbf{Reprints and permissions information} is available at \url{www.nature.com/reprints}.

\section*{Extended Data Tables}

\begin{table*}[!htbp]
    \centering
    \small
    \caption{\textbf{General Circulation Model simulations used in this investigation.} For cloud models, the composition of the condensates are provided, along with the prescribed mean radius of the condensed cloud particles.  For the RM-GCM cloud models, simulations with eight species of clouds include KCl, Cr$_2$O$_3$, SiO$_2$, Mg$_2$SiO$_4$, VO, Ca$_2$SiO$_4$, CaTiO$_2$, and Al$_2$O$_3$, while those with 13 species additionally include ZnS, Na$_2$S, MnS, Fe, and Ni; in these cases, particle sizes vary with height, and the prescribed maximum vertical extent of the clouds, expressed in pressure scale heights (H), are also provided. Further details of the models can be found in the text. Phase curve maxima and minima are expressed as the planet-to-star flux ratio in ppm, while offsets are expressed in degrees of longitude east relative to the sub-stellar longitude. Simulations featured in \ref{fig:gcmSpectra} are indicated (*).}\label{tab:gcm_list_full}
    \includegraphics[width=6.5in]{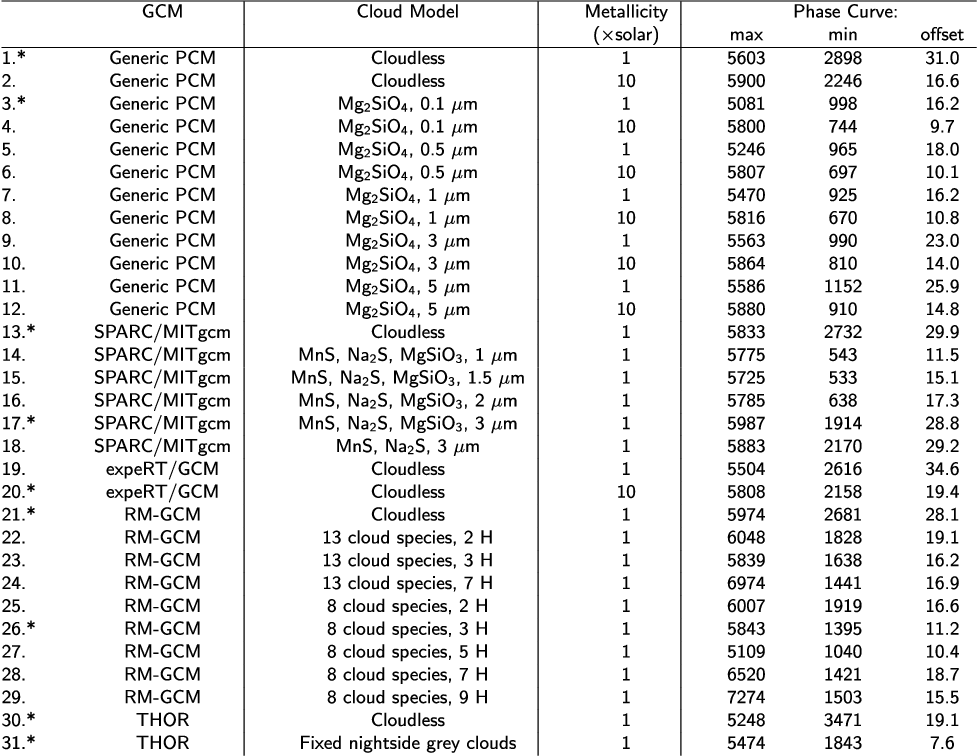}
\end{table*}

\begin{table*}
    \centering
    \small
    \caption{\textbf{Inferences on H$_2$O and CH$_4$ abundances and detection significances.} The H$_2$O volume mixing ratios and CH$_4$ upper limits inferred by each retrieval framework at each phase, in log values. H$_2$O detection significances are shown in brackets where applicable.  Codes marked with $^*$ assume equilibrium chemistry, and cannot constrain the abundances of molecules individually. For these codes, we report the abundance of water at 100 mbar.}\label{tab:ret_abundances}
    \includegraphics[width=4.7in]{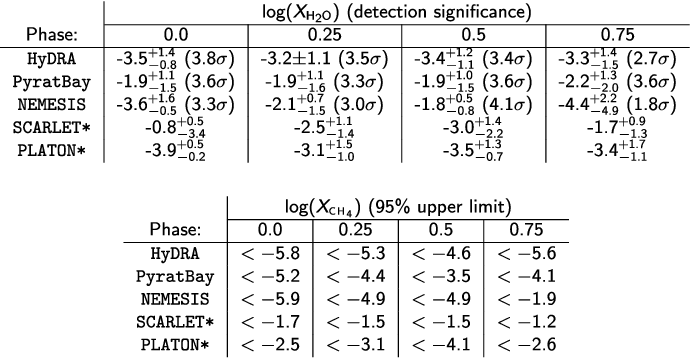}
\end{table*}

\begin{table*}
    \centering
    \small
    \caption{\textbf{Chi-squared values of equilibrium chemistry retrievals with and without opacity from water and methane.} The retrieval setup is identical aside from the inclusion or exclusion of opacity from water or methane. The results presented here do not consider the error inflation parameter (the retrievals would be free to inflate errors until a $\chi^2/N_{\rm points}$ value of 1 is reached). The models were fit to the phase-resolved spectra which each had 11 data points.  The PLATON retrieval included 9 parameters. The SCARLET retrieval included 7 parameters for phases 0.25, 0.5, and 0.75, and 6 parameters for phase 0.0.}\label{tab:ret_chi_sqr}
    \includegraphics[width=4.55in]{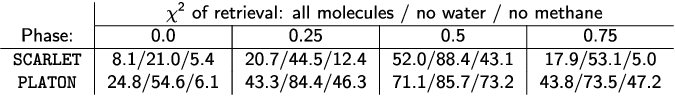}
\end{table*}

\clearpage
\section*{Extended Data Figures}

\begin{figure*}[!htbp]
    \centering
    \includegraphics[width=0.8\linewidth]{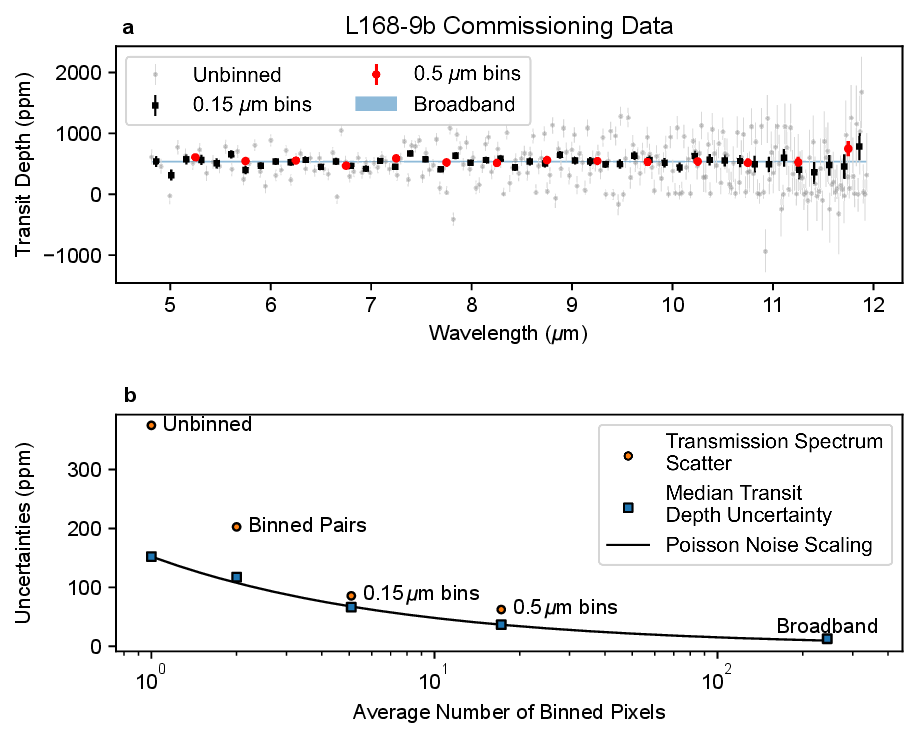}
    \caption{\textbf{The underestimation of uncertainties as a function of spectral binning for the L168-9b commissioning observations.} \textbf{a,} The observed L168-9b transmission spectrum with 1$\sigma$ error bars for spectrally unbinned data (grey circles), 0.15\,$\mu$m bins (black squares), 0.5\,$\mu$m bins (large red circles), and a 5--12\,$\mu$m broadband bin (horizontal blue shaded region). The spectrum for wavelength pairs is not shown to avoid excessive clutter. \textbf{b,} The median of the transit depth uncertainties are shown with blue squares, while the observed scatter in the transmission spectrum is shown with orange circles. For unbinned data, the transmission spectrum shows about $2.5\times$ the scatter predicted by the fits to the individual light curves. Binning pairs of wavelengths reduces the level of underestimation of the scatter in the transmission spectrum, but considerable excess noise remains. Coarser binning schemes like the constant 0.15\,$\mu$m bins those used in the MIRI time-series observation commissioning paper\cite{Bouwman2023MiriTsoCommissioning} or the 0.5\,$\mu$m bins we use in this work further reduce the level of uncertainty underestimation.}\label{fig:binningUncertainties}
\end{figure*}

\begin{figure*}
    \centering
    \includegraphics[width=0.9\linewidth]{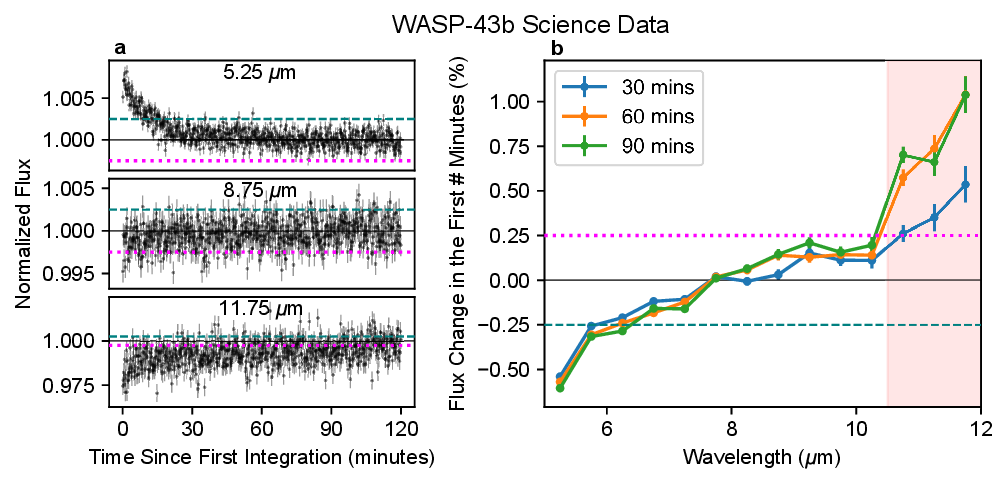}
    \caption{\textbf{A model-independent demonstration of the initial changes in flux for the \mbox{WASP-43b} observations.} \textbf{a,} The first 120 minutes of three of our spectroscopically binned light curves of \mbox{WASP-43b} (with 1$\sigma$ uncertainties) showing the initial settling behaviour as a function of wavelength. A teal dashed line shows the amplitude of a $-$0.25\% change in flux compared to the values around 120 minutes, and a magenta dotted line shows a $+$0.25\% change. \textbf{b,} A summary of the ramp amplitudes, signs, and timescales for each of our wavelength bins (with 1$\sigma$ uncertainties). The teal and magenta horizontal lines are the same as those in panel \textbf{a} to aid in translating between the two figures. At short wavelengths, the flux sharply drops by ${\sim}$0.5\% within the first 30 minutes and then largely settles but does continue to significantly decrease with time. With increasing wavelength, the strength of this initial ramp decreases and eventually changes sign, becoming an upward ramp. Within the ``shadowed region'' (marked in red), the light curves show a very strong upward ramp that takes much longer ($\gtrsim$\,60 minutes) to appreciably decay. It is important to note that the data in this figure also includes a small amount of astrophysical phase variations which should result in a small increase in flux of less than 0.05\% per hour.}\label{fig:rampAmplitudes}
\end{figure*}

\begin{figure*}
    \centering
    \includegraphics[width=0.95\textwidth]{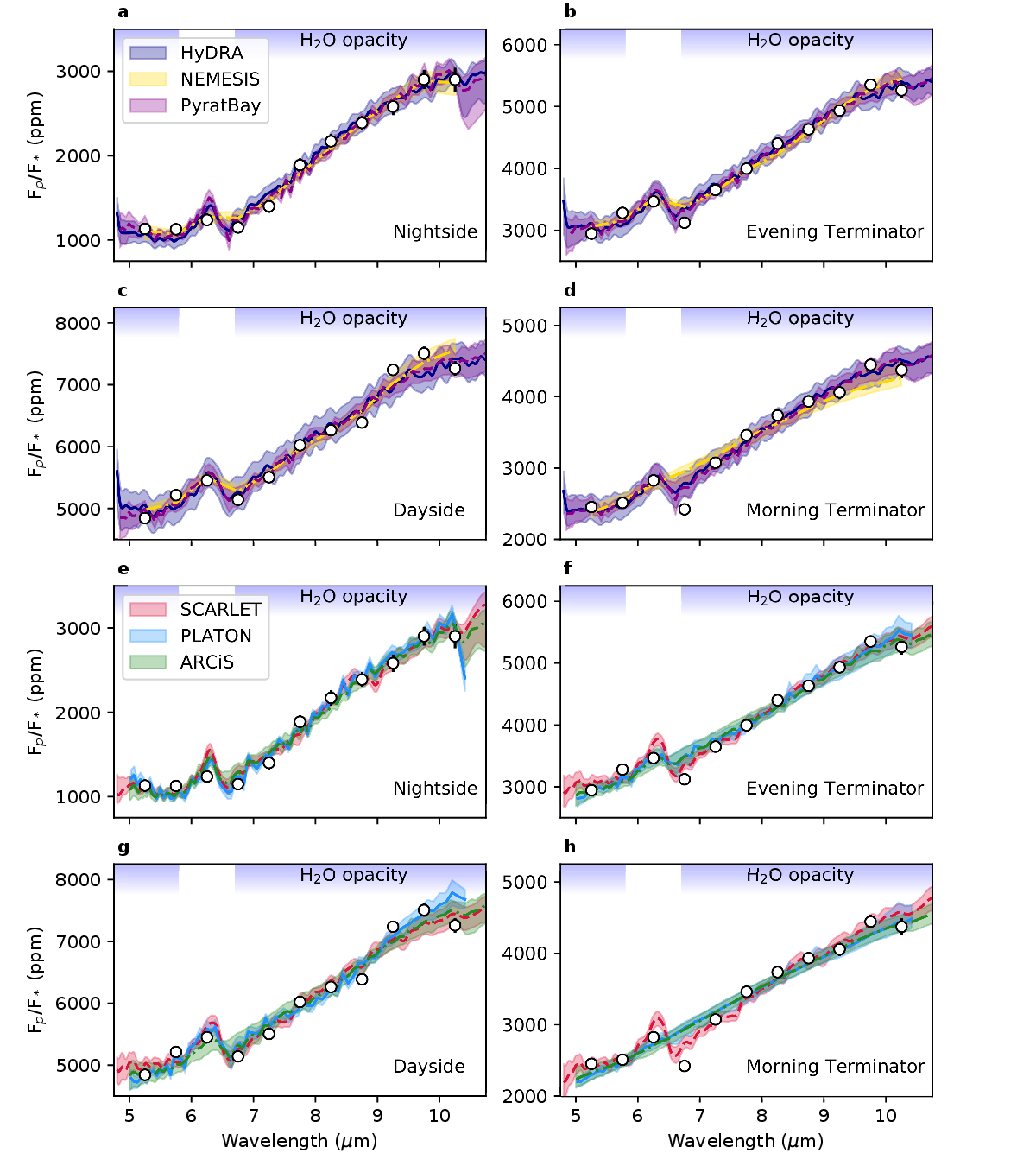}
    \caption{{\bf Retrieved spectra from the six retrievals. a,} Median retrieved nightside spectra for the HyDRA (dark blue line), NEMESIS (dash-dotted gold line), and PyratBay (dashed magenta line) and their 1-$\sigma$ contour. The regions of higher water opacity are indicated by the purple shading at the top of the panel, with the observed rise in flux at 6.3~$\mu$m being caused by a drop in opacity. \textbf{b, c, and d,} Same as \textbf{a} for the evening terminator, dayside, and morning terminator respectively. \textbf{e, f, g, and h,} Same as \textbf{a, b, c, and d,} for the SCARLET (dashed red line), PLATON (blue line), and ARCiS (dash-dotted green line).}\label{fig:retrieved_spectra}
\end{figure*}

\begin{figure*}[!htbp]
    \centering
    \includegraphics[width=1.0\linewidth]{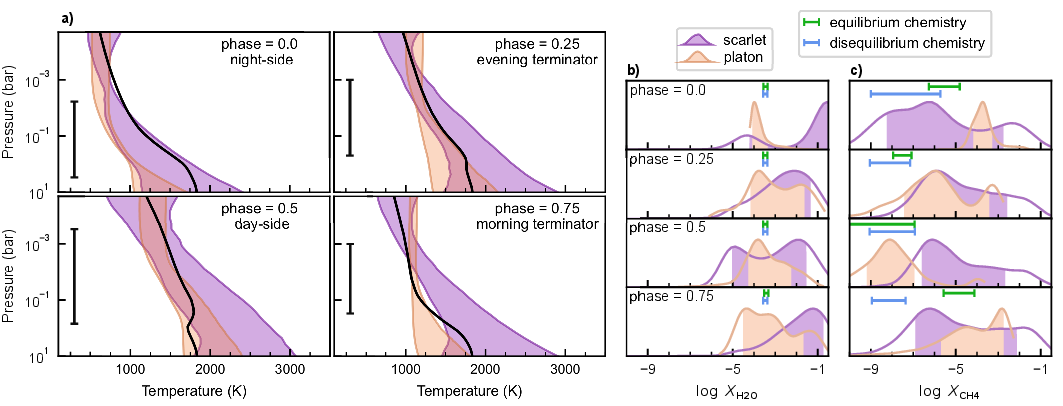}
    \caption{{\bf Chemically-consistent atmospheric retrievals.} Same as \ref{fig:retrievals} but for retrievals assuming thermochemical-equilibrium abundances consistent with the pressure-temperature profiles. {\bf a,} 1-$\sigma$ credible interval contours of the temperature profiles. The black curves show the predicted temperature profile from a 2D radiative-transport model\cite{venot20}. The vertical bars show the range of pressures probed by the observations. {\bf b and c,} probability posterior distributions for H$_{2}$O and CH$_{4}$ abundances, respectively. The shaded area for each curve denotes the 1-$\sigma$ credible interval of each posterior. The green and blue bars denote the abundances predicted by equilibrium and disequilibrium-chemistry models with solar abundances, respectively, at the pressures probed by the observations.  Compared to the free-chemistry retrievals, the thermochemical-equilibrium retrievals at the night-side hemisphere produced worse fits, this is driven particularly by the higher amount of methane expected under equilibrium chemistry.}\label{fig:equilibrium_retrievals}
\end{figure*}

\begin{figure*}
    \centering
    \includegraphics[width=1.0\linewidth]{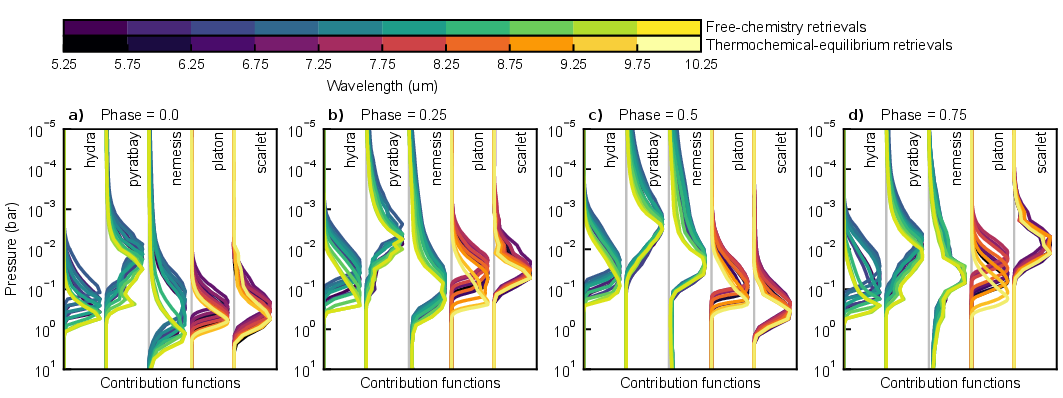}
    \caption{
    {\bf Retrieval contribution functions.}
Contribution functions integrated over the data point spectral bins, at each phase (\textbf{a--d}), and for each retrieval framework. These curves show the range of pressures probed by the observation according to the atmospheric models. The enhanced opacity from the water band around 7--9 $\mu$m makes these wavelengths probe lower pressures and hence colder temperatures, whereas the rest of the observing window probes higher pressures and higher temperatures.}\label{fig:contribution_functions}
\end{figure*}

\begin{figure*}
    \centering
    \hspace{-5pt}{\includegraphics[width=0.6\linewidth]{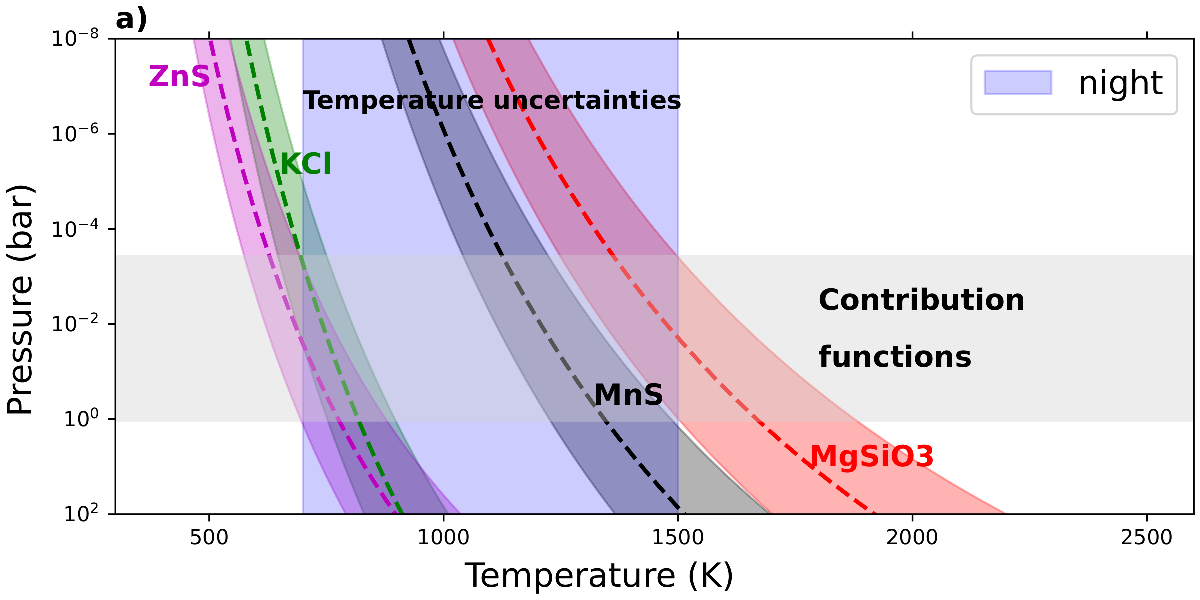}}
    \subfigure{\raisebox{1.1cm}{\includegraphics[width=0.13\linewidth]{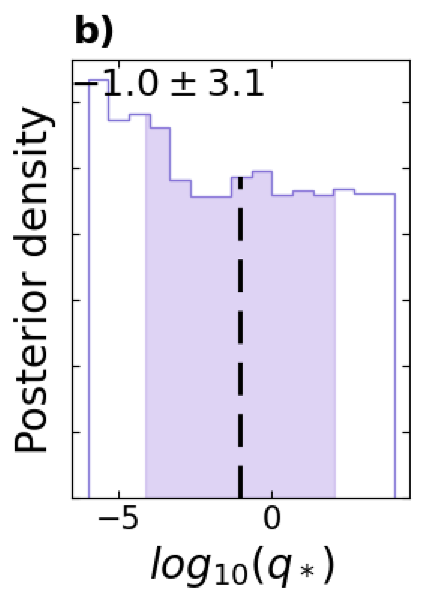}}}
    \subfigure{\raisebox{1.1cm}{\includegraphics[width=0.13\linewidth]{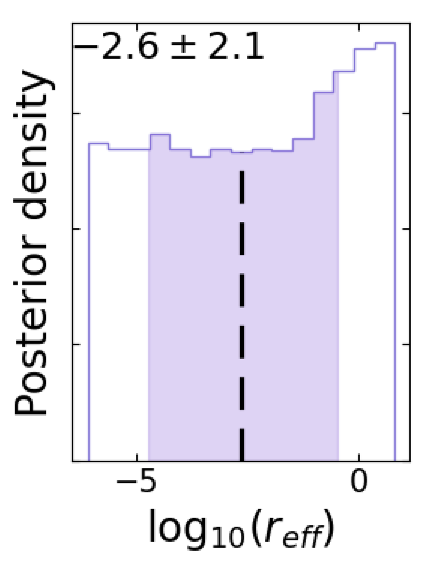}}}
    \subfigure{\raisebox{1.1cm}{\includegraphics[width=0.13\linewidth]{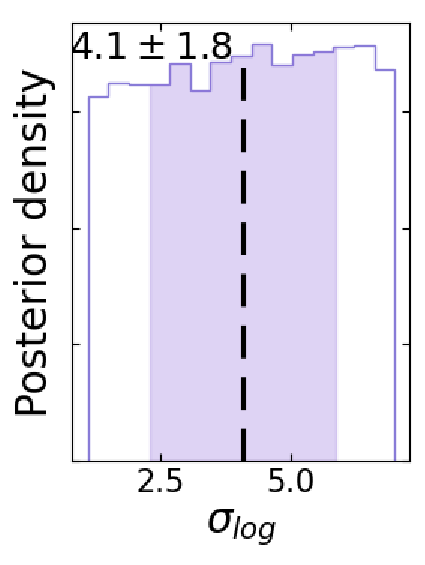}}}
    \caption{\textbf{PyratBay clouds exploration.}
    {\bf a,} Cloud species that condense on the temperature regimes expected for the \mbox{WASP-43b} nightside. Dashed lines represent vapour pressure curves\cite{MorleyEtal2012} for each species assuming solar composition, while the coloured ranges denote the corresponding extent of the vapour pressure curves assuming 100$\times$ sub- and super-solar atmospheric composition. The extent of the retrieved nightside contribution functions is shown in grey, and the extent of the retrieved temperature uncertainties is shown in light purple. The intersection between the contribution function and temperature ranges indicates the pressures at which we could observe cloud condensation and potentially detect their spectral features, if present in the observations.
    {\bf b,} Panels display the retrieved posterior density plots for the explored cloud parameters of the \textsc{TSC} model (cloud number density, $q^*$, effective particle size, $r_{eff}$, and the standard deviation of the log-normal distribution, $\sigma_{log}$) for the MnS clouds. The black vertical line denotes the parameter's median value, while the extent of the purple region denotes the 1$\sigma$ uncertainties, both  given at the top left corner of the panel. Similar, fully non-constrained posteriors are retrieved for other explored cloud species, MgSiO$_{3}$, ZnS, and KCl, suggesting the lack of observable spectral characteristics from clouds in the observed data.}\label{fig:nightside_clouds}
\end{figure*}

\begin{figure*}[!htbp]
    \centering
    \includegraphics[width=1.0\linewidth]{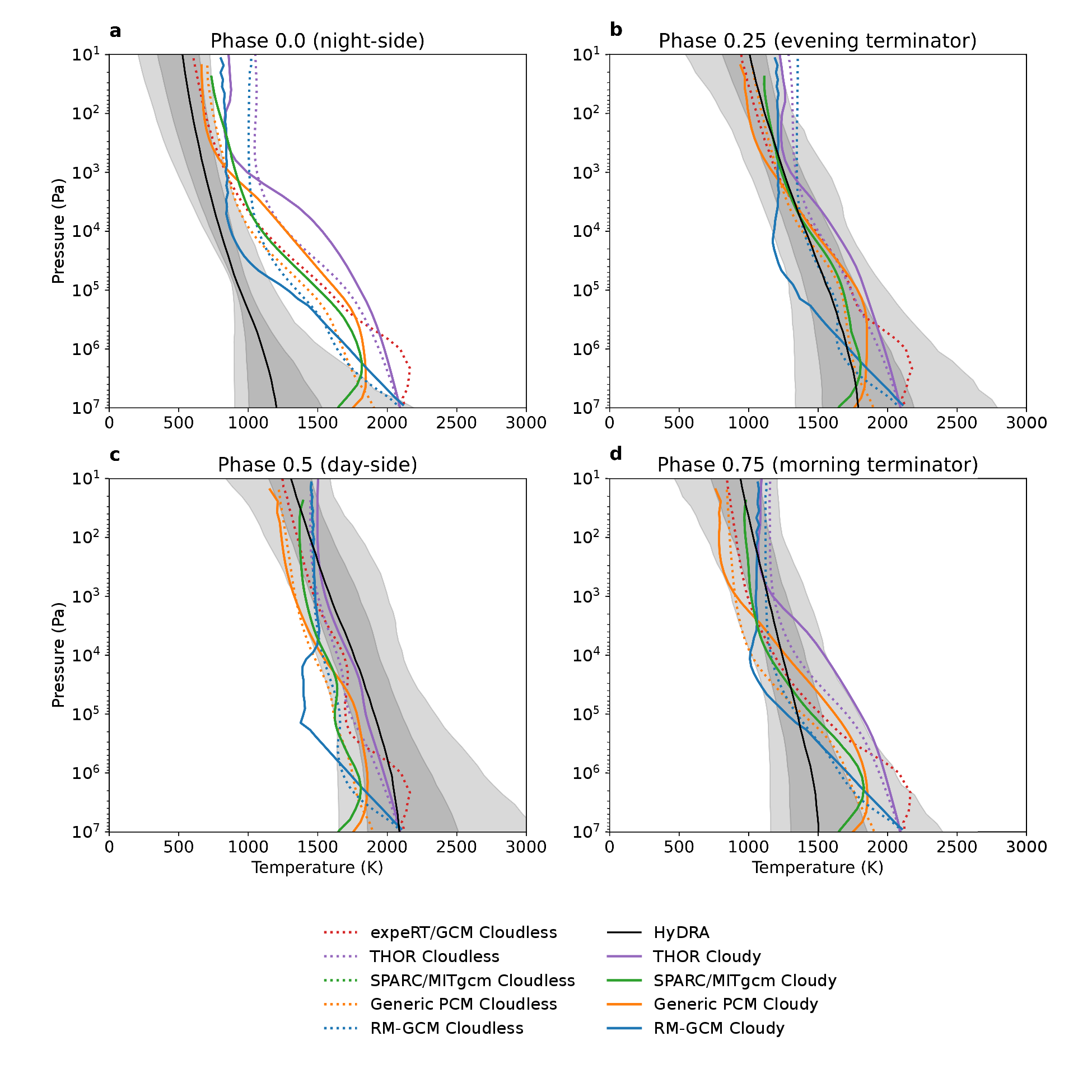}
    \caption{{\bf A comparison of the retrieved temperature-pressure profiles to the GCM simulations.} %
    Each of \textbf{a--d} shows the temperature profile retrieved by \texttt{HyDRA}, compared to the GCM simulations highlighted in \ref{fig:gcmSpectra} and listed in \ref{tab:gcm_list_full}. The GCM temperature profiles are calculated at phases 0.0, 0.25, 0.5, and 0.75 by averaging over the visible hemisphere by viewing angle, to produce a 1D profile that is comparable to the retrieved profile. The GCM simulations are generally warmer on the nightside than the retrieved temperatures; cloudy simulations emit from lower pressures and so match the observed lower brightness temperatures better (see the contribution functions in \ref{fig:contribution_functions}).}\label{fig:compare_gcm_retrieval}
\end{figure*}

\end{methods}

\clearpage
\section*{References}
\bibliography{ms.bib}

\begin{thebibliography}{100}
\expandafter\ifx\csname url\endcsname\relax
  \def\url#1{\texttt{#1}}\fi
\expandafter\ifx\csname urlprefix\endcsname\relax\def\urlprefix{URL }\fi
\providecommand{\bibinfo}[2]{#2}
\providecommand{\eprint}[2][]{\url{#2}}

\bibitem{Keating2019}
\bibinfo{author}{{Keating}, D.}, \bibinfo{author}{{Cowan}, N.~B.} \&
  \bibinfo{author}{{Dang}, L.}
\newblock \bibinfo{title}{{Uniformly hot nightside temperatures on short-period
  gas giants}}.
\newblock \emph{\bibinfo{journal}{\nastro}} \textbf{\bibinfo{volume}{3}},
  \bibinfo{pages}{1092--1098} (\bibinfo{year}{2019}).
\newblock \eprint{1809.00002}.

\bibitem{Beatty2019}
\bibinfo{author}{{Beatty}, T.~G.} \emph{et~al.}
\newblock \bibinfo{title}{{Spitzer Phase Curves of KELT-1b and the Signatures
  of Nightside Clouds in Thermal Phase Observations}}.
\newblock \emph{\bibinfo{journal}{\aj}} \textbf{\bibinfo{volume}{158}},
  \bibinfo{pages}{166} (\bibinfo{year}{2019}).
\newblock \eprint{1808.09575}.

\bibitem{Kataria2015}
\bibinfo{author}{{Kataria}, T.} \emph{et~al.}
\newblock \bibinfo{title}{{The Atmospheric Circulation of the Hot Jupiter
  WASP-43b: Comparing Three-dimensional Models to Spectrophotometric Data}}.
\newblock \emph{\bibinfo{journal}{\apj}} \textbf{\bibinfo{volume}{801}},
  \bibinfo{pages}{86} (\bibinfo{year}{2015}).
\newblock \eprint{1410.2382}.

\bibitem{mendonca2018a}
\bibinfo{author}{{Mendon{\c{c}}a}, J.~M.}, \bibinfo{author}{{Malik}, M.},
  \bibinfo{author}{{Demory}, B.-O.} \& \bibinfo{author}{{Heng}, K.}
\newblock \bibinfo{title}{{Revisiting the Phase Curves of WASP-43b: Confronting
  Re-analyzed Spitzer Data with Cloudy Atmospheres}}.
\newblock \emph{\bibinfo{journal}{\aj}} \textbf{\bibinfo{volume}{155}},
  \bibinfo{pages}{150} (\bibinfo{year}{2018}).
\newblock \eprint{1802.03047}.

\bibitem{Parmentier2018a}
\bibinfo{author}{{Parmentier}, V.} \& \bibinfo{author}{{Crossfield}, I. J.~M.}
\newblock \bibinfo{title}{{Exoplanet Phase Curves: Observations and Theory}}.
\newblock In \bibinfo{editor}{{Deeg}, H.~J.} \& \bibinfo{editor}{{Belmonte},
  J.~A.} (eds.) \emph{\bibinfo{booktitle}{Handbook of Exoplanets}},
  \bibinfo{pages}{116} (\bibinfo{publisher}{Springer}, \bibinfo{year}{2018}).

\bibitem{hu15}
\bibinfo{author}{{Hu}, R.}, \bibinfo{author}{{Demory}, B.-O.},
  \bibinfo{author}{{Seager}, S.}, \bibinfo{author}{{Lewis}, N.} \&
  \bibinfo{author}{{Showman}, A.~P.}
\newblock \bibinfo{title}{{A Semi-analytical Model of Visible-wavelength Phase
  Curves of Exoplanets and Applications to Kepler- 7 b and Kepler- 10 b}}.
\newblock \emph{\bibinfo{journal}{\apj}} \textbf{\bibinfo{volume}{802}},
  \bibinfo{pages}{51} (\bibinfo{year}{2015}).
\newblock \eprint{1501.03876}.

\bibitem{shporer15}
\bibinfo{author}{{Shporer}, A.} \& \bibinfo{author}{{Hu}, R.}
\newblock \bibinfo{title}{{Studying Atmosphere-dominated Hot Jupiter Kepler
  Phase Curves: Evidence that Inhomogeneous Atmospheric Reflection Is Common}}.
\newblock \emph{\bibinfo{journal}{\aj}} \textbf{\bibinfo{volume}{150}},
  \bibinfo{pages}{112} (\bibinfo{year}{2015}).
\newblock \eprint{1504.00498}.

\bibitem{Roman2021}
\bibinfo{author}{{Roman}, M.~T.} \emph{et~al.}
\newblock \bibinfo{title}{{Clouds in Three-dimensional Models of Hot Jupiters
  over a Wide Range of Temperatures. I. Thermal Structures and Broadband
  Phase-curve Predictions}}.
\newblock \emph{\bibinfo{journal}{\apj}} \textbf{\bibinfo{volume}{908}},
  \bibinfo{pages}{101} (\bibinfo{year}{2021}).
\newblock \eprint{2010.06936}.

\bibitem{tan2021a}
\bibinfo{author}{{Tan}, X.} \& \bibinfo{author}{{Showman}, A.~P.}
\newblock \bibinfo{title}{{Atmospheric circulation of brown dwarfs and directly
  imaged exoplanets driven by cloud radiative feedback: effects of rotation}}.
\newblock \emph{\bibinfo{journal}{\mnras}} \textbf{\bibinfo{volume}{502}},
  \bibinfo{pages}{678--699} (\bibinfo{year}{2021}).
\newblock \eprint{2005.12152}.

\bibitem{Parmentier2021}
\bibinfo{author}{{Parmentier}, V.}, \bibinfo{author}{{Showman}, A.~P.} \&
  \bibinfo{author}{{Fortney}, J.~J.}
\newblock \bibinfo{title}{{The cloudy shape of hot Jupiter thermal phase
  curves}}.
\newblock \emph{\bibinfo{journal}{\mnras}} \textbf{\bibinfo{volume}{501}},
  \bibinfo{pages}{78--108} (\bibinfo{year}{2021}).
\newblock \eprint{2010.06934}.

\bibitem{perna10}
\bibinfo{author}{{Perna}, R.}, \bibinfo{author}{{Menou}, K.} \&
  \bibinfo{author}{{Rauscher}, E.}
\newblock \bibinfo{title}{{Magnetic Drag on Hot Jupiter Atmospheric Winds}}.
\newblock \emph{\bibinfo{journal}{\apj}} \textbf{\bibinfo{volume}{719}},
  \bibinfo{pages}{1421--1426} (\bibinfo{year}{2010}).
\newblock \eprint{1003.3838}.

\bibitem{Komacek2017}
\bibinfo{author}{{Komacek}, T.~D.} \& \bibinfo{author}{{Youdin}, A.~N.}
\newblock \bibinfo{title}{{Structure and Evolution of Internally Heated Hot
  Jupiters}}.
\newblock \emph{\bibinfo{journal}{\apj}} \textbf{\bibinfo{volume}{844}},
  \bibinfo{pages}{94} (\bibinfo{year}{2017}).
\newblock \eprint{1706.07605}.

\bibitem{Rogers2017}
\bibinfo{author}{Rogers, T.~M.}
\newblock \bibinfo{title}{Constraints on the magnetic field strength of
  {HAT}-{P}-7 b and other hot giant exoplanets}.
\newblock \emph{\bibinfo{journal}{\nastro}} \textbf{\bibinfo{volume}{1}},
  \bibinfo{pages}{0131} (\bibinfo{year}{2017}).
\newblock \bibinfo{note}{\_eprint: 1704.06271}.

\bibitem{kataria14}
\bibinfo{author}{{Kataria}, T.}, \bibinfo{author}{{Showman}, A.~P.},
  \bibinfo{author}{{Fortney}, J.~J.}, \bibinfo{author}{{Marley}, M.~S.} \&
  \bibinfo{author}{{Freedman}, R.~S.}
\newblock \bibinfo{title}{{The Atmospheric Circulation of the Super Earth GJ
  1214b: Dependence on Composition and Metallicity}}.
\newblock \emph{\bibinfo{journal}{\apj}} \textbf{\bibinfo{volume}{785}},
  \bibinfo{pages}{92} (\bibinfo{year}{2014}).
\newblock \eprint{1401.1898}.

\bibitem{zhang17}
\bibinfo{author}{{Zhang}, X.} \& \bibinfo{author}{{Showman}, A.~P.}
\newblock \bibinfo{title}{{Effects of Bulk Composition on the Atmospheric
  Dynamics on Close-in Exoplanets}}.
\newblock \emph{\bibinfo{journal}{\apj}} \textbf{\bibinfo{volume}{836}},
  \bibinfo{pages}{73} (\bibinfo{year}{2017}).
\newblock \eprint{1607.04260}.

\bibitem{Carone2020}
\bibinfo{author}{{Carone}, L.} \emph{et~al.}
\newblock \bibinfo{title}{{Equatorial retrograde flow in WASP-43b elicited by
  deep wind jets?}}
\newblock \emph{\bibinfo{journal}{\mnras}} \textbf{\bibinfo{volume}{496}},
  \bibinfo{pages}{3582--3614} (\bibinfo{year}{2020}).
\newblock \eprint{1904.13334}.

\bibitem{Showman2020}
\bibinfo{author}{Showman, A.~P.}, \bibinfo{author}{Tan, X.} \&
  \bibinfo{author}{Parmentier, V.}
\newblock \bibinfo{title}{Atmospheric {Dynamics} of {Hot} {Giant} {Planets} and
  {Brown} {Dwarfs}}.
\newblock \emph{\bibinfo{journal}{\ssr}} \textbf{\bibinfo{volume}{216}},
  \bibinfo{pages}{139} (\bibinfo{year}{2020}).
\newblock \urlprefix\url{http://link.springer.com/10.1007/s11214-020-00758-8}.

\bibitem{Helling2021}
\bibinfo{author}{{Helling}, C.} \emph{et~al.}
\newblock \bibinfo{title}{{Cloud property trends in hot and ultra-hot giant gas
  planets (WASP-43b, WASP-103b, WASP-121b, HAT-P-7b, and WASP-18b)}}.
\newblock \emph{\bibinfo{journal}{\aap}} \textbf{\bibinfo{volume}{649}},
  \bibinfo{pages}{A44} (\bibinfo{year}{2021}).
\newblock \eprint{2102.11688}.

\bibitem{Hellier2011}
\bibinfo{author}{{Hellier}, C.} \emph{et~al.}
\newblock \bibinfo{title}{{WASP-43b: the closest-orbiting hot Jupiter}}.
\newblock \emph{\bibinfo{journal}{\aap}} \textbf{\bibinfo{volume}{535}},
  \bibinfo{pages}{L7} (\bibinfo{year}{2011}).
\newblock \eprint{1104.2823}.

\bibitem{Scandariato2022}
\bibinfo{author}{{Scandariato}, G.} \emph{et~al.}
\newblock \bibinfo{title}{{Phase curve and geometric albedo of WASP-43b
  measured with CHEOPS, TESS, and HST WFC3/UVIS}}.
\newblock \emph{\bibinfo{journal}{\aap}} \textbf{\bibinfo{volume}{668}},
  \bibinfo{pages}{A17} (\bibinfo{year}{2022}).
\newblock \eprint{2209.05303}.

\bibitem{Stevenson2014}
\bibinfo{author}{{Stevenson}, K.~B.} \emph{et~al.}
\newblock \bibinfo{title}{{Thermal structure of an exoplanet atmosphere from
  phase-resolved emission spectroscopy}}.
\newblock \emph{\bibinfo{journal}{\sci}} \textbf{\bibinfo{volume}{346}},
  \bibinfo{pages}{838--841} (\bibinfo{year}{2014}).
\newblock \eprint{1410.2241}.

\bibitem{Stevenson2017}
\bibinfo{author}{{Stevenson}, K.~B.} \emph{et~al.}
\newblock \bibinfo{title}{{Spitzer Phase Curve Constraints for WASP-43b at 3.6
  and 4.5 {\ensuremath{\mu}}m}}.
\newblock \emph{\bibinfo{journal}{\aj}} \textbf{\bibinfo{volume}{153}},
  \bibinfo{pages}{68} (\bibinfo{year}{2017}).
\newblock \eprint{1608.00056}.

\bibitem{VenotEtal2020-JWST-WASP-43b}
\bibinfo{author}{{Venot}, O.} \emph{et~al.}
\newblock \bibinfo{title}{{Global Chemistry and Thermal Structure Models for
  the Hot Jupiter WASP-43b and Predictions for JWST}}.
\newblock \emph{\bibinfo{journal}{\apj}} \textbf{\bibinfo{volume}{890}},
  \bibinfo{pages}{176} (\bibinfo{year}{2020}).
\newblock \eprint{2001.04759}.

\bibitem{helling2020}
\bibinfo{author}{{Helling}, C.} \emph{et~al.}
\newblock \bibinfo{title}{{Mineral cloud and hydrocarbon haze particles in the
  atmosphere of the hot Jupiter JWST target WASP-43b}}.
\newblock \emph{\bibinfo{journal}{\aap}} \textbf{\bibinfo{volume}{641}},
  \bibinfo{pages}{A178} (\bibinfo{year}{2020}).
\newblock \eprint{2005.14595}.

\bibitem{Morello2019}
\bibinfo{author}{{Morello}, G.}, \bibinfo{author}{{Danielski}, C.},
  \bibinfo{author}{{Dickens}, D.}, \bibinfo{author}{{Tremblin}, P.} \&
  \bibinfo{author}{{Lagage}, P.~O.}
\newblock \bibinfo{title}{{An Independent Analysis of the Spitzer/IRAC Phase
  Curves of WASP43 b}}.
\newblock \emph{\bibinfo{journal}{\aj}} \textbf{\bibinfo{volume}{157}},
  \bibinfo{pages}{205} (\bibinfo{year}{2019}).
\newblock \eprint{1908.06741}.

\bibitem{murphy2023}
\bibinfo{author}{{Murphy}, M.~M.} \emph{et~al.}
\newblock \bibinfo{title}{{A Lack of Variability between Repeated Spitzer Phase
  Curves of WASP-43b}}.
\newblock \emph{\bibinfo{journal}{\aj}} \textbf{\bibinfo{volume}{165}},
  \bibinfo{pages}{107} (\bibinfo{year}{2023}).
\newblock \eprint{2212.03240}.

\bibitem{Wright_2023}
\bibinfo{author}{Wright, G.~S.} \emph{et~al.}
\newblock \bibinfo{title}{The mid-infrared instrument for jwst and its
  in-flight performance}.
\newblock \emph{\bibinfo{journal}{\pasp}} \textbf{\bibinfo{volume}{135}},
  \bibinfo{pages}{048003} (\bibinfo{year}{2023}).
\newblock \urlprefix\url{https://dx.doi.org/10.1088/1538-3873/acbe66}.

\bibitem{Kendrew2015}
\bibinfo{author}{{Kendrew}, S.} \emph{et~al.}
\newblock \bibinfo{title}{{The Mid-Infrared Instrument for the James Webb Space
  Telescope, IV: The Low-Resolution Spectrometer}}.
\newblock \emph{\bibinfo{journal}{\pasp}} \textbf{\bibinfo{volume}{127}},
  \bibinfo{pages}{623} (\bibinfo{year}{2015}).
\newblock \eprint{1512.03000}.

\bibitem{Bouwman2023MiriTsoCommissioning}
\bibinfo{author}{{Bouwman}, J.} \emph{et~al.}
\newblock \bibinfo{title}{{Spectroscopic Time Series Performance of the
  Mid-infrared Instrument on the JWST}}.
\newblock \emph{\bibinfo{journal}{\pasp}} \textbf{\bibinfo{volume}{135}},
  \bibinfo{pages}{038002} (\bibinfo{year}{2023}).
\newblock \eprint{2211.16123}.

\bibitem{Esposito2017GapsWASP43b}
\bibinfo{author}{{Esposito}, M.} \emph{et~al.}
\newblock \bibinfo{title}{{The GAPS Programme with HARPS-N at TNG. XIII. The
  orbital obliquity of three close-in massive planets hosted by dwarf K-type
  stars: WASP-43, HAT-P-20 and Qatar-2}}.
\newblock \emph{\bibinfo{journal}{\aap}} \textbf{\bibinfo{volume}{601}},
  \bibinfo{pages}{A53} (\bibinfo{year}{2017}).
\newblock \eprint{1702.03136}.

\bibitem{May2020_wasp43b}
\bibinfo{author}{{May}, E.~M.} \& \bibinfo{author}{{Stevenson}, K.~B.}
\newblock \bibinfo{title}{{Introducing a New Spitzer Master BLISS Map to Remove
  the Instrument Systematic Phase-curve-parameter Degeneracy, as Demonstrated
  by a Reanalysis of the 4.5 {\ensuremath{\mu}}m WASP-43b Phase Curve}}.
\newblock \emph{\bibinfo{journal}{\aj}} \textbf{\bibinfo{volume}{160}},
  \bibinfo{pages}{140} (\bibinfo{year}{2020}).
\newblock \eprint{2007.06618}.

\bibitem{Bell2021}
\bibinfo{author}{{Bell}, T.~J.} \emph{et~al.}
\newblock \bibinfo{title}{{A comprehensive reanalysis of Spitzer's 4.5
  {\ensuremath{\mu}}m phase curves, and the phase variations of the ultra-hot
  Jupiters MASCARA-1b and KELT-16b}}.
\newblock \emph{\bibinfo{journal}{\mnras}} \textbf{\bibinfo{volume}{504}},
  \bibinfo{pages}{3316--3337} (\bibinfo{year}{2021}).
\newblock \eprint{2010.00687}.

\bibitem{showman2009}
\bibinfo{author}{{Showman}, A.~P.} \emph{et~al.}
\newblock \bibinfo{title}{{Atmospheric Circulation of Hot Jupiters: Coupled
  Radiative-Dynamical General Circulation Model Simulations of HD 189733b and
  HD 209458b}}.
\newblock \emph{\bibinfo{journal}{\apj}} \textbf{\bibinfo{volume}{699}},
  \bibinfo{pages}{564--584} (\bibinfo{year}{2009}).
\newblock \eprint{0809.2089}.

\bibitem{Showman2011}
\bibinfo{author}{{Showman}, A.~P.} \& \bibinfo{author}{{Polvani}, L.~M.}
\newblock \bibinfo{title}{{Equatorial Superrotation on Tidally Locked
  Exoplanets}}.
\newblock \emph{\bibinfo{journal}{\apj}} \textbf{\bibinfo{volume}{738}},
  \bibinfo{pages}{71} (\bibinfo{year}{2011}).
\newblock \eprint{1103.3101}.

\bibitem{Roman&Rauscher2017}
\bibinfo{author}{{Roman}, M.} \& \bibinfo{author}{{Rauscher}, E.}
\newblock \bibinfo{title}{{Modeling the Effects of Inhomogeneous Aerosols on
  the Hot Jupiter Kepler-7b{\textquoteright}s Atmospheric Circulation}}.
\newblock \emph{\bibinfo{journal}{\apj}} \textbf{\bibinfo{volume}{850}},
  \bibinfo{pages}{17} (\bibinfo{year}{2017}).
\newblock \eprint{1709.07459}.

\bibitem{Roman&Rauscher2019}
\bibinfo{author}{{Roman}, M.} \& \bibinfo{author}{{Rauscher}, E.}
\newblock \bibinfo{title}{{Modeled Temperature-dependent Clouds with Radiative
  Feedback in Hot Jupiter Atmospheres}}.
\newblock \emph{\bibinfo{journal}{\apj}} \textbf{\bibinfo{volume}{872}},
  \bibinfo{pages}{1} (\bibinfo{year}{2019}).
\newblock \eprint{1807.08890}.

\bibitem{cowan2012thermal}
\bibinfo{author}{Cowan, N.~B.} \emph{et~al.}
\newblock \bibinfo{title}{Thermal phase variations of wasp-12b: Defying
  predictions}.
\newblock \emph{\bibinfo{journal}{\apj}} \textbf{\bibinfo{volume}{747}},
  \bibinfo{pages}{82} (\bibinfo{year}{2012}).

\bibitem{mendonca2018b}
\bibinfo{author}{{Mendon{\c{c}}a}, J.~M.}, \bibinfo{author}{{Tsai}, S.-m.},
  \bibinfo{author}{{Malik}, M.}, \bibinfo{author}{{Grimm}, S.~L.} \&
  \bibinfo{author}{{Heng}, K.}
\newblock \bibinfo{title}{{Three-dimensional Circulation Driving Chemical
  Disequilibrium in WASP-43b}}.
\newblock \emph{\bibinfo{journal}{\apj}} \textbf{\bibinfo{volume}{869}},
  \bibinfo{pages}{107} (\bibinfo{year}{2018}).
\newblock \eprint{1808.00501}.

\bibitem{Agundez2014}
\bibinfo{author}{{Ag{\'u}ndez}, M.}, \bibinfo{author}{{Parmentier}, V.},
  \bibinfo{author}{{Venot}, O.}, \bibinfo{author}{{Hersant}, F.} \&
  \bibinfo{author}{{Selsis}, F.}
\newblock \bibinfo{title}{{Pseudo 2D chemical model of hot-Jupiter atmospheres:
  application to HD 209458b and HD 189733b}}.
\newblock \emph{\bibinfo{journal}{\aap}} \textbf{\bibinfo{volume}{564}},
  \bibinfo{pages}{A73} (\bibinfo{year}{2014}).
\newblock \eprint{1403.0121}.

\bibitem{tsai21k218b}
\bibinfo{author}{{Tsai}, S.-M.} \emph{et~al.}
\newblock \bibinfo{title}{{Inferring Shallow Surfaces on Sub-Neptune Exoplanets
  with JWST}}.
\newblock \emph{\bibinfo{journal}{\apjl}} \textbf{\bibinfo{volume}{922}},
  \bibinfo{pages}{L27} (\bibinfo{year}{2021}).
\newblock \eprint{2111.06429}.

\bibitem{moses22}
\bibinfo{author}{{Moses}, J.~I.}, \bibinfo{author}{{Tremblin}, P.},
  \bibinfo{author}{{Venot}, O.} \& \bibinfo{author}{{Miguel}, Y.}
\newblock \bibinfo{title}{{Chemical variation with altitude and longitude on
  exo-Neptunes: Predictions for Ariel phase-curve observations}}.
\newblock \emph{\bibinfo{journal}{Experimental Astronomy}}
  \textbf{\bibinfo{volume}{53}}, \bibinfo{pages}{279--322}
  (\bibinfo{year}{2022}).
\newblock \eprint{2103.07023}.

\bibitem{baeyens22}
\bibinfo{author}{{Baeyens}, R.}, \bibinfo{author}{{Konings}, T.},
  \bibinfo{author}{{Venot}, O.}, \bibinfo{author}{{Carone}, L.} \&
  \bibinfo{author}{{Decin}, L.}
\newblock \bibinfo{title}{{Grid of pseudo-2D chemistry models for tidally
  locked exoplanets - II. The role of photochemistry}}.
\newblock \emph{\bibinfo{journal}{\mnras}} \textbf{\bibinfo{volume}{512}},
  \bibinfo{pages}{4877--4892} (\bibinfo{year}{2022}).
\newblock \eprint{2203.11233}.

\bibitem{allard1995}
\bibinfo{author}{{Allard}, F.} \& \bibinfo{author}{{Hauschildt}, P.~H.}
\newblock \bibinfo{title}{{Model Atmospheres for M (Sub)Dwarf Stars. I. The
  Base Model Grid}}.
\newblock \emph{\bibinfo{journal}{\apj}} \textbf{\bibinfo{volume}{445}},
  \bibinfo{pages}{433} (\bibinfo{year}{1995}).
\newblock \eprint{astro-ph/9601150}.

\bibitem{hauschildt1999}
\bibinfo{author}{{Hauschildt}, P.~H.}, \bibinfo{author}{{Allard}, F.} \&
  \bibinfo{author}{{Baron}, E.}
\newblock \bibinfo{title}{{The NextGen Model Atmosphere Grid for
  3000$<$=T$_{eff}$$<$=10,000 K}}.
\newblock \emph{\bibinfo{journal}{\apj}} \textbf{\bibinfo{volume}{512}},
  \bibinfo{pages}{377--385} (\bibinfo{year}{1999}).
\newblock \eprint{astro-ph/9807286}.

\bibitem{husser2013}
\bibinfo{author}{{Husser}, T.~O.} \emph{et~al.}
\newblock \bibinfo{title}{{A new extensive library of PHOENIX stellar
  atmospheres and synthetic spectra}}.
\newblock \emph{\bibinfo{journal}{\aap}} \textbf{\bibinfo{volume}{553}},
  \bibinfo{pages}{A6} (\bibinfo{year}{2013}).
\newblock \eprint{1303.5632}.

\bibitem{venot20}
\bibinfo{author}{{Venot}, O.} \emph{et~al.}
\newblock \bibinfo{title}{{Global Chemistry and Thermal Structure Models for
  the Hot Jupiter WASP-43b and Predictions for JWST}}.
\newblock \emph{\bibinfo{journal}{\apj}} \textbf{\bibinfo{volume}{890}},
  \bibinfo{pages}{176} (\bibinfo{year}{2020}).
\newblock \eprint{2001.04759}.

\bibitem{Andras2021}
\bibinfo{author}{{G{\'a}sp{\'a}r}, A.} \emph{et~al.}
\newblock \bibinfo{title}{{The Quantum Efficiency and Diffractive Image
  Artifacts of Si:As IBC mid-IR Detector Arrays at 5-10 {\ensuremath{\mu}}m:
  Implications for the JWST/MIRI Detectors}}.
\newblock \emph{\bibinfo{journal}{\pasp}} \textbf{\bibinfo{volume}{133}},
  \bibinfo{pages}{014504} (\bibinfo{year}{2021}).
\newblock \eprint{2011.11908}.

\bibitem{bell2023arxiv}
\bibinfo{author}{{Bell}, T.~J.} \emph{et~al.}
\newblock \bibinfo{title}{{A First Look at the JWST MIRI/LRS Phase Curve of
  WASP-43b}}.
\newblock \emph{\bibinfo{journal}{arXiv e-prints}}
  \bibinfo{pages}{arXiv:2301.06350} (\bibinfo{year}{2023}).
\newblock \eprint{2301.06350}.

\bibitem{Bell2022Eureka}
\bibinfo{author}{{Bell}, T.} \emph{et~al.}
\newblock \bibinfo{title}{{Eureka!: An End-to-End Pipeline for JWST Time-Series
  Observations}}.
\newblock \emph{\bibinfo{journal}{\joss}} \textbf{\bibinfo{volume}{7}},
  \bibinfo{pages}{4503} (\bibinfo{year}{2022}).
\newblock \eprint{2207.03585}.

\bibitem{stsci2022jwst}
\bibinfo{author}{Bushouse, H.} \emph{et~al.}
\newblock \bibinfo{title}{{JWST Calibration Pipeline}} (\bibinfo{year}{2022}).
\newblock \urlprefix\url{https://github.com/spacetelescope/jwst}.

\bibitem{gaspar2021}
\bibinfo{author}{{G{\'a}sp{\'a}r}, A.} \emph{et~al.}
\newblock \bibinfo{title}{{The Quantum Efficiency and Diffractive Image
  Artifacts of Si:As IBC mid-IR Detector Arrays at 5-10 {\ensuremath{\mu}}m:
  Implications for the JWST/MIRI Detectors}}.
\newblock \emph{\bibinfo{journal}{\pasp}} \textbf{\bibinfo{volume}{133}},
  \bibinfo{pages}{014504} (\bibinfo{year}{2021}).
\newblock \eprint{2011.11908}.

\bibitem{Schlawin2020OneOverF}
\bibinfo{author}{{Schlawin}, E.} \emph{et~al.}
\newblock \bibinfo{title}{{JWST Noise Floor. I. Random Error Sources in JWST
  NIRCam Time Series}}.
\newblock \emph{\bibinfo{journal}{\aj}} \textbf{\bibinfo{volume}{160}},
  \bibinfo{pages}{231} (\bibinfo{year}{2020}).
\newblock \eprint{2010.03564}.

\bibitem{kempton2023}
\bibinfo{author}{{Kempton}, E. M.-R.} \emph{et~al.}
\newblock \bibinfo{title}{{A reflective, metal-rich atmosphere for GJ 1214b
  from its JWST phase curve}}.
\newblock \emph{\bibinfo{journal}{\nat}} \bibinfo{pages}{1476--4687}
  (\bibinfo{year}{2023}).
\newblock \eprint{2305.06240}.

\bibitem{dynesty_v1_2_3}
\bibinfo{author}{Koposov, S.} \emph{et~al.}
\newblock \bibinfo{title}{joshspeagle/dynesty: v1.2.3} (\bibinfo{year}{2022}).
\newblock \urlprefix\url{https://doi.org/10.5281/zenodo.6609296}.

\bibitem{cowan2008}
\bibinfo{author}{{Cowan}, N.~B.} \& \bibinfo{author}{{Agol}, E.}
\newblock \bibinfo{title}{{Inverting Phase Functions to Map Exoplanets}}.
\newblock \emph{\bibinfo{journal}{\apjl}} \textbf{\bibinfo{volume}{678}},
  \bibinfo{pages}{L129} (\bibinfo{year}{2008}).
\newblock \eprint{0803.3622}.

\bibitem{cowan2017}
\bibinfo{author}{{Cowan}, N.~B.}, \bibinfo{author}{{Chayes}, V.},
  \bibinfo{author}{{Bouffard}, {\'E}.}, \bibinfo{author}{{Meynig}, M.} \&
  \bibinfo{author}{{Haggard}, H.~M.}
\newblock \bibinfo{title}{{Odd Harmonics in Exoplanet Photometry: Weather or
  Artifact?}}
\newblock \emph{\bibinfo{journal}{\mnras}} \textbf{\bibinfo{volume}{467}},
  \bibinfo{pages}{747--757} (\bibinfo{year}{2017}).
\newblock \eprint{1611.05925}.

\bibitem{Trotta2008}
\bibinfo{author}{{Trotta}, R.}
\newblock \bibinfo{title}{{Bayes in the sky: Bayesian inference and model
  selection in cosmology}}.
\newblock \emph{\bibinfo{journal}{Contemporary Physics}}
  \textbf{\bibinfo{volume}{49}}, \bibinfo{pages}{71--104}
  (\bibinfo{year}{2008}).
\newblock \eprint{0803.4089}.

\bibitem{Welbanks2021}
\bibinfo{author}{{Welbanks}, L.} \& \bibinfo{author}{{Madhusudhan}, N.}
\newblock \bibinfo{title}{{Aurora: A Generalized Retrieval Framework for
  Exoplanetary Transmission Spectra}}.
\newblock \emph{\bibinfo{journal}{\apj}} \textbf{\bibinfo{volume}{913}},
  \bibinfo{pages}{114} (\bibinfo{year}{2021}).
\newblock \eprint{2103.08600}.

\bibitem{starry_v1.2.0}
\bibinfo{author}{{Luger}, R.} \emph{et~al.}
\newblock \bibinfo{title}{{rodluger/starry: v1.2.0}}.
\newblock \bibinfo{howpublished}{Zenodo} (\bibinfo{year}{2021}).

\bibitem{Bonomo2017}
\bibinfo{author}{{Bonomo}, A.~S.} \emph{et~al.}
\newblock \bibinfo{title}{{The GAPS Programme with HARPS-N at TNG . XIV.
  Investigating giant planet migration history via improved eccentricity and
  mass determination for 231 transiting planets}}.
\newblock \emph{\bibinfo{journal}{\aap}} \textbf{\bibinfo{volume}{602}},
  \bibinfo{pages}{A107} (\bibinfo{year}{2017}).
\newblock \eprint{1704.00373}.

\bibitem{Kipping2013LimbDarkening}
\bibinfo{author}{{Kipping}, D.~M.}
\newblock \bibinfo{title}{{Efficient, uninformative sampling of limb darkening
  coefficients for two-parameter laws}}.
\newblock \emph{\bibinfo{journal}{\mnras}} \textbf{\bibinfo{volume}{435}},
  \bibinfo{pages}{2152--2160} (\bibinfo{year}{2013}).
\newblock \eprint{1308.0009}.

\bibitem{pymc3}
\bibinfo{author}{Salvatier, J.}, \bibinfo{author}{Wiecki, T.~V.} \&
  \bibinfo{author}{Fonnesbeck, C.}
\newblock \bibinfo{title}{Probabilistic programming in python using pymc3}.
\newblock \emph{\bibinfo{journal}{PeerJ Computer Science}}
  \textbf{\bibinfo{volume}{2}}, \bibinfo{pages}{e55} (\bibinfo{year}{2016}).
\newblock \urlprefix\url{https://doi.org/10.7717/peerj-cs.55}.

\bibitem{GelmanRubin1992}
\bibinfo{author}{{Gelman}, A.} \& \bibinfo{author}{{Rubin}, D.~B.}
\newblock \bibinfo{title}{{Inference from Iterative Simulation Using Multiple
  Sequences}}.
\newblock \emph{\bibinfo{journal}{Statistical Science}}
  \textbf{\bibinfo{volume}{7}}, \bibinfo{pages}{457--472}
  (\bibinfo{year}{1992}).

\bibitem{Kreidberg2015Batman}
\bibinfo{author}{{Kreidberg}, L.}
\newblock \bibinfo{title}{{batman: BAsic Transit Model cAlculatioN in Python}}.
\newblock \emph{\bibinfo{journal}{\pasp}} \textbf{\bibinfo{volume}{127}},
  \bibinfo{pages}{1161} (\bibinfo{year}{2015}).
\newblock \eprint{1507.08285}.

\bibitem{foreman-mackey_emcee_2013}
\bibinfo{author}{Foreman-Mackey, D.}, \bibinfo{author}{Hogg, D.~W.},
  \bibinfo{author}{Lang, D.} \& \bibinfo{author}{Goodman, J.}
\newblock \bibinfo{title}{emcee: {The} {MCMC} {Hammer}}.
\newblock \emph{\bibinfo{journal}{arXiv:1202.3665 [astro-ph, physics:physics,
  stat]}}  (\bibinfo{year}{2013}).
\newblock \urlprefix\url{http://arxiv.org/abs/1202.3665}.
\newblock \bibinfo{note}{ArXiv: 1202.3665}.

\bibitem{exoplanet:joss}
\bibinfo{author}{{Foreman-Mackey}, D.} \emph{et~al.}
\newblock \bibinfo{title}{{exoplanet: Gradient-based probabilistic inference
  for exoplanet data \& other astronomical time series}}.
\newblock \emph{\bibinfo{journal}{\joss}} \textbf{\bibinfo{volume}{6}},
  \bibinfo{pages}{3285} (\bibinfo{year}{2021}).
\newblock \eprint{2105.01994}.

\bibitem{Ivshina2022}
\bibinfo{author}{{Ivshina}, E.~S.} \& \bibinfo{author}{{Winn}, J.~N.}
\newblock \bibinfo{title}{{TESS Transit Timing of Hundreds of Hot Jupiters}}.
\newblock \emph{\bibinfo{journal}{\apjs}} \textbf{\bibinfo{volume}{259}},
  \bibinfo{pages}{62} (\bibinfo{year}{2022}).
\newblock \eprint{2202.03401}.

\bibitem{Gillon2012}
\bibinfo{author}{{Gillon}, M.} \emph{et~al.}
\newblock \bibinfo{title}{{The TRAPPIST survey of southern transiting planets.
  I. Thirty eclipses of the ultra-short period planet WASP-43 b}}.
\newblock \emph{\bibinfo{journal}{\aap}} \textbf{\bibinfo{volume}{542}},
  \bibinfo{pages}{A4} (\bibinfo{year}{2012}).
\newblock \eprint{1201.2789}.

\bibitem{Turbet2021}
\bibinfo{author}{{Turbet}, M.} \emph{et~al.}
\newblock \bibinfo{title}{{Day-night cloud asymmetry prevents early oceans on
  Venus but not on Earth}}.
\newblock \emph{\bibinfo{journal}{\nat}} \textbf{\bibinfo{volume}{598}},
  \bibinfo{pages}{276--280} (\bibinfo{year}{2021}).
\newblock \eprint{2110.08801}.

\bibitem{Spiga2020}
\bibinfo{author}{{Spiga}, A.} \emph{et~al.}
\newblock \bibinfo{title}{{Global climate modeling of Saturn's atmosphere. Part
  II: Multi-annual high-resolution dynamical simulations}}.
\newblock \emph{\bibinfo{journal}{\icarus}} \textbf{\bibinfo{volume}{335}},
  \bibinfo{pages}{113377} (\bibinfo{year}{2020}).
\newblock \eprint{1811.01250}.

\bibitem{charnay14}
\bibinfo{author}{{Charnay}, B.}, \bibinfo{author}{{Forget}, F.},
  \bibinfo{author}{{Tobie}, G.}, \bibinfo{author}{{Sotin}, C.} \&
  \bibinfo{author}{{Wordsworth}, R.}
\newblock \bibinfo{title}{{Titan's past and future: 3D modeling of a pure
  nitrogen atmosphere and geological implications}}.
\newblock \emph{\bibinfo{journal}{\icarus}} \textbf{\bibinfo{volume}{241}},
  \bibinfo{pages}{269--279} (\bibinfo{year}{2014}).
\newblock \eprint{1407.1791}.

\bibitem{turbet_habitability_2016}
\bibinfo{author}{Turbet, M.} \emph{et~al.}
\newblock \bibinfo{title}{The habitability of {Proxima} {Centauri} b: {II}.
  {Possible} climates and observability}.
\newblock \emph{\bibinfo{journal}{\aap}} \textbf{\bibinfo{volume}{596}},
  \bibinfo{pages}{A112} (\bibinfo{year}{2016}).

\bibitem{charnay_3d_2015}
\bibinfo{author}{Charnay, B.}, \bibinfo{author}{Meadows, V.},
  \bibinfo{author}{Misra, A.}, \bibinfo{author}{Leconte, J.} \&
  \bibinfo{author}{Arney, G.}
\newblock \bibinfo{title}{{3D Modeling of GJ1214b{\textquoteright}s Atmosphere:
  Formation of Inhomogeneous High Clouds and Observational Implications}}.
\newblock \emph{\bibinfo{journal}{\apj}} \textbf{\bibinfo{volume}{813}},
  \bibinfo{pages}{L1} (\bibinfo{year}{2015}).

\bibitem{Teinturier_2023_pcm}
\bibinfo{author}{{Teinturier}, L.} \emph{et~al.}
\newblock \bibinfo{title}{{The radiative and dynamical impact of clouds in the
  atmosphere of the Hot Jupiter WASP-43 b}}.
\newblock \emph{\bibinfo{journal}{\aap}}  (\bibinfo{year}{2023}).

\bibitem{blain_2021}
\bibinfo{author}{{Blain}, D.}, \bibinfo{author}{{Charnay}, B.} \&
  \bibinfo{author}{{B{\'e}zard}, B.}
\newblock \bibinfo{title}{{1D atmospheric study of the temperate sub-Neptune
  K2-18b}}.
\newblock \emph{\bibinfo{journal}{\aap}} \textbf{\bibinfo{volume}{646}},
  \bibinfo{pages}{A15} (\bibinfo{year}{2021}).

\bibitem{falco_toward_2022}
\bibinfo{author}{Falco, A.}, \bibinfo{author}{Zingales, T.},
  \bibinfo{author}{Pluriel, W.} \& \bibinfo{author}{Leconte, J.}
\newblock \bibinfo{title}{Toward a multidimensional analysis of transmission
  spectroscopy: {I}. {Computation} of transmission spectra using a {1D}, {2D},
  or {3D} atmosphere structure}.
\newblock \emph{\bibinfo{journal}{\aap}} \textbf{\bibinfo{volume}{658}},
  \bibinfo{pages}{A41} (\bibinfo{year}{2022}).

\bibitem{Adcroft2004}
\bibinfo{author}{{Adcroft}, A.}, \bibinfo{author}{{Campin}, J.-M.},
  \bibinfo{author}{{Hill}, C.} \& \bibinfo{author}{{Marshall}, J.}
\newblock \bibinfo{title}{{Implementation of an Atmosphere Ocean General
  Circulation Model on the Expanded Spherical Cube}}.
\newblock \emph{\bibinfo{journal}{Monthly Weather Review}}
  \textbf{\bibinfo{volume}{132}}, \bibinfo{pages}{2845} (\bibinfo{year}{2004}).

\bibitem{Marley1999}
\bibinfo{author}{{Marley}, M.~S.}, \bibinfo{author}{{Gelino}, C.},
  \bibinfo{author}{{Stephens}, D.}, \bibinfo{author}{{Lunine}, J.~I.} \&
  \bibinfo{author}{{Freedman}, R.}
\newblock \bibinfo{title}{{Reflected Spectra and Albedos of Extrasolar Giant
  Planets. I. Clear and Cloudy Atmospheres}}.
\newblock \emph{\bibinfo{journal}{\apj}} \textbf{\bibinfo{volume}{513}},
  \bibinfo{pages}{879--893} (\bibinfo{year}{1999}).
\newblock \eprint{astro-ph/9810073}.

\bibitem{Freedman2008}
\bibinfo{author}{{Freedman}, R.~S.}, \bibinfo{author}{{Marley}, M.~S.} \&
  \bibinfo{author}{{Lodders}, K.}
\newblock \bibinfo{title}{{Line and Mean Opacities for Ultracool Dwarfs and
  Extrasolar Planets}}.
\newblock \emph{\bibinfo{journal}{\apjs}} \textbf{\bibinfo{volume}{174}},
  \bibinfo{pages}{504--513} (\bibinfo{year}{2008}).
\newblock \eprint{0706.2374}.

\bibitem{Freedman2014}
\bibinfo{author}{{Freedman}, R.~S.} \emph{et~al.}
\newblock \bibinfo{title}{{Gaseous Mean Opacities for Giant Planet and
  Ultracool Dwarf Atmospheres over a Range of Metallicities and Temperatures}}.
\newblock \emph{\bibinfo{journal}{\apjs}} \textbf{\bibinfo{volume}{214}},
  \bibinfo{pages}{25} (\bibinfo{year}{2014}).
\newblock \eprint{1409.0026}.

\bibitem{Marley2021}
\bibinfo{author}{{Marley}, M.~S.} \emph{et~al.}
\newblock \bibinfo{title}{{The Sonora Brown Dwarf Atmosphere and Evolution
  Models. I. Model Description and Application to Cloudless Atmospheres in
  Rainout Chemical Equilibrium}}.
\newblock \emph{\bibinfo{journal}{\apj}} \textbf{\bibinfo{volume}{920}},
  \bibinfo{pages}{85} (\bibinfo{year}{2021}).
\newblock \eprint{2107.07434}.

\bibitem{Visscher2010}
\bibinfo{author}{{Visscher}, C.}, \bibinfo{author}{{Lodders}, K.} \&
  \bibinfo{author}{{Fegley}, J., Bruce}.
\newblock \bibinfo{title}{{Atmospheric Chemistry in Giant Planets, Brown
  Dwarfs, and Low-mass Dwarf Stars. III. Iron, Magnesium, and Silicon}}.
\newblock \emph{\bibinfo{journal}{\apj}} \textbf{\bibinfo{volume}{716}},
  \bibinfo{pages}{1060--1075} (\bibinfo{year}{2010}).
\newblock \eprint{1001.3639}.

\bibitem{parmentier2013}
\bibinfo{author}{Parmentier, V.}, \bibinfo{author}{Showman, A.~P.} \&
  \bibinfo{author}{Lian, Y.}
\newblock \bibinfo{title}{3d mixing in hot jupiters atmospheres-i. application
  to the day/night cold trap in hd 209458b}.
\newblock \emph{\bibinfo{journal}{\aap}} \textbf{\bibinfo{volume}{558}},
  \bibinfo{pages}{A91} (\bibinfo{year}{2013}).

\bibitem{tan2021b}
\bibinfo{author}{{Tan}, X.} \& \bibinfo{author}{{Showman}, A.~P.}
\newblock \bibinfo{title}{{Atmospheric circulation of brown dwarfs and directly
  imaged exoplanets driven by cloud radiative feedback: global and equatorial
  dynamics}}.
\newblock \emph{\bibinfo{journal}{\mnras}} \textbf{\bibinfo{volume}{502}},
  \bibinfo{pages}{2198--2219} (\bibinfo{year}{2021}).
\newblock \eprint{2101.04417}.

\bibitem{Komacek2022}
\bibinfo{author}{{Komacek}, T.~D.}, \bibinfo{author}{{Tan}, X.},
  \bibinfo{author}{{Gao}, P.} \& \bibinfo{author}{{Lee}, E. K.~H.}
\newblock \bibinfo{title}{{Patchy Nightside Clouds on Ultra-hot Jupiters:
  General Circulation Model Simulations with Radiatively Active Cloud
  Tracers}}.
\newblock \emph{\bibinfo{journal}{\apj}} \textbf{\bibinfo{volume}{934}},
  \bibinfo{pages}{79} (\bibinfo{year}{2022}).
\newblock \eprint{2205.07834}.

\bibitem{AckermanMarley2001apj}
\bibinfo{author}{{Ackerman}, A.~S.} \& \bibinfo{author}{{Marley}, M.~S.}
\newblock \bibinfo{title}{{Precipitating Condensation Clouds in Substellar
  Atmospheres}}.
\newblock \emph{\bibinfo{journal}{\apj}} \textbf{\bibinfo{volume}{556}},
  \bibinfo{pages}{872--884} (\bibinfo{year}{2001}).
\newblock \eprint{astro-ph/0103423}.

\bibitem{Lee2022}
\bibinfo{author}{Lee, E.~K.} \emph{et~al.}
\newblock \bibinfo{title}{3d radiative transfer for exoplanet atmospheres.
  gcmcrt: A gpu-accelerated mcrt code}.
\newblock \emph{\bibinfo{journal}{\apj}} \textbf{\bibinfo{volume}{929}},
  \bibinfo{pages}{180} (\bibinfo{year}{2022}).

\bibitem{Lee2021}
\bibinfo{author}{Lee, E.~K.} \emph{et~al.}
\newblock \bibinfo{title}{Simulating gas giant exoplanet atmospheres with
  exo-fms: comparing semigrey, picket fence, and correlated-k
  radiative-transfer schemes}.
\newblock \emph{\bibinfo{journal}{\mnras}} \textbf{\bibinfo{volume}{506}},
  \bibinfo{pages}{2695--2711} (\bibinfo{year}{2021}).

\bibitem{Grimm2021}
\bibinfo{author}{Grimm, S.~L.} \emph{et~al.}
\newblock \bibinfo{title}{Helios-k 2.0 opacity calculator and open-source
  opacity database for exoplanetary atmospheres}.
\newblock \emph{\bibinfo{journal}{\apjs}} \textbf{\bibinfo{volume}{253}},
  \bibinfo{pages}{30} (\bibinfo{year}{2021}).

\bibitem{Gharib2021}
\bibinfo{author}{Gharib-Nezhad, E.} \emph{et~al.}
\newblock \bibinfo{title}{Exoplines: molecular absorption cross-section
  database for brown dwarf and giant exoplanet atmospheres}.
\newblock \emph{\bibinfo{journal}{\apjs}} \textbf{\bibinfo{volume}{254}},
  \bibinfo{pages}{34} (\bibinfo{year}{2021}).

\bibitem{Schneider2022}
\bibinfo{author}{{Schneider}, A.~D.} \emph{et~al.}
\newblock \bibinfo{title}{{Exploring the deep atmospheres of HD 209458b and
  WASP-43b using a non-gray general circulation model}}.
\newblock \emph{\bibinfo{journal}{\aap}} \textbf{\bibinfo{volume}{664}},
  \bibinfo{pages}{A56} (\bibinfo{year}{2022}).
\newblock \eprint{2202.09183}.

\bibitem{Molliere2019}
\bibinfo{author}{{Molli{\`e}re}, P.} \emph{et~al.}
\newblock \bibinfo{title}{{petitRADTRANS. A Python radiative transfer package
  for exoplanet characterization and retrieval}}.
\newblock \emph{\bibinfo{journal}{\aap}} \textbf{\bibinfo{volume}{627}},
  \bibinfo{pages}{A67} (\bibinfo{year}{2019}).
\newblock \eprint{1904.11504}.

\bibitem{Chubb2021}
\bibinfo{author}{{Chubb}, K.~L.} \emph{et~al.}
\newblock \bibinfo{title}{{The ExoMolOP database: Cross sections and k-tables
  for molecules of interest in high-temperature exoplanet atmospheres}}.
\newblock \emph{\bibinfo{journal}{\aap}} \textbf{\bibinfo{volume}{646}},
  \bibinfo{pages}{A21} (\bibinfo{year}{2021}).
\newblock \eprint{2009.00687}.

\bibitem{Polyansky2018}
\bibinfo{author}{{Polyansky}, O.~L.} \emph{et~al.}
\newblock \bibinfo{title}{{ExoMol molecular line lists XXX: a complete
  high-accuracy line list for water}}.
\newblock \emph{\bibinfo{journal}{\mnras}} \textbf{\bibinfo{volume}{480}},
  \bibinfo{pages}{2597--2608} (\bibinfo{year}{2018}).
\newblock \eprint{1807.04529}.

\bibitem{yurchenko17}
\bibinfo{author}{{Yurchenko}, S.~N.}, \bibinfo{author}{{Amundsen}, D.~S.},
  \bibinfo{author}{{Tennyson}, J.} \& \bibinfo{author}{{Waldmann}, I.~P.}
\newblock \bibinfo{title}{{A hybrid line list for CH$_{4}$ and hot methane
  continuum}}.
\newblock \emph{\bibinfo{journal}{\aap}} \textbf{\bibinfo{volume}{605}},
  \bibinfo{pages}{A95} (\bibinfo{year}{2017}).
\newblock \eprint{1706.05724}.

\bibitem{20YuMeFr.co2}
\bibinfo{author}{Yurchenko, S.~N.}, \bibinfo{author}{Mellor, T.~M.},
  \bibinfo{author}{Freedman, R.~S.} \& \bibinfo{author}{Tennyson, J.}
\newblock \bibinfo{title}{{ExoMol line lists – XXXIX. Ro-vibrational
  molecular line list for CO$_2$}}.
\newblock \emph{\bibinfo{journal}{\mnras}} \textbf{\bibinfo{volume}{496}},
  \bibinfo{pages}{5282--5291} (\bibinfo{year}{2020}).

\bibitem{Coles_2019_NH3}
\bibinfo{author}{Coles, P.~A.}, \bibinfo{author}{Yurchenko, S.~N.} \&
  \bibinfo{author}{Tennyson, J.}
\newblock \bibinfo{title}{{ExoMol} molecular line lists {\textendash} {XXXV}. a
  rotation-vibration line list for hot ammonia}.
\newblock \emph{\bibinfo{journal}{\mnras}} \textbf{\bibinfo{volume}{490}},
  \bibinfo{pages}{4638--4647} (\bibinfo{year}{2019}).
\newblock \urlprefix\url{https://doi.org/10.1093\%2Fmnras\%2Fstz2778}.

\bibitem{li2015}
\bibinfo{author}{{Li}, G.} \emph{et~al.}
\newblock \bibinfo{title}{{Rovibrational Line Lists for Nine Isotopologues of
  the {CO} Molecule in the {X}$^{1}${\ensuremath{\Sigma}}$^{+}$ Ground
  Electronic State}}.
\newblock \emph{\bibinfo{journal}{\apjs}} \textbf{\bibinfo{volume}{216}},
  \bibinfo{pages}{15} (\bibinfo{year}{2015}).

\bibitem{azzam2016}
\bibinfo{author}{{Azzam}, A. A.~A.}, \bibinfo{author}{{Tennyson}, J.},
  \bibinfo{author}{{Yurchenko}, S.~N.} \& \bibinfo{author}{{Naumenko}, O.~V.}
\newblock \bibinfo{title}{{ExoMol molecular line lists - XVI. The
  rotation-vibration spectrum of hot H$_{2}$S}}.
\newblock \emph{\bibinfo{journal}{\mnras}} \textbf{\bibinfo{volume}{460}},
  \bibinfo{pages}{4063--4074} (\bibinfo{year}{2016}).
\newblock \eprint{1607.00499}.

\bibitem{Barber2014}
\bibinfo{author}{{Barber}, R.~J.} \emph{et~al.}
\newblock \bibinfo{title}{{ExoMol line lists - III. An improved hot
  rotation-vibration line list for HCN and HNC}}.
\newblock \emph{\bibinfo{journal}{\mnras}} \textbf{\bibinfo{volume}{437}},
  \bibinfo{pages}{1828--1835} (\bibinfo{year}{2014}).
\newblock \eprint{1311.1328}.

\bibitem{Sousa_Silva_2014}
\bibinfo{author}{Sousa-Silva, C.}, \bibinfo{author}{Al-Refaie, A.~F.},
  \bibinfo{author}{Tennyson, J.} \& \bibinfo{author}{Yurchenko, S.~N.}
\newblock \bibinfo{title}{{ExoMol} line lists {\textendash} {VII}. the
  rotation{\textendash}vibration spectrum of phosphine up to 1500~k}.
\newblock \emph{\bibinfo{journal}{\mnras}} \textbf{\bibinfo{volume}{446}},
  \bibinfo{pages}{2337--2347} (\bibinfo{year}{2014}).
\newblock \urlprefix\url{https://doi.org/10.1093\%2Fmnras\%2Fstu2246}.

\bibitem{Wende2010}
\bibinfo{author}{{Wende}, S.}, \bibinfo{author}{{Reiners}, A.},
  \bibinfo{author}{{Seifahrt}, A.} \& \bibinfo{author}{{Bernath}, P.~F.}
\newblock \bibinfo{title}{{CRIRES spectroscopy and empirical line-by-line
  identification of FeH molecular absorption in an M dwarf}}.
\newblock \emph{\bibinfo{journal}{\aap}} \textbf{\bibinfo{volume}{523}},
  \bibinfo{pages}{A58} (\bibinfo{year}{2010}).
\newblock \eprint{1007.4116}.

\bibitem{Piskunov1995}
\bibinfo{author}{{Piskunov}, N.~E.}, \bibinfo{author}{{Kupka}, F.},
  \bibinfo{author}{{Ryabchikova}, T.~A.}, \bibinfo{author}{{Weiss}, W.~W.} \&
  \bibinfo{author}{{Jeffery}, C.~S.}
\newblock \bibinfo{title}{{VALD: The Vienna Atomic Line Data Base.}}
\newblock \emph{\bibinfo{journal}{\aaps}} \textbf{\bibinfo{volume}{112}},
  \bibinfo{pages}{525} (\bibinfo{year}{1995}).

\bibitem{Dalgarno1962}
\bibinfo{author}{{Dalgarno}, A.} \& \bibinfo{author}{{Williams}, D.~A.}
\newblock \bibinfo{title}{{Rayleigh Scattering by Molecular Hydrogen.}}
\newblock \emph{\bibinfo{journal}{\apj}} \textbf{\bibinfo{volume}{136}},
  \bibinfo{pages}{690--692} (\bibinfo{year}{1962}).

\bibitem{Chan1965}
\bibinfo{author}{{Chan}, Y.~M.} \& \bibinfo{author}{{Dalgarno}, A.}
\newblock \bibinfo{title}{{The refractive index of helium}}.
\newblock \emph{\bibinfo{journal}{\pps}} \textbf{\bibinfo{volume}{85}},
  \bibinfo{pages}{227--230} (\bibinfo{year}{1965}).

\bibitem{Richard2012}
\bibinfo{author}{Richard, C.} \emph{et~al.}
\newblock \bibinfo{title}{New section of the {HITRAN} database:
  Collision-induced absorption {(CIA)}}.
\newblock \emph{\bibinfo{journal}{{\jqsrt}}} \textbf{\bibinfo{volume}{113}},
  \bibinfo{pages}{1276 -- 1285} (\bibinfo{year}{2012}).
\newblock
  \urlprefix\url{http://www.sciencedirect.com/science/article/pii/S0022407311003773}.
\newblock \bibinfo{note}{Three Leaders in Spectroscopy}.

\bibitem{Hoskins1975}
\bibinfo{author}{Hoskins, B.} \& \bibinfo{author}{Simmons, A.}
\newblock \bibinfo{title}{A multi-layer spectral model and the semi-implicit
  method}.
\newblock \emph{\bibinfo{journal}{\qjrms}} \textbf{\bibinfo{volume}{101}},
  \bibinfo{pages}{637--655} (\bibinfo{year}{1975}).

\bibitem{MenouRauscher2009}
\bibinfo{author}{{Menou}, K.} \& \bibinfo{author}{{Rauscher}, E.}
\newblock \bibinfo{title}{{Atmospheric Circulation of Hot Jupiters: A Shallow
  Three-Dimensional Model}}.
\newblock \emph{\bibinfo{journal}{\apj}} \textbf{\bibinfo{volume}{700}},
  \bibinfo{pages}{887--897} (\bibinfo{year}{2009}).
\newblock \eprint{0809.1671}.

\bibitem{RauscherMenou2010}
\bibinfo{author}{Rauscher, E.} \& \bibinfo{author}{Menou, K.}
\newblock \bibinfo{title}{Three-dimensional modeling of hot jupiter atmospheric
  flows}.
\newblock \emph{\bibinfo{journal}{\apj}} \textbf{\bibinfo{volume}{714}},
  \bibinfo{pages}{1334} (\bibinfo{year}{2010}).

\bibitem{RauscherMenou2012}
\bibinfo{author}{{Rauscher}, E.} \& \bibinfo{author}{{Menou}, K.}
\newblock \bibinfo{title}{{A General Circulation Model for Gaseous Exoplanets
  with Double-gray Radiative Transfer}}.
\newblock \emph{\bibinfo{journal}{\apj}} \textbf{\bibinfo{volume}{750}},
  \bibinfo{pages}{96} (\bibinfo{year}{2012}).
\newblock \eprint{1112.1658}.

\bibitem{rauscher2014atmospheric}
\bibinfo{author}{Rauscher, E.} \& \bibinfo{author}{Kempton, E.~M.}
\newblock \bibinfo{title}{The atmospheric circulation and observable properties
  of non-synchronously rotating hot jupiters}.
\newblock \emph{\bibinfo{journal}{\apj}} \textbf{\bibinfo{volume}{790}},
  \bibinfo{pages}{79} (\bibinfo{year}{2014}).

\bibitem{may2020super}
\bibinfo{author}{May, E.~M.} \& \bibinfo{author}{Rauscher, E.}
\newblock \bibinfo{title}{From super-earths to mini-neptunes: Implications of a
  surface on atmospheric circulation}.
\newblock \emph{\bibinfo{journal}{\apj}} \textbf{\bibinfo{volume}{893}},
  \bibinfo{pages}{161} (\bibinfo{year}{2020}).

\bibitem{beltz2022magnetic}
\bibinfo{author}{Beltz, H.} \emph{et~al.}
\newblock \bibinfo{title}{Magnetic drag and 3d effects in theoretical
  high-resolution emission spectra of ultrahot jupiters: the case of wasp-76b}.
\newblock \emph{\bibinfo{journal}{\aj}} \textbf{\bibinfo{volume}{164}},
  \bibinfo{pages}{140} (\bibinfo{year}{2022}).

\bibitem{toon1989}
\bibinfo{author}{Toon, O.~B.}, \bibinfo{author}{McKay, C.},
  \bibinfo{author}{Ackerman, T.} \& \bibinfo{author}{Santhanam, K.}
\newblock \bibinfo{title}{Rapid calculation of radiative heating rates and
  photodissociation rates in inhomogeneous multiple scattering atmospheres}.
\newblock \emph{\bibinfo{journal}{Journal of Geophysical Research:
  Atmospheres}} \textbf{\bibinfo{volume}{94}}, \bibinfo{pages}{16287--16301}
  (\bibinfo{year}{1989}).

\bibitem{Gao2020}
\bibinfo{author}{{Gao}, P.} \emph{et~al.}
\newblock \bibinfo{title}{{Aerosol composition of hot giant exoplanets
  dominated by silicates and hydrocarbon hazes}}.
\newblock \emph{\bibinfo{journal}{\nastro}} \textbf{\bibinfo{volume}{4}},
  \bibinfo{pages}{951--956} (\bibinfo{year}{2020}).
\newblock \eprint{2005.11939}.

\bibitem{Zhang2017}
\bibinfo{author}{{Zhang}, J.}, \bibinfo{author}{{Kempton}, E. M.~R.} \&
  \bibinfo{author}{{Rauscher}, E.}
\newblock \bibinfo{title}{{Constraining Hot Jupiter Atmospheric Structure and
  Dynamics through Doppler-shifted Emission Spectra}}.
\newblock \emph{\bibinfo{journal}{\apj}} \textbf{\bibinfo{volume}{851}},
  \bibinfo{pages}{84} (\bibinfo{year}{2017}).
\newblock \eprint{1711.02684}.

\bibitem{Malsky2021}
\bibinfo{author}{{Malsky}, I.} \emph{et~al.}
\newblock \bibinfo{title}{{Modeling the High-resolution Emission Spectra of
  Clear and Cloudy Nontransiting Hot Jupiters}}.
\newblock \emph{\bibinfo{journal}{\apj}} \textbf{\bibinfo{volume}{923}},
  \bibinfo{pages}{62} (\bibinfo{year}{2021}).
\newblock \eprint{2110.05593}.

\bibitem{mendonca2016}
\bibinfo{author}{{Mendon{\c{c}}a}, J.~M.}, \bibinfo{author}{{Grimm}, S.~L.},
  \bibinfo{author}{{Grosheintz}, L.} \& \bibinfo{author}{{Heng}, K.}
\newblock \bibinfo{title}{{THOR: A New and Flexible Global Circulation Model to
  Explore Planetary Atmospheres}}.
\newblock \emph{\bibinfo{journal}{\apj}} \textbf{\bibinfo{volume}{829}},
  \bibinfo{pages}{115} (\bibinfo{year}{2016}).
\newblock \eprint{1607.05535}.

\bibitem{deitrick2020}
\bibinfo{author}{{Deitrick}, R.} \emph{et~al.}
\newblock \bibinfo{title}{{THOR 2.0: Major Improvements to the Open-source
  General Circulation Model}}.
\newblock \emph{\bibinfo{journal}{\apjs}} \textbf{\bibinfo{volume}{248}},
  \bibinfo{pages}{30} (\bibinfo{year}{2020}).
\newblock \eprint{1911.13158}.

\bibitem{tomita2004}
\bibinfo{author}{{Tomita}, H.} \& \bibinfo{author}{{Satoh}, M.}
\newblock \bibinfo{title}{{A new dynamical framework of nonhydrostatic global
  model using the icosahedral grid}}.
\newblock \emph{\bibinfo{journal}{Fluid Dynamics Research}}
  \textbf{\bibinfo{volume}{34}}, \bibinfo{pages}{357--400}
  (\bibinfo{year}{2004}).

\bibitem{mendonca2022}
\bibinfo{author}{{Mendon{\c{c}}a}, J.~M.}
\newblock \bibinfo{title}{{Mass transport in a moist planetary climate model}}.
\newblock \emph{\bibinfo{journal}{\aap}} \textbf{\bibinfo{volume}{659}},
  \bibinfo{pages}{A43} (\bibinfo{year}{2022}).
\newblock \eprint{2110.03719}.

\bibitem{mendonca2020b}
\bibinfo{author}{{Mendon{\c{c}}a}, J.~M.} \& \bibinfo{author}{{Buchhave},
  L.~A.}
\newblock \bibinfo{title}{{Modelling the 3D climate of Venus with OASIS}}.
\newblock \emph{\bibinfo{journal}{\mnras}} \textbf{\bibinfo{volume}{496}},
  \bibinfo{pages}{3512--3530} (\bibinfo{year}{2020}).
\newblock \eprint{2002.09506}.

\bibitem{shao2022}
\bibinfo{author}{{Shao}, W.~D.}, \bibinfo{author}{{Zhang}, X.},
  \bibinfo{author}{{Mendon{\c{c}}a}, J.} \& \bibinfo{author}{{Encrenaz}, T.}
\newblock \bibinfo{title}{{Local-time Dependence of Chemical Species in the
  Venusian Mesosphere}}.
\newblock \emph{\bibinfo{journal}{\psj}} \textbf{\bibinfo{volume}{3}},
  \bibinfo{pages}{3} (\bibinfo{year}{2022}).
\newblock \eprint{2112.07037}.

\bibitem{deitrick2022}
\bibinfo{author}{{Deitrick}, R.} \emph{et~al.}
\newblock \bibinfo{title}{{The THOR + HELIOS general circulation model:
  multiwavelength radiative transfer with accurate scattering by
  clouds/hazes}}.
\newblock \emph{\bibinfo{journal}{\mnras}} \textbf{\bibinfo{volume}{512}},
  \bibinfo{pages}{3759--3787} (\bibinfo{year}{2022}).
\newblock \eprint{2203.02293}.

\bibitem{tsai2018}
\bibinfo{author}{{Tsai}, S.-M.} \emph{et~al.}
\newblock \bibinfo{title}{{Toward Consistent Modeling of Atmospheric Chemistry
  and Dynamics in Exoplanets: Validation and Generalization of the Chemical
  Relaxation Method}}.
\newblock \emph{\bibinfo{journal}{\apj}} \textbf{\bibinfo{volume}{862}},
  \bibinfo{pages}{31} (\bibinfo{year}{2018}).
\newblock \eprint{1711.08492}.

\bibitem{malik2017}
\bibinfo{author}{{Malik}, M.} \emph{et~al.}
\newblock \bibinfo{title}{{HELIOS: An Open-source, GPU-accelerated Radiative
  Transfer Code for Self-consistent Exoplanetary Atmospheres}}.
\newblock \emph{\bibinfo{journal}{\aj}} \textbf{\bibinfo{volume}{153}},
  \bibinfo{pages}{56} (\bibinfo{year}{2017}).
\newblock \eprint{1606.05474}.

\bibitem{charnay2022}
\bibinfo{author}{{Charnay}, B.} \emph{et~al.}
\newblock \bibinfo{title}{{A survey of exoplanet phase curves with Ariel}}.
\newblock \emph{\bibinfo{journal}{Experimental Astronomy}}
  \textbf{\bibinfo{volume}{53}}, \bibinfo{pages}{417--446}
  (\bibinfo{year}{2022}).
\newblock \eprint{2102.06523}.

\bibitem{mendonca2015}
\bibinfo{author}{{Mendon{\c{c}}a}, J.~M.}, \bibinfo{author}{{Read}, P.~L.},
  \bibinfo{author}{{Wilson}, C.~F.} \& \bibinfo{author}{{Lee}, C.}
\newblock \bibinfo{title}{{A new, fast and flexible radiative transfer method
  for Venus general circulation models}}.
\newblock \emph{\bibinfo{journal}{\planss}} \textbf{\bibinfo{volume}{105}},
  \bibinfo{pages}{80--93} (\bibinfo{year}{2015}).

\bibitem{Yurchenko2014}
\bibinfo{author}{{Yurchenko}, S.~N.} \& \bibinfo{author}{{Tennyson}, J.}
\newblock \bibinfo{title}{{ExoMol line lists - IV. The rotation-vibration
  spectrum of methane up to 1500 K}}.
\newblock \emph{\bibinfo{journal}{\mnras}} \textbf{\bibinfo{volume}{440}},
  \bibinfo{pages}{1649--1661} (\bibinfo{year}{2014}).
\newblock \eprint{1401.4852}.

\bibitem{Yurchenko2011}
\bibinfo{author}{{Yurchenko}, S.~N.}, \bibinfo{author}{{Barber}, R.~J.} \&
  \bibinfo{author}{{Tennyson}, J.}
\newblock \bibinfo{title}{{A variationally computed line list for hot
  NH$_{3}$}}.
\newblock \emph{\bibinfo{journal}{\mnras}} \textbf{\bibinfo{volume}{413}},
  \bibinfo{pages}{1828--1834} (\bibinfo{year}{2011}).
\newblock \eprint{1011.1569}.

\bibitem{Harris2006}
\bibinfo{author}{{Harris}, G.~J.}, \bibinfo{author}{{Tennyson}, J.},
  \bibinfo{author}{{Kaminsky}, B.~M.}, \bibinfo{author}{{Pavlenko}, Y.~V.} \&
  \bibinfo{author}{{Jones}, H.~R.~A.}
\newblock \bibinfo{title}{{Improved HCN/HNC linelist, model atmospheres and
  synthetic spectra for WZ Cas}}.
\newblock \emph{\bibinfo{journal}{\mnras}} \textbf{\bibinfo{volume}{367}},
  \bibinfo{pages}{400--406} (\bibinfo{year}{2006}).
\newblock \eprint{astro-ph/0512363}.

\bibitem{Rothman2010}
\bibinfo{author}{{Rothman}, L.~S.} \emph{et~al.}
\newblock \bibinfo{title}{{HITEMP, the high-temperature molecular spectroscopic
  database}}.
\newblock \emph{\bibinfo{journal}{\jqsrt}} \textbf{\bibinfo{volume}{111}},
  \bibinfo{pages}{2139--2150} (\bibinfo{year}{2010}).

\bibitem{rothman2013}
\bibinfo{author}{{Rothman}, L.~S.} \emph{et~al.}
\newblock \bibinfo{title}{{The HITRAN2012 molecular spectroscopic database}}.
\newblock \emph{\bibinfo{journal}{\jqsrt}} \textbf{\bibinfo{volume}{130}},
  \bibinfo{pages}{4--50} (\bibinfo{year}{2013}).

\bibitem{draine2011}
\bibinfo{author}{{Draine}, B.~T.}
\newblock \emph{\bibinfo{title}{{Physics of the Interstellar and Intergalactic
  Medium}}} (\bibinfo{publisher}{Princeton University Press},
  \bibinfo{year}{2011}).

\bibitem{stock2018}
\bibinfo{author}{{Stock}, J.~W.}, \bibinfo{author}{{Kitzmann}, D.},
  \bibinfo{author}{{Patzer}, A. B.~C.} \& \bibinfo{author}{{Sedlmayr}, E.}
\newblock \bibinfo{title}{{FastChem: A computer program for efficient complex
  chemical equilibrium calculations in the neutral/ionized gas phase with
  applications to stellar and planetary atmospheres}}.
\newblock \emph{\bibinfo{journal}{\mnras}} \textbf{\bibinfo{volume}{479}},
  \bibinfo{pages}{865--874} (\bibinfo{year}{2018}).
\newblock \eprint{1804.05010}.

\bibitem{Gandhi2018}
\bibinfo{author}{{Gandhi}, S.} \& \bibinfo{author}{{Madhusudhan}, N.}
\newblock \bibinfo{title}{{Retrieval of exoplanet emission spectra with
  HyDRA}}.
\newblock \emph{\bibinfo{journal}{\mnras}} \textbf{\bibinfo{volume}{474}},
  \bibinfo{pages}{271--288} (\bibinfo{year}{2018}).
\newblock \eprint{1710.06433}.

\bibitem{Feroz2009}
\bibinfo{author}{{Feroz}, F.}, \bibinfo{author}{{Hobson}, M.~P.} \&
  \bibinfo{author}{{Bridges}, M.}
\newblock \bibinfo{title}{{MULTINEST: an efficient and robust Bayesian
  inference tool for cosmology and particle physics}}.
\newblock \emph{\bibinfo{journal}{\mnras}} \textbf{\bibinfo{volume}{398}},
  \bibinfo{pages}{1601--1614} (\bibinfo{year}{2009}).
\newblock \eprint{0809.3437}.

\bibitem{buchner14}
\bibinfo{author}{{Buchner}, J.} \emph{et~al.}
\newblock \bibinfo{title}{{X-ray spectral modelling of the AGN obscuring region
  in the CDFS: Bayesian model selection and catalogue}}.
\newblock \emph{\bibinfo{journal}{\aap}} \textbf{\bibinfo{volume}{564}},
  \bibinfo{pages}{A125} (\bibinfo{year}{2014}).
\newblock \eprint{1402.0004}.

\bibitem{Skilling2006}
\bibinfo{author}{Skilling, J.}
\newblock \bibinfo{title}{Nested sampling for general bayesian computation}.
\newblock \emph{\bibinfo{journal}{Bayesian Anal.}}
  \textbf{\bibinfo{volume}{1}}, \bibinfo{pages}{833--859}
  (\bibinfo{year}{2006}).
\newblock \urlprefix\url{https://doi.org/10.1214/06-BA127}.

\bibitem{Gandhi2020}
\bibinfo{author}{{Gandhi}, S.}, \bibinfo{author}{{Madhusudhan}, N.} \&
  \bibinfo{author}{{Mandell}, A.}
\newblock \bibinfo{title}{{H- and Dissociation in Ultra-hot Jupiters: A
  Retrieval Case Study of WASP-18b}}.
\newblock \emph{\bibinfo{journal}{\aj}} \textbf{\bibinfo{volume}{159}},
  \bibinfo{pages}{232} (\bibinfo{year}{2020}).
\newblock \eprint{2004.07252}.

\bibitem{Piette2020b}
\bibinfo{author}{{Piette}, A. A.~A.} \& \bibinfo{author}{{Madhusudhan}, N.}
\newblock \bibinfo{title}{{Considerations for atmospheric retrieval of
  high-precision brown dwarf spectra}}.
\newblock \emph{\bibinfo{journal}{\mnras}} \textbf{\bibinfo{volume}{497}},
  \bibinfo{pages}{5136--5154} (\bibinfo{year}{2020}).
\newblock \eprint{2007.15004}.

\bibitem{Piette2022}
\bibinfo{author}{{Piette}, A. A.~A.}, \bibinfo{author}{{Madhusudhan}, N.} \&
  \bibinfo{author}{{Mandell}, A.~M.}
\newblock \bibinfo{title}{{HyDRo: atmospheric retrieval of rocky exoplanets in
  thermal emission}}.
\newblock \emph{\bibinfo{journal}{\mnras}} \textbf{\bibinfo{volume}{511}},
  \bibinfo{pages}{2565--2584} (\bibinfo{year}{2022}).
\newblock \eprint{2112.05059}.

\bibitem{Gandhi2019}
\bibinfo{author}{{Gandhi}, S.}, \bibinfo{author}{{Madhusudhan}, N.},
  \bibinfo{author}{{Hawker}, G.} \& \bibinfo{author}{{Piette}, A.}
\newblock \bibinfo{title}{{HyDRA-H: Simultaneous Hybrid Retrieval of
  Exoplanetary Emission Spectra}}.
\newblock \emph{\bibinfo{journal}{\aj}} \textbf{\bibinfo{volume}{158}},
  \bibinfo{pages}{228} (\bibinfo{year}{2019}).
\newblock \eprint{1910.14042}.

\bibitem{Gandhi2022}
\bibinfo{author}{{Gandhi}, S.} \emph{et~al.}
\newblock \bibinfo{title}{{Spatially resolving the terminator: variation of Fe,
  temperature, and winds in WASP-76 b across planetary limbs and orbital
  phase}}.
\newblock \emph{\bibinfo{journal}{\mnras}} \textbf{\bibinfo{volume}{515}},
  \bibinfo{pages}{749--766} (\bibinfo{year}{2022}).
\newblock \eprint{2206.11268}.

\bibitem{Madhusudhan2009}
\bibinfo{author}{{Madhusudhan}, N.} \& \bibinfo{author}{{Seager}, S.}
\newblock \bibinfo{title}{{A Temperature and Abundance Retrieval Method for
  Exoplanet Atmospheres}}.
\newblock \emph{\bibinfo{journal}{\apj}} \textbf{\bibinfo{volume}{707}},
  \bibinfo{pages}{24--39} (\bibinfo{year}{2009}).
\newblock \eprint{0910.1347}.

\bibitem{Taylor_2020}
\bibinfo{author}{{Taylor}, J.} \emph{et~al.}
\newblock \bibinfo{title}{{Understanding and mitigating biases when studying
  inhomogeneous emission spectra with JWST}}.
\newblock \emph{\bibinfo{journal}{\mnras}} \textbf{\bibinfo{volume}{493}},
  \bibinfo{pages}{4342--4354} (\bibinfo{year}{2020}).
\newblock \eprint{2002.00773}.

\bibitem{Yurchenko2013}
\bibinfo{author}{{Yurchenko}, S.~N.}, \bibinfo{author}{{Tennyson}, J.},
  \bibinfo{author}{{Barber}, R.~J.} \& \bibinfo{author}{{Thiel}, W.}
\newblock \bibinfo{title}{{Vibrational transition moments of CH$_{4}$ from
  first principles}}.
\newblock \emph{\bibinfo{journal}{Journal of Molecular Spectroscopy}}
  \textbf{\bibinfo{volume}{291}}, \bibinfo{pages}{69--76}
  (\bibinfo{year}{2013}).
\newblock \eprint{1302.1720}.

\bibitem{Gordon2017}
\bibinfo{author}{{Gordon}, I.~E.} \emph{et~al.}
\newblock \bibinfo{title}{{The HITRAN2016 molecular spectroscopic database}}.
\newblock \emph{\bibinfo{journal}{\jqsrt}} \textbf{\bibinfo{volume}{203}},
  \bibinfo{pages}{3--69} (\bibinfo{year}{2017}).

\bibitem{Underwood2016mnras}
\bibinfo{author}{Underwood, D.~S.} \emph{et~al.}
\newblock \bibinfo{title}{{ExoMol molecular line lists – XIV. The
  rotation–vibration spectrum of hot SO2}}.
\newblock \emph{\bibinfo{journal}{\mnras}} \textbf{\bibinfo{volume}{459}},
  \bibinfo{pages}{3890--3899} (\bibinfo{year}{2016}).
\newblock \urlprefix\url{https://doi.org/10.1093/mnras/stw849}.
\newblock
  \eprint{https://academic.oup.com/mnras/article-pdf/459/4/3890/8191566/stw849.pdf}.

\bibitem{Chubb2018}
\bibinfo{author}{Chubb, K.~L.} \emph{et~al.}
\newblock \bibinfo{title}{Marvel analysis of the measured high-resolution
  rovibrational spectra of {H$_2^{32}$S}}.
\newblock \emph{\bibinfo{journal}{\jqsrt}} \textbf{\bibinfo{volume}{218}},
  \bibinfo{pages}{178--186} (\bibinfo{year}{2018}).
\newblock
  \urlprefix\url{https://www.sciencedirect.com/science/article/pii/S0022407318302565}.

\bibitem{Pinhas2017}
\bibinfo{author}{{Pinhas}, A.} \& \bibinfo{author}{{Madhusudhan}, N.}
\newblock \bibinfo{title}{{On signatures of clouds in exoplanetary transit
  spectra}}.
\newblock \emph{\bibinfo{journal}{\mnras}} \textbf{\bibinfo{volume}{471}},
  \bibinfo{pages}{4355--4373} (\bibinfo{year}{2017}).
\newblock \eprint{1705.08893}.

\bibitem{CubillosBlecic2021-PyratBay}
\bibinfo{author}{{Cubillos}, P.~E.} \& \bibinfo{author}{{Blecic}, J.}
\newblock \bibinfo{title}{{The Pyrat Bay Framework for Exoplanet Atmospheric
  Modeling: A Population Study of Hubble/WFC3 Transmission Spectra}}.
\newblock \emph{\bibinfo{journal}{\mnras}}  (\bibinfo{year}{2021}).
\newblock \eprint{2105.05598}.

\bibitem{BurrowsEtal2000apjBDspectra}
\bibinfo{author}{{Burrows}, A.}, \bibinfo{author}{{Marley}, M.~S.} \&
  \bibinfo{author}{{Sharp}, C.~M.}
\newblock \bibinfo{title}{{The Near-Infrared and Optical Spectra of Methane
  Dwarfs and Brown Dwarfs}}.
\newblock \emph{\bibinfo{journal}{\apj}} \textbf{\bibinfo{volume}{531}},
  \bibinfo{pages}{438--446} (\bibinfo{year}{2000}).
\newblock \eprint{astro-ph/9908078}.

\bibitem{Kurucz1970saorsAtlas}
\bibinfo{author}{{Kurucz}, R.~L.}
\newblock \bibinfo{title}{{Atlas: a Computer Program for Calculating Model
  Stellar Atmospheres}}.
\newblock \emph{\bibinfo{journal}{SAO Special Report}}
  \textbf{\bibinfo{volume}{309}} (\bibinfo{year}{1970}).

\bibitem{LecavelierEtal2008aaRayleighHD189733b}
\bibinfo{author}{{Lecavelier Des Etangs}, A.}, \bibinfo{author}{{Pont}, F.},
  \bibinfo{author}{{Vidal-Madjar}, A.} \& \bibinfo{author}{{Sing}, D.}
\newblock \bibinfo{title}{{Rayleigh scattering in the transit spectrum of HD
  189733b}}.
\newblock \emph{\bibinfo{journal}{\aap}} \textbf{\bibinfo{volume}{481}},
  \bibinfo{pages}{L83--L86} (\bibinfo{year}{2008}).
\newblock \eprint{0802.3228}.

\bibitem{TennysonEtal2016jmsExomol}
\bibinfo{author}{{Tennyson}, J.} \emph{et~al.}
\newblock \bibinfo{title}{{The ExoMol database: Molecular line lists for
  exoplanet and other hot atmospheres}}.
\newblock \emph{\bibinfo{journal}{Journal of Molecular Spectroscopy}}
  \textbf{\bibinfo{volume}{327}}, \bibinfo{pages}{73--94}
  (\bibinfo{year}{2016}).
\newblock \eprint{1603.05890}.

\bibitem{RothmanEtal2010jqsrtHITEMP}
\bibinfo{author}{{Rothman}, L.~S.} \emph{et~al.}
\newblock \bibinfo{title}{{HITEMP, the high-temperature molecular spectroscopic
  database}}.
\newblock \emph{\bibinfo{journal}{\jqsrt}} \textbf{\bibinfo{volume}{111}},
  \bibinfo{pages}{2139--2150} (\bibinfo{year}{2010}).

\bibitem{Cubillos2017apjRepack}
\bibinfo{author}{{Cubillos}, P.~E.}
\newblock \bibinfo{title}{{An Algorithm to Compress Line-transition Data for
  Radiative-transfer Calculations}}.
\newblock \emph{\bibinfo{journal}{\apj}} \textbf{\bibinfo{volume}{850}},
  \bibinfo{pages}{32} (\bibinfo{year}{2017}).
\newblock \eprint{1710.02556}.

\bibitem{BorysowEtal2001jqsrtH2H2highT}
\bibinfo{author}{{Borysow}, A.}, \bibinfo{author}{{Jorgensen}, U.~G.} \&
  \bibinfo{author}{{Fu}, Y.}
\newblock \bibinfo{title}{{High-temperature (1000-7000 K) collision-induced
  absorption of H$_2$ pairs computed from the first principles, with
  application to cool and dense stellar atmospheres}}.
\newblock \emph{\bibinfo{journal}{\jqsrt}} \textbf{\bibinfo{volume}{68}},
  \bibinfo{pages}{235--255} (\bibinfo{year}{2001}).

\bibitem{ParmentierGuillot2014aapTmodel}
\bibinfo{author}{{Parmentier}, V.} \& \bibinfo{author}{{Guillot}, T.}
\newblock \bibinfo{title}{{A non-grey analytical model for irradiated
  atmospheres. I. Derivation}}.
\newblock \emph{\bibinfo{journal}{\aap}} \textbf{\bibinfo{volume}{562}},
  \bibinfo{pages}{A133} (\bibinfo{year}{2014}).
\newblock \eprint{1311.6597}.

\bibitem{MadhusudhanSeager2009apjRetrieval}
\bibinfo{author}{{Madhusudhan}, N.} \& \bibinfo{author}{{Seager}, S.}
\newblock \bibinfo{title}{{A Temperature and Abundance Retrieval Method for
  Exoplanet Atmospheres}}.
\newblock \emph{\bibinfo{journal}{\apj}} \textbf{\bibinfo{volume}{707}},
  \bibinfo{pages}{24--39} (\bibinfo{year}{2009}).
\newblock \eprint{0910.1347}.

\bibitem{BlecicEtal2016apsjTEA}
\bibinfo{author}{{Blecic}, J.}, \bibinfo{author}{{Harrington}, J.} \&
  \bibinfo{author}{{Bowman}, M.~O.}
\newblock \bibinfo{title}{{TEA: A Code Calculating Thermochemical Equilibrium
  Abundances}}.
\newblock \emph{\bibinfo{journal}{\apjs}} \textbf{\bibinfo{volume}{225}},
  \bibinfo{pages}{4} (\bibinfo{year}{2016}).
\newblock \eprint{1505.06392}.

\bibitem{LineParmentier2016-patchy}
\bibinfo{author}{{Line}, M.~R.} \& \bibinfo{author}{{Parmentier}, V.}
\newblock \bibinfo{title}{{The Influence of Nonuniform Cloud Cover on Transit
  Transmission Spectra}}.
\newblock \emph{\bibinfo{journal}{\apj}} \textbf{\bibinfo{volume}{820}},
  \bibinfo{pages}{78} (\bibinfo{year}{2016}).
\newblock \eprint{1511.09443}.

\bibitem{benneke_strict_2015}
\bibinfo{author}{{Benneke}, B.}
\newblock \bibinfo{title}{{Strict Upper Limits on the Carbon-to-Oxygen Ratios
  of Eight Hot Jupiters from Self-Consistent Atmospheric Retrieval}}.
\newblock \emph{\bibinfo{journal}{arXiv e-prints}}
  \bibinfo{pages}{arXiv:1504.07655} (\bibinfo{year}{2015}).
\newblock \eprint{1504.07655}.

\bibitem{Benneke2013}
\bibinfo{author}{{Benneke}, B.} \& \bibinfo{author}{{Seager}, S.}
\newblock \bibinfo{title}{{How to Distinguish between Cloudy Mini-Neptunes and
  Water/Volatile-dominated Super-Earths}}.
\newblock \emph{\bibinfo{journal}{\apj}} \textbf{\bibinfo{volume}{778}},
  \bibinfo{pages}{153} (\bibinfo{year}{2013}).
\newblock \eprint{1306.6325}.

\bibitem{MorleyEtal2012}
\bibinfo{author}{{Morley}, C.~V.} \emph{et~al.}
\newblock \bibinfo{title}{{Neglected Clouds in T and Y Dwarf Atmospheres}}.
\newblock \emph{\bibinfo{journal}{\apj}} \textbf{\bibinfo{volume}{756}},
  \bibinfo{pages}{172} (\bibinfo{year}{2012}).
\newblock \eprint{1206.4313}.

\bibitem{terBraak2008SnookerDEMC}
\bibinfo{author}{{ter Braak}, C. J.~F.} \& \bibinfo{author}{{Vrugt}, J.~A.}
\newblock \bibinfo{title}{Differential evolution markov chain with snooker
  updater and fewer chains}.
\newblock \emph{\bibinfo{journal}{Statistics and Computing}}
  \textbf{\bibinfo{volume}{18}}, \bibinfo{pages}{435--446}
  (\bibinfo{year}{2008}).
\newblock \urlprefix\url{http://dx.doi.org/10.1007/s11222-008-9104-9}.

\bibitem{CubillosEtal2017apjRednoise}
\bibinfo{author}{{Cubillos}, P.} \emph{et~al.}
\newblock \bibinfo{title}{{On Correlated-noise Analyses Applied to Exoplanet
  Light Curves}}.
\newblock \emph{\bibinfo{journal}{\aj}} \textbf{\bibinfo{volume}{153}},
  \bibinfo{pages}{3} (\bibinfo{year}{2017}).
\newblock \eprint{1610.01336}.

\bibitem{feroz09}
\bibinfo{author}{{Feroz}, F.}, \bibinfo{author}{{Hobson}, M.~P.} \&
  \bibinfo{author}{{Bridges}, M.}
\newblock \bibinfo{title}{{MULTINEST: an efficient and robust Bayesian
  inference tool for cosmology and particle physics}}.
\newblock \emph{\bibinfo{journal}{\mnras}} \textbf{\bibinfo{volume}{398}},
  \bibinfo{pages}{1601--1614} (\bibinfo{year}{2009}).
\newblock \eprint{0809.3437}.

\bibitem{Irwin2008}
\bibinfo{author}{{Irwin}, P.~G.~J.} \emph{et~al.}
\newblock \bibinfo{title}{{The NEMESIS planetary atmosphere radiative transfer
  and retrieval tool}}.
\newblock \emph{\bibinfo{journal}{\jqsrt}} \textbf{\bibinfo{volume}{109}},
  \bibinfo{pages}{1136--1150} (\bibinfo{year}{2008}).

\bibitem{kt2018}
\bibinfo{author}{{Krissansen-Totton}, J.}, \bibinfo{author}{{Garland}, R.},
  \bibinfo{author}{{Irwin}, P.} \& \bibinfo{author}{{Catling}, D.~C.}
\newblock \bibinfo{title}{{Detectability of Biosignatures in Anoxic Atmospheres
  with the James Webb Space Telescope: A TRAPPIST-1e Case Study}}.
\newblock \emph{\bibinfo{journal}{\aj}} \textbf{\bibinfo{volume}{156}},
  \bibinfo{pages}{114} (\bibinfo{year}{2018}).
\newblock \eprint{1808.08377}.

\bibitem{lacisoinas}
\bibinfo{author}{{Lacis}, A.~A.} \& \bibinfo{author}{{Oinas}, V.}
\newblock \bibinfo{title}{{A description of the correlated-k distribution
  method for modelling nongray gaseous absorption, thermal emission, and
  multiple scattering in vertically inhomogeneous atmospheres}}.
\newblock \emph{\bibinfo{journal}{\jgr}} \textbf{\bibinfo{volume}{96}},
  \bibinfo{pages}{9027--9064} (\bibinfo{year}{1991}).

\bibitem{barstow20}
\bibinfo{author}{{Barstow}, J.~K.}
\newblock \bibinfo{title}{{Unveiling cloudy exoplanets: the influence of cloud
  model choices on retrieval solutions}}.
\newblock \emph{\bibinfo{journal}{\mnras}} \textbf{\bibinfo{volume}{497}},
  \bibinfo{pages}{4183--4195} (\bibinfo{year}{2020}).
\newblock \eprint{2002.02945}.

\bibitem{Irwin2020}
\bibinfo{author}{{Irwin}, P. G.~J.} \emph{et~al.}
\newblock \bibinfo{title}{{2.5D retrieval of atmospheric properties from
  exoplanet phase curves: application to WASP-43b observations}}.
\newblock \emph{\bibinfo{journal}{\mnras}} \textbf{\bibinfo{volume}{493}},
  \bibinfo{pages}{106--125} (\bibinfo{year}{2020}).

\bibitem{Borysow2002}
\bibinfo{author}{{Borysow}, A.}
\newblock \bibinfo{title}{{Collision-induced absorption coefficients of H$_{2}$
  pairs at temperatures from 60 K to 1000 K}}.
\newblock \emph{\bibinfo{journal}{\aap}} \textbf{\bibinfo{volume}{390}},
  \bibinfo{pages}{779--782} (\bibinfo{year}{2002}).

\bibitem{benneke_water_2019}
\bibinfo{author}{{Benneke}, B.} \emph{et~al.}
\newblock \bibinfo{title}{{Water Vapor and Clouds on the Habitable-zone
  Sub-Neptune Exoplanet K2-18b}}.
\newblock \emph{\bibinfo{journal}{\apjl}} \textbf{\bibinfo{volume}{887}},
  \bibinfo{pages}{L14} (\bibinfo{year}{2019}).
\newblock \eprint{1909.04642}.

\bibitem{pelletier_where_2021}
\bibinfo{author}{{Pelletier}, S.} \emph{et~al.}
\newblock \bibinfo{title}{{Where Is the Water? Jupiter-like C/H Ratio but
  Strong H$_{2}$O Depletion Found on {\ensuremath{\tau}} Bo{\"o}tis b Using
  SPIRou}}.
\newblock \emph{\bibinfo{journal}{\aj}} \textbf{\bibinfo{volume}{162}},
  \bibinfo{pages}{73} (\bibinfo{year}{2021}).
\newblock \eprint{2105.10513}.

\bibitem{fastchem2}
\bibinfo{author}{{Stock}, J.~W.}, \bibinfo{author}{{Kitzmann}, D.} \&
  \bibinfo{author}{{Patzer}, A. B.~C.}
\newblock \bibinfo{title}{{FASTCHEM 2 : an improved computer program to
  determine the gas-phase chemical equilibrium composition for arbitrary
  element distributions}}.
\newblock \emph{\bibinfo{journal}{\mnras}} \textbf{\bibinfo{volume}{517}},
  \bibinfo{pages}{4070--4080} (\bibinfo{year}{2022}).
\newblock \eprint{2206.08247}.

\bibitem{bell1987}
\bibinfo{author}{{Bell}, K.~L.} \& \bibinfo{author}{{Berrington}, K.~A.}
\newblock \bibinfo{title}{{Free-free absorption coefficient of the negative
  hydrogen ion}}.
\newblock \emph{\bibinfo{journal}{Journal of Physics B Atomic Molecular
  Physics}} \textbf{\bibinfo{volume}{20}}, \bibinfo{pages}{801--806}
  (\bibinfo{year}{1987}).

\bibitem{john1988}
\bibinfo{author}{{John}, T.~L.}
\newblock \bibinfo{title}{{Continuous absorption by the negative hydrogen ion
  reconsidered}}.
\newblock \emph{\bibinfo{journal}{\aap}} \textbf{\bibinfo{volume}{193}},
  \bibinfo{pages}{189--192} (\bibinfo{year}{1988}).

\bibitem{Chubb_2020}
\bibinfo{author}{Chubb, K.~L.}, \bibinfo{author}{Tennyson, J.} \&
  \bibinfo{author}{Yurchenko, S.~N.}
\newblock \bibinfo{title}{{ExoMol} molecular line lists {\textendash} {XXXVII}.
  spectra of acetylene}.
\newblock \emph{\bibinfo{journal}{\mnras}} \textbf{\bibinfo{volume}{493}},
  \bibinfo{pages}{1531--1545} (\bibinfo{year}{2020}).
\newblock \urlprefix\url{https://doi.org/10.1093\%2Fmnras\%2Fstaa229}.

\bibitem{McKemmish2019}
\bibinfo{author}{{McKemmish}, L.~K.} \emph{et~al.}
\newblock \bibinfo{title}{{ExoMol molecular line lists - XXXIII. The spectrum
  of Titanium Oxide}}.
\newblock \emph{\bibinfo{journal}{\mnras}} \textbf{\bibinfo{volume}{488}},
  \bibinfo{pages}{2836--2854} (\bibinfo{year}{2019}).
\newblock \eprint{1905.04587}.

\bibitem{McKemmish2016}
\bibinfo{author}{{McKemmish}, L.~K.}, \bibinfo{author}{{Yurchenko}, S.~N.} \&
  \bibinfo{author}{{Tennyson}, J.}
\newblock \bibinfo{title}{{ExoMol line lists - XVIII. The high-temperature
  spectrum of VO}}.
\newblock \emph{\bibinfo{journal}{\mnras}} \textbf{\bibinfo{volume}{463}},
  \bibinfo{pages}{771--793} (\bibinfo{year}{2016}).
\newblock \eprint{1609.06120}.

\bibitem{zhang_2020}
\bibinfo{author}{{Zhang}, M.}, \bibinfo{author}{{Chachan}, Y.},
  \bibinfo{author}{{Kempton}, E. M.~R.}, \bibinfo{author}{{Knutson}, H.~A.} \&
  \bibinfo{author}{{Chang}, W.~H.}
\newblock \bibinfo{title}{{PLATON II: New Capabilities and a Comprehensive
  Retrieval on HD 189733b Transit and Eclipse Data}}.
\newblock \emph{\bibinfo{journal}{\apj}} \textbf{\bibinfo{volume}{899}},
  \bibinfo{pages}{27} (\bibinfo{year}{2020}).
\newblock \eprint{2004.09513}.

\bibitem{line_2013}
\bibinfo{author}{{Line}, M.~R.} \& \bibinfo{author}{{Yung}, Y.~L.}
\newblock \bibinfo{title}{{A Systematic Retrieval Analysis of Secondary Eclipse
  Spectra. III. Diagnosing Chemical Disequilibrium in Planetary Atmospheres}}.
\newblock \emph{\bibinfo{journal}{\apj}} \textbf{\bibinfo{volume}{779}},
  \bibinfo{pages}{3} (\bibinfo{year}{2013}).
\newblock \eprint{1309.6679}.

\bibitem{18OrMi}
\bibinfo{author}{Ormel, C.~W.} \& \bibinfo{author}{Min, M.}
\newblock \bibinfo{title}{{ARCiS} framework for exoplanet atmospheres - the
  cloud transport model}.
\newblock \emph{\bibinfo{journal}{\aap}} \textbf{\bibinfo{volume}{622}},
  \bibinfo{pages}{A121} (\bibinfo{year}{2019}).

\bibitem{20MiOrCh}
\bibinfo{author}{Min, M.}, \bibinfo{author}{Ormel, C.~W.},
  \bibinfo{author}{Chubb, K.}, \bibinfo{author}{Helling, C.} \&
  \bibinfo{author}{Kawashima, Y.}
\newblock \bibinfo{title}{{The ARCiS framework for Exoplanet Atmospheres:
  Modelling Philosophy and Retrieval}}.
\newblock \emph{\bibinfo{journal}{\aap}} \textbf{\bibinfo{volume}{642}},
  \bibinfo{pages}{A28} (\bibinfo{year}{2020}).

\bibitem{20ChMiKa}
\bibinfo{author}{Chubb, K.~L.}, \bibinfo{author}{Min, M.},
  \bibinfo{author}{Kawashima, Y.}, \bibinfo{author}{Helling, C.} \&
  \bibinfo{author}{Waldmann, I.}
\newblock \bibinfo{title}{{Aluminium oxide in the atmosphere of Hot Jupiter
  WASP-43b}}.
\newblock \emph{\bibinfo{journal}{\aap}} \textbf{\bibinfo{volume}{639}},
  \bibinfo{pages}{A3} (\bibinfo{year}{2020}).

\bibitem{Kreidberg2014}
\bibinfo{author}{{Kreidberg}, L.} \emph{et~al.}
\newblock \bibinfo{title}{{A Precise Water Abundance Measurement for the Hot
  Jupiter WASP-43b}}.
\newblock \emph{\bibinfo{journal}{\apjl}} \textbf{\bibinfo{volume}{793}},
  \bibinfo{pages}{L27} (\bibinfo{year}{2014}).
\newblock \eprint{1410.2255}.

\bibitem{22ChMi}
\bibinfo{author}{Chubb, K.~L.} \& \bibinfo{author}{Min, M.}
\newblock \bibinfo{title}{Exoplanet atmosphere retrievals in {3D} using phase
  curve data with {ARCiS}: Application to {WASP-43b}}.
\newblock \emph{\bibinfo{journal}{\aap}} \textbf{\bibinfo{volume}{665}},
  \bibinfo{pages}{A2} (\bibinfo{year}{2022}).

\bibitem{14BlHaMa}
\bibinfo{author}{Blecic, J.} \emph{et~al.}
\newblock \bibinfo{title}{Spitzer observations of the thermal emission from
  {WASP-43b}}.
\newblock \emph{\bibinfo{journal}{\apj}} \textbf{\bibinfo{volume}{781}},
  \bibinfo{pages}{116} (\bibinfo{year}{2014}).

\bibitem{ExoMol_AlO}
\bibinfo{author}{Patrascu, A.~T.}, \bibinfo{author}{Tennyson, J.} \&
  \bibinfo{author}{Yurchenko, S.~N.}
\newblock \bibinfo{title}{{ExoMol molecular linelists: VII: The spectrum of
  AlO}}.
\newblock \emph{\bibinfo{journal}{\mnras}} \textbf{\bibinfo{volume}{449}},
  \bibinfo{pages}{3613--3619} (\bibinfo{year}{2015}).

\bibitem{20TeYuAl}
\bibinfo{author}{Tennyson, J.} \emph{et~al.}
\newblock \bibinfo{title}{The 2020 release of the exomol database: Molecular
  line lists for exoplanet and other hot atmospheres}.
\newblock \emph{\bibinfo{journal}{\jqsrt}} \textbf{\bibinfo{volume}{255}},
  \bibinfo{pages}{107228} (\bibinfo{year}{2020}).

\bibitem{ExoMol_SiO}
\bibinfo{author}{Barton, E.~J.}, \bibinfo{author}{Yurchenko, S.~N.} \&
  \bibinfo{author}{Tennyson, J.}
\newblock \bibinfo{title}{{ExoMol Molecular linelists -- II. The ro-vibrational
  spectrum of SiO}}.
\newblock \emph{\bibinfo{journal}{\mnras}} \textbf{\bibinfo{volume}{434}},
  \bibinfo{pages}{1469--1475} (\bibinfo{year}{2013}).

\bibitem{08Trotta}
\bibinfo{author}{Trotta, R.}
\newblock \bibinfo{title}{Bayes in the sky: Bayesian inference and model
  selection in cosmology}.
\newblock \emph{\bibinfo{journal}{Contemporary Physics}}
  \textbf{\bibinfo{volume}{49}}, \bibinfo{pages}{71--104}
  (\bibinfo{year}{2008}).

\bibitem{15LiTeBu}
\bibinfo{author}{Line, M.~R.}, \bibinfo{author}{Teske, J.},
  \bibinfo{author}{Burningham, B.}, \bibinfo{author}{Fortney, J.~J.} \&
  \bibinfo{author}{Marley, M.~S.}
\newblock \bibinfo{title}{Uniform atmospheric retrieval analysis of ultracool
  dwarfs. {I. C}haracterizing benchmarks, {GI570D} and {HD3651b}}.
\newblock \emph{\bibinfo{journal}{\apj}} \textbf{\bibinfo{volume}{807}},
  \bibinfo{pages}{183} (\bibinfo{year}{2015}).

\bibitem{10Guillot}
\bibinfo{author}{Guillot, T.}
\newblock \bibinfo{title}{On the radiative equilibrium of irradiated planetary
  atmospheres}.
\newblock \emph{\bibinfo{journal}{\aap}} \textbf{\bibinfo{volume}{520}},
  \bibinfo{pages}{A27} (\bibinfo{year}{2010}).

\bibitem{malik2019}
\bibinfo{author}{{Malik}, M.} \emph{et~al.}
\newblock \bibinfo{title}{{Self-luminous and Irradiated Exoplanetary
  Atmospheres Explored with HELIOS}}.
\newblock \emph{\bibinfo{journal}{\aj}} \textbf{\bibinfo{volume}{157}},
  \bibinfo{pages}{170} (\bibinfo{year}{2019}).
\newblock \eprint{1903.06794}.

\bibitem{numpy}
\bibinfo{author}{Harris, C.~R.} \emph{et~al.}
\newblock \bibinfo{title}{Array programming with {NumPy}}.
\newblock \emph{\bibinfo{journal}{Nature}} \textbf{\bibinfo{volume}{585}},
  \bibinfo{pages}{357--362} (\bibinfo{year}{2020}).
\newblock \urlprefix\url{https://doi.org/10.1038/s41586-020-2649-2}.

\bibitem{astropy2013}
\bibinfo{author}{{Astropy Collaboration}} \emph{et~al.}
\newblock \bibinfo{title}{{Astropy: A community Python package for astronomy}}.
\newblock \emph{\bibinfo{journal}{\aap}} \textbf{\bibinfo{volume}{558}},
  \bibinfo{pages}{A33} (\bibinfo{year}{2013}).
\newblock \eprint{1307.6212}.

\bibitem{astropy2018}
\bibinfo{author}{{Astropy Collaboration}} \emph{et~al.}
\newblock \bibinfo{title}{{The Astropy Project: Building an Open-science
  Project and Status of the v2.0 Core Package}}.
\newblock \emph{\bibinfo{journal}{\aj}} \textbf{\bibinfo{volume}{156}},
  \bibinfo{pages}{123} (\bibinfo{year}{2018}).
\newblock \eprint{1801.02634}.

\bibitem{matplotlib}
\bibinfo{author}{Hunter, J.~D.}
\newblock \bibinfo{title}{Matplotlib: A 2d graphics environment}.
\newblock \emph{\bibinfo{journal}{Computing in Science \& Engineering}}
  \textbf{\bibinfo{volume}{9}}, \bibinfo{pages}{90--95} (\bibinfo{year}{2007}).

\end{thebibliography}

\clearpage
\section*{\LARGE Supplementary Information}
\renewcommand{\figurename}{\hspace{-4pt}}
\renewcommand{\thefigure}{Supplementary Fig.~\arabic{figure}}
\renewcommand{\theHfigure}{Supplementary Fig.~\arabic{figure}}
\renewcommand{\tablename}{\hspace{-4pt}}
\renewcommand{\thetable}{Supplementary Table \arabic{table}}
\renewcommand{\theHtable}{Supplementary Table \arabic{table}}
\setcounter{figure}{0}
\setcounter{table}{0}

~

\vspace{-1cm}

\begin{table*}[!htbp]
    \centering
    \small
    \caption{\textbf{Summary of the five GCMs used in this study.}}\label{tab:GCMsummaries}
    \begin{tabular}{l|l|l|l}
    GCM Name      & Radiative Transfer &  Post-processing & References  \\
    \hline
    Generic PCM               & non-grey correlated k    & Pytmosph3R     & \cite{Turbet2021,Spiga2020,charnay14,charnay_3d_2015,turbet_habitability_2016}      \\
    SPARC/MITgcm                          & non-grey correlated k      & gCMCRT              &   \cite{showman2009,parmentier2013,tan2021a,tan2021b}    \\
    expeRT/GCM          &   non-grey correlated k  & petitRADTRANS  & \cite{Adcroft2004,Carone2020,Schneider2022}      \\
    RM-GCM               & double-grey    &   unnamed\cite{Zhang2017, Malsky2021}    &   \cite{MenouRauscher2009,RauscherMenou2010,RauscherMenou2012,Roman&Rauscher2017,Roman2021}      \\
    THOR                               & double-grey   & HELIOS      & \cite{mendonca2015,mendonca2016,mendonca2018a,malik2017,malik2019,deitrick2020}
    \end{tabular}
\end{table*}

\vspace{2cm}

\begin{figure*}[!htbp]
    \centering
    \includegraphics[width=\linewidth]{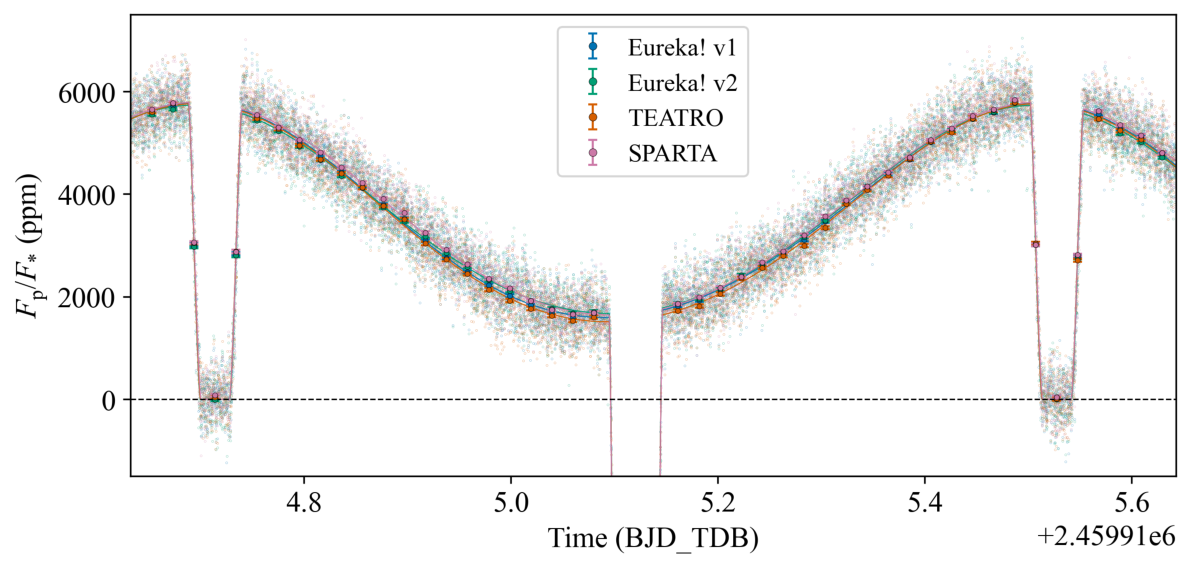}
    \caption{\textbf{Broadband light curve obtained from the four independent reductions.} Each colour indicates a different reduction. Data points at the original time sampling are shown as small open circles and a binning with 40 points per orbital period (170 integrations per bin, ${\sim}$30 minute sampling) is shown as filled circles, computed using the biweight\textunderscore location function from astropy. Thin lines show the phase curve model. The 1$\sigma$ uncertainties in each bin are obtained from the standard deviation of the residuals (data $-$ model) divided by the square root of the number of points in that bin (170 integrations). The flux measured during the eclipse, which is the stellar flux only, is used as a reference and is shown as a dashed horizontal line.}\label{fig:broadband_all}
\end{figure*}

\begin{figure*}
    \centering
    \includegraphics[width=0.8\linewidth]{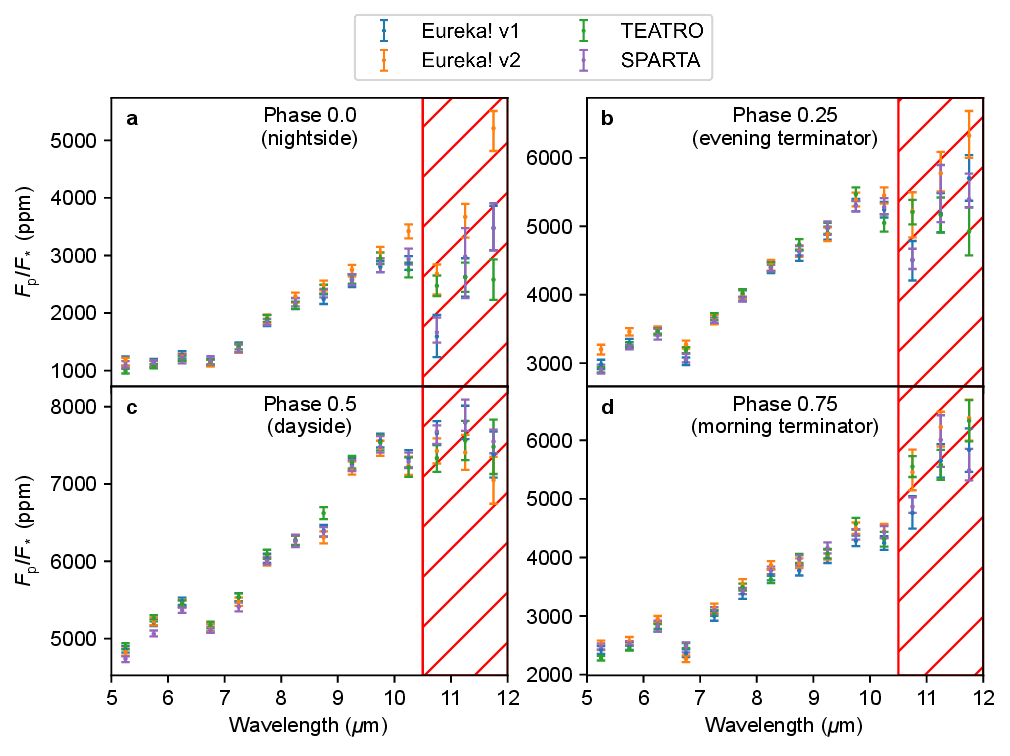}
    \includegraphics[width=0.45\linewidth, trim=0 0 0 40, clip]{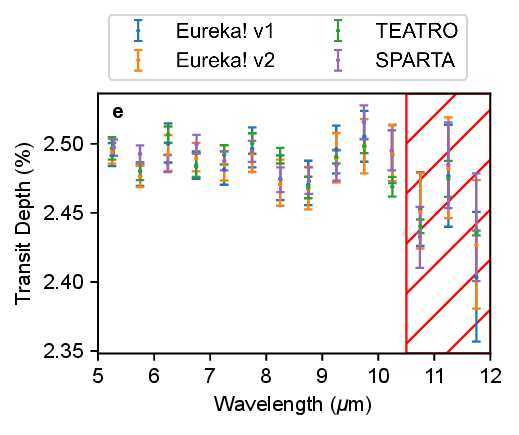}
    \caption{\textbf{Comparison of the phase-resolved and transmission spectra from different reductions.} Panels \textbf{a--d} show the phase resolved emission spectra from our four reductions with 1$\sigma$ error bars, and panel \textbf{e} shows each of our transmission spectra with 1$\sigma$ error bars. In general, there is good agreement about the phase-resolved spectra between our four semi-independent reductions. Larger differences arise $>$10.5\,$\mu$m due to the ``shadowed region effect'' (indicated with red hatching). The transmission spectrum appears flat (within uncertainties) and shows no significant differences between reduction methods.}\label{fig:spectrumComparison}
\end{figure*}

\end{document}